\newif\iffullversion
\fullversiontrue

\ifdefined\authorNote
\newcommand{\elaine}[1]{{\footnotesize\color{magenta}[Elaine: #1]}}
\newcommand{\hao}[1]{{\footnotesize\color{blue}[Hao: #1 N.K.]}}
\newcommand{\ke}[1]{{\footnotesize\color{red}[Ke: #1 QOD]}}
\else
\newcommand{\elaine}[1]{}
\newcommand{\hao}[1]{}
\newcommand{\ke}[1]{}
\fi  

\documentclass[11pt,letterpaper]{article}
\usepackage[paper=letterpaper,margin=1in]{geometry}
\usepackage[margin=1in]{geometry}
\usepackage{amsthm}

\usepackage{cite}

\usepackage[bookmarks=true,pdfstartview=FitH,colorlinks,linkcolor=darkred,filecolor=darkred,citecolor=darkred,urlcolor=darkred]{hyperref}

\usepackage{multicol}

\usepackage{colortbl}
\usepackage{mdframed}
\usepackage{color,soul}

\usepackage{paralist}
\usepackage{pdfpages}
\usepackage{enumitem}
\usepackage{float}
\usepackage{booktabs}
\usepackage[override]{cmtt}
\usepackage{color,xcolor}
\usepackage{authblk}
\usepackage{graphicx,color,eso-pic}

\usepackage[
	lambda,
	advantage,
	operators,
	sets,
	adversary,
	landau,
	probability,
	notions,
	logic,
	ff,
	mm,
	primitives,
	events,
	complexity,
	asymptotics,
	keys]
{cryptocode}

\usepackage{tikz}
\usepackage{pgfplots}
\usepgfplotslibrary{fillbetween}
\usetikzlibrary{fit, shapes, positioning, patterns, arrows,shadows}
\usetikzlibrary{decorations.pathreplacing,calligraphy}

\tikzset{
    master/.style={
        execute at end picture={
            \coordinate (lower right) at (current bounding box.south east);
            \coordinate (upper left) at (current bounding box.north west);
        }
    },
    slave/.style={
        execute at end picture={
            \pgfresetboundingbox
            \path (upper left) rectangle (lower right);
        }
    }
}

\sethlcolor{lightgray}
\newcommand{\mcal}[1]{\ensuremath{\mathcal {#1}}}

\newcommand{\fmpc}{{{\ensuremath{\mcal{F}_{\textrm{mpc}}}}}\xspace}

\newcommand{\R}{\ensuremath{\mathbb{R}}}

\newcommand{\E}{\mathbf{E}}

\newcommand{\eps}{\epsilon}

\definecolor{darkgreen}{rgb}{0,0.5,0}
\definecolor{lightblue}{RGB}{0,176,240}
\definecolor{darkblue}{RGB}{0,112,192}
\definecolor{lightpurple}{RGB}{124, 66, 168}
\definecolor{grey}{RGB}{139, 137, 137}
\definecolor{maroon}{RGB}{178, 34, 34}
\definecolor{green}{RGB}{34, 139, 34}
\definecolor{types}{RGB}{72, 61, 139}
\definecolor{gold}{rgb}{0.8, 0.33, 0.0}
\definecolor{mygreen}{HTML}{588D6A}
\definecolor{myred}{HTML}{C86733}
\definecolor{myblue}{HTML}{5B68FF}
\definecolor{myshadow}{HTML}{E6C5B4}

\definecolor{darkgray}{gray}{0.3}

\newcommand{\skiptext}[1]{}
\usepackage{boxedminipage}
\usepackage{url}
\usepackage{graphicx}
\usepackage[bf,justification=centering]{caption}
\usepackage[justification=centering]{subcaption}
\usepackage{color}
\usepackage{xspace}
\usepackage{multirow}
\usepackage{amsthm,amstext}
\usepackage{amsmath,amstext,amsfonts,amssymb,latexsym}
\usepackage{amsbsy}
\usepackage{stmaryrd}
\usepackage{pifont}

\usepackage[capitalize,noabbrev]{cleveref} %% for referencing use \cref

\usepackage{wrapfig}
\definecolor{darkred}{rgb}{0.5, 0, 0}
\definecolor{darkgreen}{rgb}{0, 0.5, 0}
\definecolor{darkblue}{rgb}{0,0,0.5}

\newcommand\markx[2]{}

\renewcommand{\path}{\ensuremath{\mathsf{path}}\xspace}

\newcommand{\ignore}[1]{}

\renewcommand{\part}{\ensuremath{\mcal{S}}\xspace}

\newcounter{task}

\newtheorem{thm}{Theorem}[section]      % A counter for Algorithms etc

\newtheorem{theorem}[thm]{Theorem}

\newtheorem{lemma}[thm]{Lemma}
\newtheorem{claim}[thm]{Claim}
\newtheorem{corollary}[thm]{Corollary}
\newtheorem{fact}[thm]{Fact}

\theoremstyle{definition}
\newtheorem{definition}[thm]{Definition}
\newtheorem{remark}[thm]{Remark}

\newtheoremstyle{boxes}% name  
{2pt}% ?Space above ? 
{0pt}% ?Space below ?
{}% ?Body font ?
{}% ?Indent amount ?
{\bfseries}% ?Theorem head font?
{}% ?Punctuation after theorem head ?
{\newline}% ?Space after theorem head ?
{\thmname{#1}\thmnumber{ #2}:  
\thmnote{#3}}

\theoremstyle{boxes}
\newcommand{\bids}{{\bf b}}

\newcommand{\m}{\ensuremath{m}}

\newcounter{cnt:challenge}

\newcommand{\bfb}{{\bf b}}

\newcommand{\bico}{\gamma}

\newcommand{\threshold}{\widehat{\bf y}}
\newcommand{\correction}{\boldsymbol{\delta}}
\newcommand{\MatA}{A}
\newcommand{\err}{{\bf e}}

\newcommand{\correct}{z}

\newcommand{\thresh}{t}
\newcommand{\target}{\overline{\mu}}
\newcommand{\environ}{(h,\rho,c,d)}

\iffullversion
\title{Maximizing Miner Revenue in
Transaction Fee Mechanism Design\footnote{Author order is randomized.}}
\author{Ke Wu\thanks{{\tt kew2@andrew.cmu.edu}}}
\author{Elaine Shi\thanks{{\tt runting@gmail.com}}}
\author{Hao Chung\thanks{\tt haochung@andrew.cmu.edu}}
\affil{Computer Science Department, Carnegie Mellon University}
\date{}

\else
\author{}
\date{}
\fi

\begin{document}

\iffullversion
\begin{titlepage}
\maketitle
\begin{abstract}
Transaction fee mechanism design is a new decentralized mechanism design
problem where users bid for space on the blockchain.  
Several recent works showed that the transaction fee mechanism
design fundamentally departs from classical mechanism design.
They then systematically explored the mathematical landscape 
of this new decentralized mechanism design problem
in two settings: in the plain setting  
where no cryptography is employed, and in a cryptography-assisted
setting where the rules of the mechanism
are enforced by a multi-party computation protocol.
Unfortunately, in both settings,  
prior works showed that if we want the mechanism to incentivize honest behavior 
for both users as well as miners (possibly colluding with users), 
then the miner revenue has to be zero.
Although adopting a relaxed, approximate notion of 
incentive compatibility gets around this zero
miner-revenue limitation, the scaling of the miner revenue   
is nonetheless poor.

In this paper, we show that if we make a mild reasonable-world assumption that there are sufficiently many honest users,  
we can circumvent the known limitations on miner revenue, 
and design auctions that  
generate asymptotically optimal miner revenue.
We also systematically explore the mathematical landscape
of transaction fee mechanism design under the new reasonable-world assumptions, 
and demonstrate how such assumptions can alter the 
feasibility and infeasibility landscape.

\end{abstract}
\thispagestyle{empty}
\end{titlepage}

\else
\maketitle
\begin{abstract}
Transaction fee mechanism design is a new decentralized mechanism design
problem where users bid for space on the blockchain.  
Several recent works showed that the transaction fee mechanism
design fundamentally departs from classical mechanism design.
They then systematically explored the mathematical landscape 
of this new decentralized mechanism design problem
in two settings: in the plain setting  
where no cryptography is employed, and in a cryptography-assisted
setting where the rules of the mechanism
are enforced by a multi-party computation protocol.
Unfortunately, in both settings,  
prior works showed that if we want the mechanism to incentivize honest behavior 
for both users as well as miners (possibly colluding with users), 
then the miner revenue has to be zero.
Although adopting a relaxed, approximate notion of 
incentive compatibility gets around this zero
miner-revenue limitation, the scaling of the miner revenue   
is nonetheless poor.

In this paper, we show that if we make a mild reasonable-world assumption that there are sufficiently many honest users,  
we can circumvent the known limitations on miner revenue, 
and design auctions that  
generate asymptotically optimal miner revenue.
We also systematically explore the mathematical landscape
of transaction fee mechanism design under the new reasonable-world assumptions, 
and demonstrate how such assumptions can alter the 
feasibility and infeasibility landscape.

\end{abstract}
\fi

\iffullversion
\tableofcontents
\clearpage
\fi

\newpage

\section{Introduction}
\label{sec:intro}
The transaction fee mechanism (TFM)~\cite{zoharfeemech,yaofeemech,functional-fee-market,eip1559,roughgardeneip1559,roughgardeneip1559-ec,dynamicpostedprice,foundation-tfm,greedy-tfm,bayesian-tfm} is a 
new decentralized mechanism design problem that arises
in a blockchain environment.  
Since the space on the blockchain is scarce, users 
must bid to get their transactions included and confirmed  
whenever a new block is minted.
Earlier works observed that 
mechanism design in a decentralized environment
departs fundamentally from classical mechanism design~\cite{zoharfeemech,yaofeemech,functional-fee-market,eip1559,roughgardeneip1559,roughgardeneip1559-ec,dynamicpostedprice,foundation-tfm,greedy-tfm,bayesian-tfm}.
The majority of classical 
mechanisms 
assume that the auctioneer is trusted 
and will honestly implement the prescribed mechanism. 
However, in a decentralized environment, the auction is implemented
by a set of miners or consensus nodes\footnote{Throughout the paper, we call
the consensus nodes ``miners'' regardless of whether
the consensus protocol uses proof-of-work or proof-of-stake.}
who are incentivized to take advantage of 
profitable deviations (if there are any) rather than implementing the mechanism honestly. 
%who may deviate from the honest mechanism
%if there exist profitable deviations.
%who may not act honestly if deviating can increase its profit.
%who are incentivized to act selfishly to maximize their own profit.
As a simple example, while the Vickrey auction~\cite{vickrey1961counterspeculation}
(a.k.a., the second-price
auction) is considered
an awesome auction by classical standards, it is not a great  
fit for a decentralized environment as explained below. 
Suppose we confirm $k$ bids and they all
pay the $(k+1)$-th price to the miner. Then the miner 
could inject a fake bid that is slightly 
less than the $k$-th price, thus causing confirmed bids to pay 
essentially the $k$-th price.
Alternatively, the same effect can also be achieved
if the miner colludes with the $(k+1)$-th bidder
and asks it to raise its bid to almost exactly the $k$-th price.
The coalition can then 
split off the additional gains off the table, using {\it binding} side contracts
that can be instantiated through 
the decentralized smart contracts that are available in blockchain environments.

This drives us to rethink what is a ``dream'' TFM in the absence of a fully trusted  
auctioneer.
Recent works~\cite{eip1559,roughgardeneip1559,roughgardeneip1559-ec,foundation-tfm}
formulated the following desiderata:
\begin{itemize}[leftmargin=5mm,itemsep=1pt] 
\item {\it User incentive compatibility (UIC)}: 
a user's best strategy is to bid truthfully, even when the user has observed others' bids.  
\item {\it Miner incentive compatibility (MIC)}: 
the miner's best strategy is to implement the mechanism honestly, even when 
the miner has observed all users' bids. 
\item {\it Side-contract-proofness (SCP)}: 
playing honestly maximizes the joint utility of a  
coalition consisting of the miner and at most $c$ users, even after having observed all others' bids.\footnote{The formal definition of joint utility is given in \cref{sec:MPC-model}.}
\end{itemize} 

%In the {\it plain model} without cryptography, 
Roughgarden showed
that Ethereum's EIP-1559 mechanism can simultaneously achieve all three properties, 
as long as the block size is {\it infinite}~\cite{eip1559,roughgardeneip1559}.
In practice, EIP-1559 tries to be in the ``infinite block size'' regime by  
estimating a reserve price based on the recent history.
The reserve price is a minimum threshold used to filter the transactions 
to avoid congestion.
For the case of finite block size (i.e., when congestions do occur),  
Chung and Shi~\cite{foundation-tfm}
showed that it is impossible to satisfy all three properties at the same time
without the use of cryptography. 
However, the subsequent work of Shi, Chung, Wu~\cite{crypto-tfm}
shows that we can use the {\it MPC-assisted model} to circumvent this impossibility.
In the MPC-assisted model, 
the TFM's rules are securely enforced
by a multi-party computation protocol among the miners, thus taking
away the miner's ability to unilaterally decide which transactions to include
in the block.
Variants of the MPC-assisted model are being developed by mainstream
blockchain projects such as Ethereum. In particular, the community has been
making an effort 
to build ``encrypted  mempools''~\cite{encmempool}, which can be viewed
as a concrete instantiation of the MPC-assisted model\footnote{Exactly whether 
encrypted mempool realizes the 
MPC ideal functionality needed by Shi, Chung, Wu~\cite{crypto-tfm} 
requires a rigorous proof.}.
Under the MPC-assisted model, \cite{crypto-tfm} showed that 
one can indeed construct a TFM that simultaneously satisfies
UIC, MIC, and SCP (for $c = 1$) under finite block size.
\vspace{-5pt}
\paragraph{Limit on miner revenue.}
Unfortunately, %all prior work
no matter whether in the plain or in the MPC-assisted model, 
all these prior works~\cite{roughgardeneip1559,foundation-tfm,crypto-tfm}
suffer from a ``zero-miner-revenue limitation'':
all the payment from the users is burnt\footnote{We stress that the miners can still get a fixed
block reward which incentivizes them to mine --- in fact, Ethereum's EIP-1559
burns all base fees and pays the miner a fixed block reward.
This fixed block reward
does not affect the game-theoretic analysis and thus
is typically ignored in the game-theoretic modeling~\cite{roughgardeneip1559,foundation-tfm,crypto-tfm}.} 
and the miner obtains zero revenue.
The works of Chung and Shi~\cite{foundation-tfm}
and Shi, Chung, Wu~\cite{crypto-tfm}
proved that this limitation is
in fact inherent.
Specifically, any TFM (either in the plain model or MPC-assisted model)
that simultaneously satisfies UIC, MIC, and SCP (even for $c = 1$)
must have zero miner revenue.
%, i.e., all payment from the users
%must be burnt\footnote{We stress that the miners can still get a fixed
%block reward which incentivizes them to mine --- in fact, Ethereum's EIP-1559
%burns all base fees and pays the miner a fixed block reward. 
%This fixed block reward
%does not affect the game-theoretic analysis and thus 
%is typically ignored in the game-theoretic modeling~\cite{roughgarden-eip1559,foundation-tfm,crypto-tfm}.} rather than paid to the miner.
In both the plain and the MPC-assisted models, the zero miner-revenue limitation
holds in a very strong sense: 
regardless of whether the block size is finite or infinite, and even  
when the miner colludes with at most $c = 1$ user.

Moreover, \cite{crypto-tfm}
additionally explored whether relaxing the {\it strict} incentive 
compatibility notion to {\it approximate} incentive compatibility
can increase the miner revenue. 
They show a somewhat pessimistic, 
{\it unscalability of miner revenue} result. Specifically,
with $\epsilon$-incentive compatibility, 
%obtain only some function of $\epsilon$ revenue from each bidder --- in this sense, 
the miner revenue cannot enjoy linear scaling w.r.t.~the  
magnitude of the bids.
Given the landscape, 
we ask the following natural question:
\begin{itemize}[leftmargin=6mm]
\item[] 
{\it 
%Can we introduce some reasonable-world assumptions to 
Can we circumvent
the severe limitation on miner revenue, under some reasonable-world assumptions?
%How can we maximize the miner revenue in transaction fee mechanism
%design under reasonable assumptions?
}
\end{itemize}

\subsection{Our Results and Contributions}
We are inspired by the philosophy 
adopted by a line of work at the intersection of cryptography
and game theory~\cite{gtcrypto00,gtcrypto01,gtcrypto02,gtcrypto03,giladutilityindjournal,giladgtcrypto,rdp00,rdp01,rdp02,katzgametheory,gtcrypto06,seqrationalcrypto,gt-fair-cointoss,gt-fair-coin-complete,gt-leader-shi,fruitchain,logstar-gt-leader,credibleauction-comm00,credibleauction-comm01}.
In these works, the game theoretic properties hold 
as long as sufficiently many players are honest. 
Because the game theoretic guarantees ensure that 
honest behavior is an equilibrium, and that players are incentivized to behave honestly, this in turn  
reinforces the ``sufficient honesty'' assumption.

Therefore, we ask whether 
we can overcome the severe limitation on miner revenue
also under some type of ``sufficient honesty'' assumption.
Phrasing the precise ``sufficient honesty'' assumption, however, turns out to be 
technically subtle,  
partly because TFMs must work in an {\it open} setting
where anyone can post a bid,  
and the mechanism is unaware of the number of bids a-priori. 
%Figuring out what reasonable assumptions 
%can circumvent the severe limitation on miner revenue 
%turns out to be rather challenging and technically subtle, 
%partly because the zero miner revenue limitation holds
%in a very strong sense. 
One na\"ive attempt is to assume that among the 
bids posted, half of them come from honest users. 
Unfortunately, this approach does not work. 
In \cref{section:revenue-bound-honest-majority}, we show that 
even under such an ``honest majority bids'' assumption, 
we would still suffer from an $O(1)$-miner revenue limitation.
%(see \cref{section:revenue-bound-honest-majority} \elaine{refer} for an explanation).
\elaine{TODO: write this}

\paragraph{Reasonable-world assumption: known lower bound on the number of honest users.}
Instead of the ``honest majority bids'' assumption, we make a subtly different assumption ---  
we assume that there is an a-priori known 
lower bound $h$ on the number of honest users. 
Note that this assumption also promises that at least $h$ users 
will show up.
%because
%the TFM must work in an open setting without knowing the number of bids a-priori,
%the known-$h$ assumption is different from assuming honest majority of posted bids. 
We refer to this as the \emph{known-$h$ model}.
In this model, we first observe that the zero miner-revenue
limitation no longer holds. 
Instead, we can  prove an $O(h)$-limit
on the miner revenue as stated in the following theorem.

\begin{theorem}[Informal: limit on miner revenue in the known-$h$ model]
\label{thm:theta-h-minerrev-informal}
In the known-$h$ model, 
no MPC-assisted mechanism that simultaneously satisfies UIC, MIC, and SCP 
(even in the Bayesian setting)
can achieve more  
than $h \cdot \E(\mcal{D})$ expected miner revenue where 
$\E(\mcal{D})$ denotes the expectation of the value distribution $\mcal{D}$.

More generally, in the known-$h$ model,  
if the number of users is $n$,
no MPC-assisted mechanism that simultaneously satisfies $\epsilon$-UIC, $\epsilon$-MIC, and $\epsilon$-SCP (even in the Bayesian setting)
can achieve more  
than $h \cdot \E(\mcal{D}) + \frac{2(n-h)}{\rho}\left(\epsilon+C_{\mcal{D}}\sqrt{\epsilon}\right)$ expected miner revenue,
% than $h \cdot \E(\mcal{D}) + 2(n-h)\left(\epsilon+C'_{\mcal{D}}\sqrt{\epsilon}\right)$ expected miner revenue,
where 
$\rho$ is an upper bound on the fraction of miners 
controlled by the strategic 
coalition, and 
$C_{\mcal{D}} = \E_{X\sim\mcal{D}}[\sqrt{X}]$.

Furthermore, the above limitation holds no matter when the block size is finite or infinite,
and even when the miners collude 
with at most $c = 1$ user.
\label{thm:intro-limit-rev}
%\elaine{FILL: state also for approx IC too}
\end{theorem}

In the above theorem, $\epsilon \geq 0$
is a parameter that measures the slack in the incentive compatibility notion.
When $\epsilon = 0$, there is no slack, and we achieve strict incentive compatibility.
One informal interpretation of the above theorem is the following: 
for $\epsilon$ incentive compatibility,
\Cref{thm:intro-limit-rev}
allows us to  hope for a mechanism where  
roughly speaking, 
from each of $h$ users, the miners can hope to get 
$\E(\mcal{D})$ revenue
which scales
proportionally w.r.t.~to the bid distribution $\mcal{D}$. 
For each of the remaining users, the miners can 
potentially get some function that depends on $\epsilon$ and
the bid distribution $\mcal{D}$, but the term does not scale linearly w.r.t.
the magnitude of the bid distribution for natural distributions.

\ignore{
the miner can potentially get $\Theta(\E[\mcal{D}])$ revenue 
from each of $h$ users, 
and from each of the remaining users, the miner 
can potentially get some function that depends on $\epsilon$ and  
the bid distribution $\mcal{D}$, but it does not scale linearly w.r.t.
the magnitude of the bid distribution.
Importantly, the former part (where the miner gets $\Theta(\E[\mcal{D}])$ from
each of $h$ users) 
scales w.r.t.~both the bid magnitude and $h$ itself.
}

\ignore{
So, can we indeed design a mechanism in the known-$h$ model 
that achieves asymptotically optimal miner revenue matching
\Cref{thm:intro-limit-rev}?
}
The above \Cref{thm:intro-limit-rev} allows us to hope
for a TFM in the known-$h$ model that achieves
revenue that scales with $h$ as well as the magnitude of the bid distribution $\mcal{D}$. 
So can we indeed design a mechanism with 
asymptotically optimal mine revenue 
matching 
\Cref{thm:intro-limit-rev}?

\vspace{-5pt}
\paragraph{Mechanisms for infinite block size.}
%We answer the question affirmatively by presenting two mechanisms 
For the infinite block size regime, we propose two 
mechanisms in the MPC-assisted model:
\begin{itemize}[leftmargin=5mm,itemsep=1pt]
\item 
The first one, called {\it threshold-based mechanism}, is a simple
and practical mechanism 
that satisfies almost-strict incentive compatibility
except for a tiny slack $\epsilon$
that is exponentially small in $h$. 
\item 
The second one, called {\it LP-based mechanism} (since it uses linear programming), is 
a result of theoretical interest. It achieves
{\it strict} incentive compatibility, but under one extra assumption (besides
a-priori knowledge of $h$),
that the number of fake bids injected by the strategic coalition 
is bounded.
\end{itemize}
Both mechanisms achieve asymptotically optimal miner revenue\footnote{We achieve asymptotic optimal miner revenue w.r.t. $h$ assuming that the expectation and median of the distribution $\mcal{D}$ is a constant independent of $h$.}
w.r.t.~\Cref{thm:intro-limit-rev}.
We assume that only honest users' true values are i.i.d. sampled from some distribution $\mcal{D}$, whereas the strategic users' true values can be \emph{arbitrary} non-negative real numbers.
Next, we state the corresponding theorems 
for the two mechanisms below:

\begin{theorem}[Informal: threshold-based mechanism]
%\elaine{FILL}
Suppose that honest users' values are sampled i.i.d.~from some distribution $\mcal{D}$.
Then, 
there exists an MPC-assisted TFM in the known-$h$ model that satisfies 
ex post UIC, Bayesian $\epsilon$-MIC, and 
Bayesian $\epsilon$-SCP (for any number of colluding users)
for $\epsilon = O_\mcal{D}(\exp(-\Omega(h)))$
where $O_\mcal{D}(\cdot)$ 
hides terms related to the value distribution $\mcal{D}$.
Furthermore, the expected total miner revenue 
$\Theta(h) \cdot {\sf median}(\mcal{D})$.
\label{thm:intro-thresh}
\end{theorem}

Essentially, from each of $h$ users, 
the miners can obtain revenue that scales
linearly w.r.t.~both the bid magnitude.
%We show that such miner revenue is asymptotically optimal 
%except for factors negligibly small in $h$ in \elaine{refer}. 
By contrast, 
without the known-$h$ assumption, 
for our choice of $\epsilon$ which is exponentially small in $h$, 
the miner revenue must be exponentially small in $h$ as shown
in prior work~\cite{crypto-tfm}.
Observe also that the miner revenue is asymptotically optimal
up to additive factors that are exponentially small in $h$ 
due to \Cref{thm:intro-limit-rev}.

\begin{theorem}[Informal: LP-based mechanism]
\label{thm:lp-based-informal}
Suppose that honest users' values are sampled i.i.d.~from some distribution $\mcal{D}$.
Then, 
there exists an MPC-assisted TFM that satisfies 
ex post UIC, Bayesian MIC, and Bayesian SCP 
where the MIC and SCP guarantees hold as long as 
the total number of bids $d$
contributed by the strategic coalition  
satisfies  $d \leq \frac18\sqrt{\frac{h}{2\log h}}$.
Further, the expected total miner revenue is $\Theta(h) \cdot {\sf median}(\mcal{D})$.
\label{thm:intro-LP}
\end{theorem}

In the above theorem, 
we need the extra assumption that the strategic coalition does not control
too many bids. Effectively, this is assuming that the coalition cannot inject
too many fake bids.
Currently, we do not know whether this extra 
assumption (besides known-$h$) is needed
to overcome the zero miner-revenue limitation. 
We leave this as an interesting open question.

\ignore{
Below we elaborate on these results.

\paragraph{Threshold-based mechanism.}
We propose a threshold-based mechanism in the MPC-assisted model,
where the rules of the mechanism are securely enforced 
by a multi-party computation protocol among the miners. 
The mechanism is simple. 
Suppose that honest users' bids 
are sampled i.i.d.~from some distribution $\mcal{D}$
whose median is $m$.
Then, all transactions that bid at 
$m$ will get confirmed 
}

\paragraph{Mechanisms for finite block size.}
For the finite block size case, 
we propose two mechanisms:
\begin{itemize}[leftmargin=5mm]
\item 
We propose a simple mechanism called %\elaine{FILL}
{\it diluted threshold-based mechanism} 
that achieves 
{\it approximate} incentive compatibility. Further, for sufficiently large $h$, the 
mechanism achieves  
asymptotically optimal 
miner revenue. 

\item 
For theoretical interest, we propose another mechanism called {\it LP-based mechanism with random selection} 
which achieves {\it strict} incentive compatibility
and asymptotically optimal miner revenue --- %but we additionally need to assume
but under the additional assumptions that 
the coalition cannot inject too many fake bids, and moreover, 
the miners collude with at most $c = 1$ user.
Jumping ahead, the $c=1$ assumption will be later justified in \cref{thm:zero-user-welfare-informal}.
\end{itemize}

We state the corresponding theorems 
for the two mechanisms below:

\begin{theorem}[Informal: diluted threshold-based mechanism]
\label{thm:diluted-informal}
  Suppose the block size is $k$,
  and that honest users' values are sampled i.i.d.~from some 
bounded distribution $\mcal{D}$.
  Then, 
  there exists an MPC-assisted TFM in the known-$h$ model that satisfies 
  ex post UIC, Bayesian $\epsilon$-MIC, and 
  Bayesian $\epsilon$-SCP (for any number of colluding users)
  for $\epsilon = O_\mcal{D}(\exp(-\Omega(h)))$.
  %where $O_\mcal{D}(\cdot)$ 
  %hides terms related to the value distribution $\mcal{D}$.
Furthermore, for sufficiently large $h$, 
the mechanism achieves
  expected total miner revenue 
  $\Theta(k) \cdot {\sf median}(\mcal{D})$.
%, where $\Theta_\mcal{D}(\cdot)$ 
%  hides terms related to the value distribution $\mcal{D}$.
\label{thm:intro-diluted-thresh}
\end{theorem}

\begin{theorem}[Informal: LP-based mechanism with random selection]
\label{thm:lp-random-sel-informal}
  Suppose the block size is k,
  and suppose that honest users' values are sampled i.i.d.~from some distribution $\mcal{D}$.
  Then, 
  there exists an MPC-assisted TFM that satisfies 
  ex post UIC, Bayesian MIC, and Bayesian SCP, where the MIC and SCP guarantees hold
when 1) at most $c = 1$ user colludes when miners, and 2)  
  the total number of bids $d$
  contributed by the strategic coalition  
  satisfies  $d \leq \frac18\sqrt{\frac{h}{2\log h}}$.
  Further, 
the expected total miner revenue is $\Theta(\min\{h,k\}) \cdot {\sf median}(\mcal{D})$.
\label{thm:intro-diluted-LP}
\end{theorem}

We justify the $c=1$ assumption 
in the LP-based mechanism with random selection
by proving the following impossibility result: for finite block size,
no ``interesting'' mechanism can simultaneously achieve
UIC, MIC, and SCP 
for $c \geq 2$ even in the MPC-assisted model. Specifically,
\begin{theorem}[Informal: finite block, $c \geq 2$]
  \label{thm:zero-user-welfare-informal}
%  Suppose that the block size is finite,
%  and fix any $h \geq 1$, any $c \geq 2$.
Even in the known-$h$ model, 
any MPC-assisted TFM that
  simultaneously satisfies Bayesian UIC, Bayesian MIC, and Bayesian SCP for $c \geq 2$ 
must suffer from $0$ 
expected social welfare for the users under a bid vector $\bids \sim \mcal{D}^\ell$ 
where $\ell > h$.
\end{theorem}

\paragraph{Necessity of Bayesian equilibrium.}
All of our feasibility results, namely, \cref{thm:intro-thresh,thm:intro-LP,thm:intro-diluted-thresh,thm:intro-diluted-LP}, 
rely on a Bayesian notion of equilibrium (for the MIC and SCP guarantees).
As argued by \cite{crypto-tfm}, the Bayesian notion of equilibrium 
is suitable for the MPC-assisted model since the users
cannot observe others' bids before submitting 
their own.

We show that the reliance on Bayesian 
notions of equilibrium is necessary (see \cref{sec:expost}) --- 
had we insisted on an {\it ex post} notion of equilibrium in the MPC-assisted model,
our additional reasonable-world assumptions
would not help us overcome the 
previously known impossibility results. 
More specifically, we show that 
any MPC-assisted mechanism that simultaneously achieves ex post UIC 
and SCP 
must suffer from zero miner revenue even in the known-$h$ model. 
%in an $(h, \rho, c, d)$-environment must suffer from $0$ 
%miner revenue even for $d = c = 1$
%and an arbitrarily small positive $\rho$.
Similarly, 
for approximate but ex post notions of  incentive compatibility, 
the same miner revenue limitation 
stated in \cite{crypto-tfm} still applies
even in the known-$h$ model.
Further, the above restrictions on miner revenue hold
no matter whether the block size is finite or infinite.

%\paragraph{Philosophical discussions about Our assumptions.}
\subsection{Philosophical Discussions about Our Assumptions and Modeling}

\paragraph{Known-$h$ assumption.}
Our assumption about a known upper bound $h$ on the number of honest users has the following justifications:
\begin{itemize}[leftmargin=5mm,itemsep=0pt]
\item 
First, as mentioned, the zero miner-revenue limitation  
in earlier works holds 
in a very strong sense, and even when we make a cryptographic style
assumptions such as ``a majority of the users or bids are honest'' --- 
see \cref{section:revenue-bound-honest-majority} for more details.
\item 
Second, 
TFMs must work in an {\it open setting} where anyone can post a bid and the mechanism
or players do not know the number of bids in advance.
This is also an important reason why TFMs depart from classical mechanism design.
The precise assumption we need for circumventing the zero miner revenue limitation
is an absolute lower bound $h$ on the number of honest users. 
Simply assuming that the majority of the bids are honest is 
not sufficient (see \Cref{section:revenue-bound-honest-majority}). 
% specifically, 
% if there are only three bids, then only two of them are guaranteed to be honest.
%if only three users show up, then only two of them are guaranteed to be honest. 
\item 
Finally, our definitional framework ensures 
that {\it honest behavior is an equilibrium}, and thus players are incentivized
to behave honestly. This, in turn, reinforces the $h$-honest users assumption  
(as long as enough users show up). As mentioned earlier, the same philosophy 
has been adopted in a line of prior works 
at the intersection of game theory and cryptography~\cite{gtcrypto00,gtcrypto01,gtcrypto02,gtcrypto03,giladutilityindjournal,giladgtcrypto,rdp00,rdp01,rdp02,katzgametheory,gtcrypto06,seqrationalcrypto,gt-fair-cointoss,gt-fair-coin-complete,gt-leader-shi,fruitchain,logstar-gt-leader,credibleauction-comm00,credibleauction-comm01}.
\end{itemize} 

\vspace{-5pt}
\paragraph{Limited fake bids.}
Recall that besides the known-$h$ assumption, 
our results that achieve strict incentive compatibility
require an extra assumption that the number of fake bids
injected by the coalition is limited.
%This justifies our assumption that the number of fake bids is limited.
This assumption is motivated and justified by the following observations.
To submit fake bids, the strategic player 
or coalition 
needs to have some coin or account with a non-zero balance.
Given that the strategic player has a limited initial budget, 
it cannot control infinitely many accounts.
Moreover, if one posts multiple conflicting 
transactions double-spending the same coin or units of currency, 
they can easily be detected and suppressed. 

As mentioned, we currently do not know whether this 
limited fake bids assumption can be removed while still achieving 
strict incentive compatibility. 
We pose this as an open question.

\vspace{-5pt}
\paragraph{Robustness w.r.t.~parameter estimation.}
Among our proposed mechanisms, the ones that achieve approximate incentive
compatibility, namely, threshold-based or diluted 
threshold-based mechanisms are 
simpler and more practical.
Just like how Ethereum's EIP-1559 needs to estimate a suitable base fee, 
these mechanisms also need
to estimate some parameters a-priori.
In particular, 
our (diluted) threshold-based mechanism needs to know an estimate of $h$ 
and the median of the value distribution $\mcal{D}$ in advance. 
Just like Ethereum's EIP-1559, 
we can estimate these parameters from recent history.
For example, one can estimate the total number of bids 
$n$ from the degree of congestion observed in recent blocks. 
Now, if we are willing to assume that half of the $n$ 
anticipated bids are honest (note
that our mechanisms incentivize honest behavior),
we can get an estimate of $h$. 
Similarly, one can estimate the median of the distribution  
$\mcal{D}$ from the recent history too.

One important observation is that our threshold-based mechanism
and diluted threshold-based mechanisms are quite robust 
to errors in the estimates.
As mentioned later in \cref{remark:robustness}, 
if we set the threshold to $\widehat{h}/4$ for some estimated $\widehat{h}$, 
and let $\widehat{m}$ be the estimated median, 
then the mechanisms
will achieve approximate incentive compatibility
as long as 
$h_{\rm real} \cdot q_{\rm real} \geq \widehat{h} \cdot \frac{(1+\delta)}{4}$
for some arbitrarily small constant $\delta > 0$,
where $h_{\rm real}$
is the actual number 
of honest users, and $q_{\rm real}$
is the actual percentile of the estimate $\widehat{m}$.
For example, if $h_{\rm real} = 0.6h$, and 
$q_{\rm real} = 40\%$, 
then our mechanisms still satisfy approximate incentive compatibility
for an exponentially small $\epsilon$. 
\vspace{-5pt}
\paragraph{Independent identically distributed true values assumption.}
All the mechanisms in our paper only assume that honest users' true values are i.i.d.~sampled from the distribution $\mcal{D}$.
The strategic users can have arbitrary non-negative true values.

\subsection{Additional Related Work}
We now review some closely related recent works besides
the prior works 
on transaction mechanism 
design~\cite{zoharfeemech,yaofeemech,functional-fee-market,eip1559,roughgardeneip1559,roughgardeneip1559-ec,dynamicpostedprice,foundation-tfm} already mentioned.

\ignore{
\paragraph{SCP v.s.~OCA-proof.}

To capture the strategic behavior of a coalition formed by a miner and users, Roughgarden \cite{roughgardeneip1559-ec} defines a notion called off-chain agreement (OCA), and a mechanism is OCA-proof if for any coalition formed by a miner and any subset of users, playing honestly maximizes the \emph{social welfare} (the sum of the utilities of the miner and all users, including the users not in the coalition).
Notice that side-contract-proofness (SCP) requires that playing honestly maximizes the \emph{joint utility of the coalition} instead of the social welfare.
Since the joint utility of a coalition formed by the miner and \emph{all users} is exactly the social welfare, a mechanism satisfying SCP must also be OCA-proof.
However, the converse is not true.
Even if a mechanism always maximizes the social welfare, a coalition formed by the miner and a subset of users may still profit from the deviation by taking advantage of other honest users.
The fact that the miners can extract extra gain by taking advantage of honest users has been demonstrated in practice.
For example, since the miner can decide the order of the transactions in a block, the miner or a miner-user coalition may reorder the transactions so that the coalition can profit from other users' trades by arbitrage \cite{flashboy}, which is known as \emph{miner extractable value (MEV)}.
Therefore, we focus on SCP in this work.
}
% \hao{mentioning flash boy here.}

\paragraph{TFM in a Bayesian setting.}
The recent works of Zhao, Chen, and Zhou~\cite{bayesian-tfm} and Gafni and Yaish~\cite{greedy-tfm}
both consider TFM in a Bayesian setting. Although their works
did not explicitly define the MPC-assisted model, from a practical standpoint, their results 
are in fact only relevant in an MPC-assisted (or a similar) model. As explained in 
\Cref{sec:ic} and \Cref{fct:plainexpost}, plain-model TFMs that achieve  
{\it Bayesian} equilibrium also achieve {\it ex post} equilibrium,  
since in the plain-model game, 
the strategic player can decide its actions {\it after} having observed
honest users' bids. 

Gafni and Yaish \cite{greedy-tfm} suggest a mechanism that satisfies Bayesian UIC, while also satisfying 
MIC and OCA-proof (short for offchain-agreement-proof) even if the miner knows everyone's bid. 
Further, their mechanism works in the finite-block setting while
achieving asymptotical optimality in social welfare and revenue.
We stress that their result does not contradict
the zero miner-revenue limitation proven by \cite{crypto-tfm}
since their OCA-proofness notion (originally defined
by Roughgarden~\cite{roughgardeneip1559,roughgardeneip1559-ec} )
is of a different nature from our side-contract-proofness (SCP) notion (originally defined
by Chung and Shi~\cite{foundation-tfm}).
Roughly speaking, OCA-proofness requires that a strategic coalition
cannot enter an off-chain contract that increases {\it everyone}'s 
utility ({\it not just those in the coalition}) relative to what's achievable on-chain. 
In comparison, SCP is the notion that 
directly captures the cryptocurrency community's outpouring 
concerns about Miner Extractable Value (MEV). In particular, 
middleman platforms such as Flashbot \hao{flash boy?}
facilitate the collusion of miners and users, 
where the coalition plays strategically to 
profit themselves at the expense of other users.
This is why we choose to use the SCP notion rather than OCA-proofness.
Moreover, the reason why the cryptocurrency community is developing
encrypted mempool techniques (which can be viewed as instantiations of the MPC-assisted
model) is also because they care about SCP (i.e., resilience to MEV).

\ignore{
%  when the block size is finite.
In short, by the revelation principle \cite{myerson}, if a mechanism has a Bayesian Nash-equlibrium (BNE), one can design another mechanism that satisfies UIC in the Bayesian setting. 
It has been shown that the first price auction has a BNE, so Gafni and Yaish suggest using the modified first price auction resulting from the revelation principle, where the honest miner always chooses the top $k$ bids into the block when the block size is $k$.
% Briefly speaking, they first show that their mechanism satisfies UIC in the Bayesian setting.
Because the mechanism satisfies UIC, every individual user bids truthfully, and choosing the top $k$ bids always maximizes the social welfare.
Thus, the OCA-proof follows.
}

Zhao, Chen, and Zhou \cite{bayesian-tfm} suggest a mechanism that generates positive miner revenue while 
achieving Bayesian UIC and Bayesian 1-SCP even for the finite block setting.
Their result does not contradict %the impossibility for the finite block size in \cite{foundation-tfm} and 
the 0-miner revenue limitation of \cite{crypto-tfm}, 
since Zhao, Chen, Zhou~\cite{bayesian-tfm} consider only a restricted strategy space.
In their work, a strategic user or a miner-user coalition can only deviate 
by bidding untruthfully; the coalition cannot inject fake bids, strategic users cannot drop out,
and nor can strategic miners alter the inclusion rule.
Due to their restricted strategy space, their results are only relevant
under very stringent assumptions: 1) the TFM is implemented 
in the MPC-assisted (or similar) model; 2) 
the TFM is fully ``permissioned'' and allows only 
a set of pre-registered users to submit bids.
In particular, the latter ``permissioned'' requirement is  
unrealistic for major decentralized cryptocurrencies today where any user can join and submit transactions.
%Notice that the impossibility for the finite block size in \cite{foundation-tfm} relies on the fact that the miner can choose which transactions to be included, and the 0-miner revenue limitation in \cite{crypto-tfm} relies on that miners in the coalition can inject fake bids.

\paragraph{Cryptography meets game theory.}
Prior to the advent of cryptocurrencies, 
a line of work~\cite{gtcrypto00,gtcrypto01,gtcrypto02,gtcrypto03,giladutilityindjournal,giladgtcrypto,rdp00,rdp01,rdp02,katzgametheory,gtcrypto06,seqrationalcrypto,gt-fair-cointoss,gt-fair-coin-complete,gt-leader-shi,fruitchain,logstar-gt-leader,credibleauction-comm00,credibleauction-comm01}
investigated how cryptography and game theory can help each other.
For example, cryptography can help 
remove the trusted mediator assumption in correlated equilibria~\cite{gtcrypto06}.
Adopting game-theoretic fairness 
can allow us to circumvent lower bounds pertaining to the more
stringent cryptographic notions of fairness~\cite{gtcrypto00,gtcrypto02,gtcrypto05,gtcrypto03,gt-fair-cointoss,gt-fair-coin-complete}.
Ferreira and Weinberg~\cite{credibleauction-comm00}
and Essaidi, Ferreira and Weinberg~\cite{credibleauction-comm01}
showed that cryptographic commitments
can help us circumvent impossibilities pertaining to credible auctions.
As Chung and Shi~\cite{foundation-tfm}
explained in detail, credible auction
is of a fundamentally different nature from transaction fee mechanism design.
%Transaction fee mechanism is a new type of decentralized
%mechanism design problem,  and the new connections
%between cryptography and mechanism design
%revealed in our paper differ
%in nature from the settings in prior works.

\section{Technical Roadmap}
\elaine{TODO: define the environment here.}

\subsection{Transaction Fee Mechanism}
\label{sec:TFM}

%To help make it formal what the mechanism needs to know as input, 
\paragraph{Environment.}
For preciseness in our subsequent formal description, 
we define an $(h, \rho, c, d)$-environment, where
$h$ is the promised lower bound on the number of honest users,
$\rho \in [0,1]$ is the fraction of strategic miners, 
$c$ is the maximum number of strategic users that collude
with miners, and $d$ is the maximum number of bids contributed
by the strategic coalition, i.e.~fake bids.

\paragraph{Universality.}
We can replace 
a subset of these variables with a wildcard $*$ if the mechanism
achieves incentive compatibility no matter what the variable  
turns out to be.
For example, 
a TFM that achieves incentive compatibility in an $(*, \rho, c, *)$-environment if it works when the maximum fraction of strategic miners is $\rho$,  
and the maximum number of colluding users is at most $c$, 
and regardless of 
of how many honest users there are and how many bids
are contributed by strategic individuals or the coalition.
In this case, 
%If a mechanism satisfies incentive compatibility
%in $(*, \rho, c, *)$-environments, 
we also say
that the mechanism is universal 
in the parameters $h$ and $d$. 
Using this notation, the mechanisms described
 by Shi, Chung, Wu~\cite{crypto-tfm}
are universal in the parameters $h$ and $d$.
Similarly, the limitation on miner revenue they prove can also be interpreted as a limitation of mechanisms
that are universal in $h$ and $d$.

%is one that takes the parameters $h, \rho, c, d$ as input.
\ignore{
\item 
A mechanism in an $(*, *, *, *)$-environment
is a universal mechanism that need not know any of these parameters and provide
incentive compatibility no matter what these parameters actually turn out to be;
\item 
Some of these parameters can be concretized, e.g., 
a mechanism in an $(*, \rho, 2, *)$-environment
is one that takes the parameters $\rho$ and $c$ as input
and only provides guarantees
the number of colluding users is at most $c = 2$.
\item 
sing this notation, the MPC-assisted mechanisms in prior work~\cite{}
typically work in 
$(*, \rho, c, *)$-environments 
or $(*, \rho, *, *)$-environments 
\end{itemize}
}

\paragraph{Transaction fee mechanism in the MPC-assisted model.} 
As in previous works, we consider a single auction instance that decides which transactions can be confirmed in the next block.
In this paper, ``bids'' and ``transactions'' are used interchangeably.
A transaction fee mechanism (TFM) in the MPC-assisted model consists of the following randomized algorithms.
\begin{itemize}[leftmargin=5mm,itemsep=1pt]
    \item \emph{Confirmation rule} chooses a subset of at most $k$ bids to confirm, where $k$ denotes the block size.\footnote{In the MPC-assisted model, the mechanism is implemented by the ideal functionality, and the miners cannot decide which transactions are included and considered as the input of the mechanism. Since all mechanisms proposed in our paper work in the MPC-assisted model, we simplify the definition of TFM compared to \cite{foundation-tfm}, where the TFM in \cite{foundation-tfm} also needs to specify the \emph{inclusion rule} that tells the miner which transactions are included in the next block. However, a strategic miner can still inject some fake bids. The formal strategy space is defined in \cref{sec:model}.}
    \item \emph{Payment rule} decides how much each confirmed bid pays.
    \item \emph{Miner revenue rule} decides how much revenue the miners get.
\end{itemize}

We consider a single-parameter environment, i.e., each user has a transaction with the true value represented by a single, non-negative real number.
A user's utility equals its true value minus its payment if its transaction gets confirmed in the block.
An honest user submits one bid $b \in \mathbb{R}$ representing its true value, while an honest miner submits zero bids and follows the protocol honestly.
Strategic users or miners, however, can choose to register one or more identities and submit arbitrary bids under these identities.
Moreover, they can choose to drop out during the execution.
When a strategic user or miner submits multiple bids, they can only obtain the value when the bid with the true value is confirmed, and other bids are considered fake bids.
Fake bids have no intrinsic value to the users and the miners even when they are confirmed.

All mechanisms proposed in our paper work in the MPC-assisted model.
We consider the same, generic MPC-assisted model as \cite{crypto-tfm},
where the miners jointly execute an MPC protocol
that securely evaluates
the rules of the TFM. Further, the miners equally 
divide up the revenue among themselves.
A formal description of the model can be found in \cref{sec:model}.

To illustrate the technical highlights, 
it is convenient to abstract out the multi-party computation protocol
as an ideal functionality.  
Therefore, we can think of the game as follows:
\begin{itemize}[leftmargin=5mm,itemsep=1pt]
\item 
Each player (either user or miner)
first submit zero to multiple bids 
to the ideal functionality. 
\item 
The ideal functionality
then executes the rules of the TFM, outputs the set of confirmed bids, how much reach bid needs to pay, and how much the miners get based on the prescribed rules.
\end{itemize}

Notably, in the MPC-assisted model, players cannot see others' bids when posting
their own. In particular, in an actual instantiation of the above functionality,
the players would send bids either in a secret-shared or encrypted format.
For this reason, Bayesian notions of equilibrium make sense in the MPC-assisted model. %({\it c.f.} in the
\cite{crypto-tfm}
suggested one way to securely realize
the above ideal functionality. 
Meanwhile, efforts to build an ``encrypted mempool''
by the cryptocurrency community can also be viewed  
an alternative way to instantiate the ideal functionality --- although
whether the suggested approaches provably realize this ideal functionality 
is yet to be proven.

\cite{crypto-tfm}
argued that %from a game theoretic perspective, 
as a starting point,
it makes sense to first 
explore such a generic functionality, since 
currently, we lack understanding
how cryptography in general can help decentralized mechanism 
design from a game theoretic perspective. 
Once we understand this better, we can then focus on making customized
optimizations for the actual computational tasks that need to be securely evaluated.
Our work is motivated by the same philosophy, i.e.,
we focus on understanding the game theoretic aspects rather than concrete optimizations
for realizing the MPC.

\ignore{
\subsection{Old Text}
\elaine{TODO: elaborate on the MPC model}
In both our paper and the prior work of ~\cite{crypto-tfm}, 
the $\rho$ parameter is only needed  
to instantiate the underlying MPC protocol among the miners, since
the MPC protocol needs to resist $\rho$ fraction of corrupt miners.
}

\paragraph{Universality in $\rho$ in the idealized model.}
As mentioned, it is convenient to think of the MPC as an ideal functionality.
In this idealized model, our mechanisms can achieve universality in $\rho$.
However, when the ideal functionality is actually instantiated, e.g.,  
with an honest-majority 
MPC protocol, the resulting protocol would require $\rho < 50\%$.

\ignore{
As explained in \Cref{sec:model}, 
when we focus on the game theoretic analysis, it helps to abstract 
out the underlying MPC protocol and think of it as an ideal functionality 
(i.e., a trusted third party) whose job is to  
correctly compute the confirmation rule, payment rule, and miner revenue rule
of the TFM.
In this idealized model, 
all the mechanisms described in this paper 
as well as \cite{crypto-tfm}
actually achieve universality in the parameter $\rho$.
Henceforth throughout the  
paper, we will state the results in the idealized world with universality in $\rho$.
}
\ignore{
As mentioned in Remark~\ref{rmk:allbutone}, all of our mechanisms can actually be made
universal in the parameter $\rho$ if we instantiate
the underlying MPC protocol to be secure
against all-but-one corruptions --- even though knowing
an upper bound on $\rho$ can allow us to pick a more efficient
MPC implementation. 
Henceforth, throughout the paper, we shall make this simplifying assumption 
that the underlying MPC secures against all but one corruption,
and thus all of our 
mechanisms will be universal in $\rho$.
}

\subsection{Infinite Block Setting}
\label{sec:roadmap-inf}
For the infinite block setting, we can achieve $\Theta(h)$ miner revenue
in $(h, \rho, c, d)$-environments.
%To aid understanding, we break down the thought process
%into several intermediate steps, eventually leading to our final mechanism,
%the LP-based mechanism.
To aid understanding, we first present a simple parity-based mechanism that works for $h = 1$, and then we present our main results.

\paragraph{Glimpse of hope.}
First, consider the special case where we are promised
that there is at least $h = 1$ honest user.
In this case, the following simple parity-based mechanism satisfies 
ex-post UIC, 
Bayesian MIC, and Bayesian SCP %for an arbitrary $c$, assuming infinite block size.
in $(1, *, *, *)$-environments.

\begin{mdframed}
\begin{center}
{\bf MPC-assisted, parity-based mechanism}\\
{\it //~Let $m$ be the median of the distribution $\mcal{D}$ 
such that $\Pr_{x\sim\mcal{D}}[x \geq m] = 1/2$. }
\end{center}
\begin{itemize}[leftmargin=5mm,itemsep=1pt,topsep=1pt]
%\item 
%Given a bid vector ${\bf b} = (b_1, \ldots, b_n)$, for $i \in [n]$, let $\gamma_i$
%be $0$ if $b_i < m$, else let $\gamma_i = 1$.
\item 
All bids that are at least $m$ get confirmed and pay $m$.
\item 
If the number of confirmed bids is odd, then the total miner
revenue is $m$; else the total miner revenue is $0$.
%If $\oplus_{i \in [n]}\gamma_i = 1$, then the total miner revenue is $m$; 
%else the total miner revenue is $0$.
\end{itemize}
\end{mdframed}
In the above mechanism, as long as there is at least one honest bid, 
the expected miner revenue is always $m/2$ no matter 
how the coalition behaves. 
This is because the strategic coalition cannot predict
whether the honest bid is bidding at least the median or not.
With this key observation, it is not hard to see 
that the mechanism satisfies Bayesian MIC and Bayesian 
SCP (for an arbitrary $c$).
Further, ex post UIC follows directly
since the mechanism is a simple posted-price auction  
from a user's perspective.

Observe also the following subtlety:
when the number of confirmed bids is odd, it implies that there is  
at least one confirmed bid. Therefore, the mechanism guarantees
that the miner revenue does not exceed the total payment.

\elaine{should we defer this remark to the very end}
\begin{remark}[A note about the median assumption]
In the above, we assumed that the bid distribution 
$\mcal{D}$ 
has a median $m$ such that 
$\Pr_{x\sim\mcal{D}}[x \geq m] = 1/2$.
In case %such a median does not exist, 
the median $m$ does not exactly equally divide the probability 
mass half and half, 
then it must be that 
$\Pr_{x\sim\mcal{D}}[x > m] < 1/2$ and 
$\Pr_{x\sim\mcal{D}}[x < m] < 1/2$.
In this case, 
we can modify the above mechanism slightly as follows:
if a user's bid is strictly greater than $m$, then it is confirmed;
if a user's bid is exactly $m$, then we confirm it with some appropriate probability $q$;
else the user's bid is not confirmed.
We can always pick a $q$ such that a bid randomly sampled from $\mcal{D}$
is confirmed with probability exactly $1/2$.
Finally, the miner revenue rule is still decided the same way as before.
\label{rmk:nomedian}
\end{remark}

\vspace{-10pt}
\paragraph{Threshold-based mechanism.}
The parity-based mechanism overcomes the 0 miner-revenue limitation  
by assuming the existence of at least $h=1$ honest user.
However, the drawback is obvious:  
the total miner revenue is severely restricted and does not increase
w.r.t. the number of bids.
A natural question is whether we can 
achieve $O(h)$ expected miner revenue %if we are promised 
for general $h$.
\ignore{
Unfortunately, there does not seem to be a straightforward
way to extend the parity-based mechanism
to achieve $\Theta(h)$ revenue for 
a more general choice of $h$.
}

%To get to our final construction, 
%As a stepping stone, 
%we first consider  
%an intermediate mechanism that aims only for {\it approximate} incentive compatibility:
We give an affirmative answer. We first present a simple, practical mechanism
called the threshold-based mechanism that achieves 
almost-strict incentive compatibility except for a tiny slack
$\epsilon$ that is exponentially small in $h$.
Then, for theoretical interest, we present another mechanism
that achieves strict incentive compatibility but requires an extra assumption 
on the number of bids contributed by strategic players.

\begin{mdframed}
\begin{center}
{\bf MPC-assisted, threshold-based mechanism}\\
{\it //~Let $m$ be the median of the distribution $\mcal{D}$ such that $\Pr_{x\sim\mcal{D}}[x \geq m] = 1/2$. }
\end{center}
\begin{itemize}[leftmargin=5mm,itemsep=1pt,topsep=1pt]
\item 
All bids that are at least $m$ get confirmed and pay $m$.
\item 
If the number of confirmed bids is 
at least $h/4$, then the miner revenue is $m \cdot h/4$;
else the total miner revenue is $0$.
\end{itemize}
\end{mdframed}

Due to the standard Chernoff bound, 
%given at least $h$ honest bids, 
except with $e^{-\Omega(h)}$ probability, 
the number of confirmed bids among the $h$ (or more) honest bids 
is at least $h/4$.
Therefore, the above mechanism 
achieves at least 
$m\cdot h/4 \cdot (1 - e^{-\Omega(h)})$ expected miner revenue.
If the number of confirmed honest bids is $h/4$ or higher, then the coalition 
cannot increase the miner revenue 
no matter how it behaves. 
Only when the number of confirmed
honest bids is less than $h/4$, is it possible for 
the coalition to influence the miner revenue by 
at most $m \cdot h/4$.
Therefore, it is not hard to see
that the mechanism satisfies $\epsilon$-Bayesian
MIC and $\epsilon$-Bayesian SCP 
%for an arbitrary $c$, 
in $(h, *, *, *)$-environments, 
for $\epsilon = m \cdot h \cdot e^{-\Omega(h)}/4$.

Just like before, in case %there does not exist an $m$ such that 
the median $m$ does not exactly divide the probability mass half and half,  
%i.e., $\Pr_{x\sim\mcal{D}}[x < m] < 1/2$, 
%and $\Pr_{x\sim\mcal{D}}[x > m] <  1/2$, 
we can use the same approach
of \Cref{rmk:nomedian}
to modify the mechanism and make it work.

\ignore{
Finally, we remark that while 
in this paper, we use the threshold-based 
mechanism 
as a stepping stone towards getting strict incentive compatibility, 
the mechanism itself might be of interest in practical scenarios, due
to its simplicity and the error tolerance in choosing the parameter $m$. 
}

\begin{remark}[On the robustness of parameter estimation]
\label{remark:robustness}
The threshold-based mechanism requires the mechanism to estimate 
$h$ and the median $m$ of the distribution $\mcal{D}$.
Just like how Ethereum EIP-1559 estimates its base price,
we can estimate $h$ and the median from recent history.
In particular, from the congestion level in recent blocks,
we can estimate a lower bound on the total number 
of bids. Now, assuming that at least half of them are honest (recall that our mechanism
incentivizes honesty), we can get an estimate of $h$ correspondingly.
Similarly, we can estimate the median of the bid distribution from past history.

An advantage of the threshold-based mechanism is that 
it is quite tolerant of errors in the estimation.
For example, if the estimated $m$ 
is actually the $40$-percentile of $\mcal{D}$, 
and the actual number of honest users is 
only $0.7 h$ where $h$ is the estimate used by the mechanism, 
the expected number of users bidding 
at least $m$ is at least $0.4 \cdot 0.7 h = 0.28 h$.
In this case, we can still guarantee that except with exponentially
small in $h$ probability, at least $h/4$ users will bid at least $m$.
Thus, the resulting mechanism would still be almost strictly incentive
compatible except for 
a slack $\epsilon$ that is exponentially small in $h$.
\end{remark}

\vspace{-8pt}
\paragraph{LP-based mechanism.}
Although the $\epsilon$ slack in the 
threshold-based mechanism is exponentially small which is not a problem in practice,
it is still theoretically interesting to  
%achieves only approximate incentive compatibility. 
ask whether we can get  
$\Theta(h)$ total miner revenue
but with {\it strict} incentive compatibility.
To achieve this,  
our idea is to devise a mechanism that is ``close in distance''
to the aforementioned threshold-based mechanism, but correcting
the ``error'' such that  we can achieve strict incentive compatibility.
%We thus devise the following mechanism:

%To aid understanding, we shall take another intermediate step 
%and consider a mechanism %\elaine{FILL} 
%that removes the $\epsilon$ slack but 
%requires an extra assumption. 
Observe that the  
earlier threshold-based mechanism 
only needs an a-priori known lower bound 
on $h$, and it is universal in the parameters 
$c$ and $d$.
To achieve strict incentive compatibility, we 
additionally assume that the 
the number of bids contributed by the strategic coalition
is upper bounded by some a-priori known parameter $d$.
%Instead of assuming an a-priori known lower bound
%$h$ on the number of honest users, we now assume
%that the mechanism is promised that at most $c$ bids
%come from the strategic coalition.
\ignore{
Another way to view this assumption is that
identity creation may be expensive, 
such that a strategic miner or miner-user coalition is unable to inject
an arbitrary number of fake bids. 
As a result, we assume that at most $c$ bids are contributed 
by the strategic miner or miner-user coalition.
}
%Under this assumption, if the mechanism receives
%a bid vector of length $n$, we know that at least $n-c$ bids
%must come from honest users.

Now, consider the following mechanism
that relies on linear programming
to correct the error in the earlier threshold-based mechanism. 
For simplicity, we assume
that the honest bid distribution $\mcal{D}$
has a median $m$ such that
$\Pr_{x\sim\mcal{D}}[x \geq m] = 1/2$ --- if not, 
we can again use the technique of \Cref{rmk:nomedian}
to modify the mechanism and make it work.

\begin{mdframed}
\begin{center}
{\bf MPC-assisted, LP-based mechanism}\\
{\it //~Let $m$ be the median of the distribution $\mcal{D}$, i.e., $\Pr_{x\sim\mcal{D}}[x \geq m] = 1/2$. }
\end{center}
\begin{itemize}[leftmargin=5mm,itemsep=1pt,topsep=1pt]
\item 
All bids that are at least $m$ get confirmed and pay $m$.
\item 
Let $n$ be the length of the bid vector, let 
${\bf y} := (y_0, y_1, \ldots, y_n)$ 
be any feasible solution to the following linear program:
\begin{alignat}{2}
%\label{eqn:budget}
\forall i \in [n]: & \ \   0 \leq y_i \leq i \cdot m
\label{eqn:budget}
\\
\forall 0 \leq j \leq d:  
%& \ \ \sum_{i = 0}^{\max(h, n-d)} q_i \cdot y_{i + j} 
& \ \ \sum_{i = 0}^{n-d} q_i \cdot y_{i + j} 
= \frac{m \cdot h}{4} 
\label{eqn:expmu}
\end{alignat}
where $q_i$ is the probability 
of observing $i$ heads if we flip $n-d$ independent fair coins.
\item 
The total miner revenue
is $y_s$ 
where $s$ is the number of bids confirmed.
\end{itemize}
\end{mdframed}

In the above, %\Cref{eqn:budget} 
%\elaine{REF} 
\Cref{eqn:budget}
expresses a {\it budget feasibility} requirement, i.e., the total miner
revenue cannot exceed the total user payment.
\Cref{eqn:expmu} expresses a {\it fixed-revenue requirement}
stipulating that the miner revenue must be exactly $m \cdot h/4$ no matter
how the strategic individual or coalition behaves (as long
as it controls at most $d$ bids). 
More specifically, 
\Cref{eqn:expmu} contains
one requirement for each $j \in [0, d]$: conditioned
on the fact that  
among the (at most) $d$ bids controlled by 
the strategic individual or coalition, exactly $j$
of them are confirmed, the expected miner revenue must be exactly $m \cdot h/4$
where $h$ is an a-priori known lower bound on the number of honest users.
%\elaine{REF} 
\ignore{
\Cref{eqn:expmu}
guarantees that no matter how the strategic individual or coalition 
behaves,  
as long as it cannot control more than $c$ bids,
it cannot 
influence the expected miner revenue.
}

\begin{remark}
We know that the actual number of honest users that show
up is at least $\max(n-d, h)$. So if $n-d > h$, 
it means that 
more honest users showed up than the anticipated number $h$.
Observe that on the left-hand side of 
\Cref{eqn:expmu}, we are tossing coins for $n-d$ honest users' bids.
However, it is important that the right-hand-side of 
\Cref{eqn:expmu} 
use the a-priori known $h$ rather than the observed $n-d$;  
otherwise, injecting extra (but up to $d -c$) 
fake $0$-bids can increase the expected miner revenue,
which violates MIC and SCP.
\end{remark}

If the LP in the 
above mechanism indeed has a feasible solution, then we can prove
that the resulting mechanism satisfies ex post UIC, 
Bayesian MIC, and Bayesian SCP
in $(h, *, *, d)$-environments.
%, under
%the assumptions that 1) at most $c$ bids are controlled
%by the strategic individual or coalition, and 2) at least $h$
%users are honest. 
The formal proofs are presented in \cref{sec:lp-proof}.
\elaine{missing ref here}

The key technical challenge is to answer the question of why the LP has a feasible solution.
Intuitively, the earlier threshold-based mechanism
gives an ``approximate'' solution $\widehat{\bf y} := (\widehat{y}_0, \ldots, 
\widehat{y}_n)$ to the LP, 
where 
$\widehat{y}_i = 0$ for $i \leq (n-d)/4$ and 
$\widehat{y}_i = \frac{m \cdot h}{4}$ %for $i \leq (n-d)/4$ and 
otherwise.
With the approximate 
solution $\widehat{\bf y}$, 
the equality constraints in 
\Cref{eqn:expmu} may be satisfied with some small error.
We want to show that we can adjust the  
$\widehat{\bf y} := (y_0, y_1, \ldots, y_n)$
vector slightly such that we can correct
the error, 
and yet without violating the budget feasibility constraints (\Cref{eqn:budget}).

To achieve this, we will take a constructive approach. 
We first guess that a feasible solution 
is of the form ${\bf y} = \widehat{\bf y} + {\bf e}$
where ${\bf e}$ is a correction vector
that is zero everywhere except 
in the coordinates $\tau, \tau +1, \ldots, \tau + d$ 
for some appropriate choice of $\tau$ that is close to $(n-d)/2$.
Henceforth, let $\boldsymbol{\delta} := 
{\bf e}[\tau: \tau+d]/\left(\frac{m \cdot h}{4}\right)$ be the non-zero coordinates of
the correction vector ${\bf e}$ scaled by $\frac{m \cdot h}{4}$.

\ignore{
First, 
we solve the linear system specified
by the fixed-revenue requirement, i.e., \Cref{eqn:expmu}.
We then prove that the solution indeed satisfies
the budget feasibility constraints, i.e., \Cref{eqn:budget}.
For simplicity, in  this informal overview, we focus on the case when $n > h + d$
and $n-d$ is a multiple of $4$.  
Specifically, 
we guess that one feasible solution is of the following form
for some hopefully small correction vector
%$\boldsymbol{\delta} := \{\delta_i\}_{i \in [\frac{n-d}{2} + d, \frac{n-d}{2} + 2d)}$.
$\boldsymbol{\delta} := \{\delta_i\}_{i \in [z, z + d)}$
for some appropriate choice of $z$ that is close to $n/2$.

\begin{equation}
\forall i \in [0, n]: \ \ 
x_i = \begin{cases}
0 & i \leq \frac{n-d}{4}\\
%\frac{m \cdot h}{4} & \frac{n-d}{4} < i < \frac{n-d}{2} + d\\
\frac{m \cdot h}{4} & \frac{n-d}{4} < i < z\\
\frac{m \cdot h}{4} \cdot (1 + \delta_i) &    z \leq i <  
z + d\\
%\frac{n-d}{2} + 2d \\
%\frac{m \cdot h}{4} & i > \frac{n-d}{2} + 2d
\frac{m \cdot h}{4} & i \geq z+d
\end{cases}
\label{eqn:guessedsol}
\end{equation}
In comparison with the 
approximate solution implied by the threshold-based mechanism,
the only difference is that we now introduce a hopefully
small correction amount $\delta_i$ into the coordinates 
%from $\frac{n-d}{2} + d$
%to $\frac{n-d}{2} + 2d - 1$ respectively.
from $z$
to $z+d-1$ respectively.
}
%If the guessed solution of \Cref{eqn:guessedsol}
% is indeed a feasible solution to the LP, 
By \Cref{eqn:expmu}, we know that 
the correction vector $\boldsymbol{\delta}$
must satisfy the following system of linear equations where $t:= \frac{n-d}{4}$:
\begingroup
        \renewcommand*{\arraystretch}{1.5}
        \begin{equation}
        \begin{pmatrix}
        \binom{n-d}{\tau} & \binom{n-d}{\tau+1} &  \dots &  \binom{n-d}{\tau+d}\\
        \binom{n-d}{\tau-1} & \binom{n-d}{\tau}  & \dots &  \binom{n-d}{\tau+d-1}\\
        \vdots & \vdots & \ddots & \vdots \\
         \binom{n-d}{\tau-d} & \binom{n-d}{\tau-d+1} & \dots &  \binom{n-d}{\tau}\\
        \end{pmatrix}
\cdot        {\boldsymbol{\delta}}=
        \begin{pmatrix}
            \sum_{i=0}^{\thresh} \binom{n-d}{i}\\
            \sum_{i=0}^{\thresh-1} \binom{n-d}{i}\\
            \vdots\\
            \sum_{i=0}^{\thresh-d} \binom{n-d}{i}
        \end{pmatrix}.
%        \elaine{FILL}
        \label{eqn:correctmat}
        \end{equation}
\endgroup

\ignore{
\begin{equation}
\forall i \in \{0, 1, \ldots, d\}:
\ \ 
%\sum_{j = 0}^{d-1} p_{\frac{n-d}{2}+d + j + (d - i)} \cdot \delta_{\frac{n-d}{2}+d +j }
\sum_{j = 0}^{d-1} p_{z + j + (d - i)} \cdot \delta_{z +j }
 =   
\sum_{j = 1}^{\frac{n-d}{4} - i}
p_j 
\label{eqn:correcteqs}
\end{equation}
where $p_k$ is the 
probability of observing $k$ heads when we flip $n-d$ independent fair coins.
\elaine{TODO: give a graphical illustration of one of the above equations}
\Cref{eqn:correcteqs}
can be further rewritten as:
}

In \cref{lem:delta-sol}, we prove that  
as long as $d \leq \frac18 \sqrt{\frac{h}{2\log h}}$, 
and that $\tau$ is an appropriate choice close to $n/2$, then the solution 
$\boldsymbol{\delta}$ to the linear system 
in \Cref{eqn:correctmat}
has a small infinity norm --- specifically, 
$\|\boldsymbol{\delta}\|_{\infty} \leq 1$ ---
such that the resulting ${\bf y}$ vector 
will respect the 
budget feasibility constraints, i.e., \Cref{eqn:budget}.
The actual proof of this bound is somewhat involved and thus
deferred to \cref{sec:infinite}.
In particular, a key step is to bound the {\it smallest singular
value}
of the matrix in  
\Cref{eqn:correctmat} (henceforth denoted $A$)
appropriately --- to achieve
this, we 
first bound $A$'s determinant, and then use an inequality 
proven by \cite{yi1997note} which
relates the smallest singular value and the determinant.

\ignore{
\begin{remark}
The precise mechanism 
in the formal technical sections (\elaine{refer})
is a slight modification of the one above. In particular,
instead of using exactly the coordinates 
$\frac{n-d}{2} + d$
to $\frac{n-d}{2} + 2d - 1$ to perform correction, 
we actually select $d$ coordinates close to the center 
--- this technicality arises due
to the fact that with the extra wiggle room, we can more easily prove
that the resulting matrix \elaine{FILL} is invertible. 
\end{remark}
}

%\paragraph{Limit on miner revenue in the known-$h$ model.}
%\elaine{TODO: add}

%\paragraph{Our final mechanism: improved LP-based mechanism.}

\subsection{Finite Block Setting}

\subsubsection{Strict Incentive Compatibility}
\label{sec:roadmap-finite}

\paragraph{Feasibility for $c = 1$.}
The LP-based mechanism confirms any bid that offers to pay at least $m$. Thus, 
total number of confirmed bids may be unbounded.
Therefore, when the block size $k$ is finite, we cannot directly run
the LP-based mechanism.
We suggest the following modification to the LP-based mechanism 
such that it works for the finite-block setting:
%One na\"ive idea is the following: 
%if there are more than $k$ bids, we randomly down-select
%to $k$ bids, and then run the LP-based mechanism
%on the down-selected set.
\begin{mdframed}
\begin{center}
{\bf MPC-assisted, LP-based mechanism with random selection}
\end{center}
{\it //~Let $k$ be the block size, let $m$ be the median $\mcal{D}$ 
such that $\Pr_{x\sim\mcal{D}}[x \geq m] = 1/2$. }
\begin{itemize}[leftmargin=5mm]
\item 
All bids offering at least $m$ are candidates.
If there are more than $k$ 
candidates, randomly select $k$ of them to confirm; else confirm all candidates. 
Every confirmed bid pays $m$.
\item 
Let $n$ be the length of the bid vector, let ${\bf y} = (y_0, y_1, \ldots, y_n)$ 
be any feasible solution to the following linear program:
\begin{alignat}{2}
\forall i \in [n]: & \ \   0 \leq y_i \leq \min(i, k)\cdot m
\label{eqn:budget-finite}
\\
\forall 0 \leq j \leq d:  
& \ \ \sum_{i = 0}^{n-d} q_i \cdot y_{i + j} 
= \frac{m \cdot \min(h, k)}{4} 
\label{eqn:expmu-finite}
\end{alignat}
where $q_i$ is the probability 
of observing $i$ heads if we flip $n-d$ independent fair coins.
\item 
The total miner revenue is $y_s$ where $s$ is the number of {\it candidates}.
\end{itemize}
\end{mdframed}
In comparison with the earlier LP-based mechanism, we modify
the budget feasibility constraints (\Cref{{eqn:budget-finite}}) to make sure that
the total miner revenue 
is constrained by the actual number of confirmed bids which is now $\min(i, k)$
if the number of candidates is $i$.
Further, we modify the expected miner revenue  
(\Cref{eqn:expmu-finite})
to be $\frac{m \cdot \min(h, k)}{4}$ which takes into account the block size $k$.
In \cref{sec:finc1}, we prove
that as long as $c\leq d \leq \frac18 \sqrt{\frac{h}{2\log h}}$, 
the above LP indeed has a feasible solution and the resulting mechanism  
satisfies ex post UIC, Bayesian MIC, and Bayesian SCP
in $(h, *, 1, d)$-environments.

\paragraph{Infeasibility for $c \geq 2$.}
Unfortunately, the above approach fails for $c \geq 2$.
In this case, two users Alice and Bob may be in the same coalition.
%If Alice happens to have small true value
%and Bob has large true value, then  
Alice can now help Bob simply by dropping out and not posting
a bid, thus effectively increasing
Bob's chance of getting confirmed.
In the event that Alice's true value is very small 
and Bob's true value is sufficiently large, 
this strategic action can increase the coalition's joint utility.

%Therefore, an interesting question is whether we can achieve non-trivial miner revenue 
%for $c \geq 2$ in the finite block setting.
%Another immediate observation is the following. Consider the bid distribution $\mcal{D}$

Interestingly, it turns out that this is no accident.  
In fact, we prove that 
for any $h \geq 1$, $\rho \in (0, 1)$, and $d \geq c \geq 2$, 
no ``interesting'' mechanism 
can simultaneously achieve Bayesian UIC, MIC and SCP 
in $(h, \rho, c, d)$-environments --- 
\elaine{is this correct}
any such mechanism must suffer from $0$ total social welfare
for the users if the actual number of bids received
is greater than $h$ (see \Cref{thm:zero-usw} for the formal statement).
We can regard \Cref{thm:zero-user-welfare-informal} as a generalization
of \cite{crypto-tfm}'s 
Theorem 5.5: 
they show that any MPC-assisted mechanism that achieves 
Bayesian UIC, MIC, and SCP 
in $(*, \rho, c, *)$-environments 
for $c \geq 2$ must suffer from 
$0$ social welfare for users. 

\elaine{TODO: fix these missing refs, they should refer to thms in the new intro}

\paragraph{Proof roadmap.}
We use the following blueprint to 
prove \Cref{thm:zero-user-welfare-informal}.
Below, consider any 
TFM that satisfies Bayesian UIC, MIC and Bayesian SCP
in $(h, \rho, c, d)$-environments where $d\geq c \geq 2$. %\elaine{is it SCP?}
\begin{enumerate}[leftmargin=5mm]
\item 
First, in \cref{lem:hartline}, using techniques
inspired by Goldberg and Hartline~\cite{goldberghartline}
we prove the following:  
provided that there are at least $h$ honest users (not including $i$ and $j$) 
whose bids are sampled
at random from $\mcal{D}$, then a strategic user $i$ changing its bid should not affect the utility 
of another user $j$, if user $j$'s bid is also sampled at random from $\mcal{D}$. 
\item 
Next, in \cref{lem:hartline}, we prove 
a strategic user $i$ dropping out 
should not affect another user $j$'s utility, assuming
that 
at least $h$ bids (excluding user $i$)
sampled at random from $\mcal{D}$.
\item 
Next, in \cref{cor:uihv}, %we conclude that according to \cref{lem:hartline},
we show that in a world of at least $h$ random bids (excluding user $i$) sampled from $\mcal{D}$,  
user $i$'s expected utility when its bid is sampled randomly from $\mcal{D}$
depends only on $i$'s identity, and does
not depend on the identities of the other random bids.
Therefore, henceforth we can use $U_i$ to denote this expected utility. 
\item 
Next, in \cref{lem:equal-util}, we show that %for any fixed set $H$, 
for any two identities $i, j$, 
it must be that $U_i = U_j$, otherwise, it violates 
the assumption that the mechanism is weakly symmetric (see definition
of weak symmetry below).
\item 
%Given \cref{cor:uihv,lem:equal-util}, 
Next, we can show that $U_i = 0$: 
imagine a world with $K$ bids 
sampled independently from $\mcal{D}$ whose support is bounded. 
There must exist some user 
whose confirmation probability is upper bounded by $k/K$.
This user's expected utility 
must be arbitrarily small when $K$ is arbitrarily large.  
%With \cref{lem:equal-util}, we can conclude that for any any $i$, we have that $\E_{v\in\mathcal{D}}[U^{v}_i]$ must be $0$.
With a little more work, we can show that 
if the world consists of more than $h$ bids 
sampled independently at random from $\mcal{D}$,
it must be that every user's 
expected utility is $0$.
\end{enumerate}

One technicality that arises in the full proof (see \cref{sec:zero-usw})
is the usage of the weak symmetry assumption. 
In particular, the proof would have been much easier 
if we could instead assume 
{\it strong symmetry} which, unfortunately, is too stringent.
In strong symmetry, we assume that any two users who bid the same amount
will receive the same treatment.
While it is a good approach for gaining intuition about the
proof, it is too stringent since 
there could well be more bids offering the same value
than the block size $k$ --- in this case,
a non-trivial mechanism would treat them differently, i.e., 
confirm some while rejecting others.
Our actual proof of \cref{thm:zero-user-welfare-informal}
needs only a {\it weak symmetry} assumption which is a standard
assumption made in prior works~\cite{foundation-tfm,crypto-tfm}, that is, 
if two input bid vectors $\bfb, \bfb'$ of length $n$ are permutations of each other, then 
the joint distribution of the {\it set} of outcomes
$\{(x_i, p_i)\}_{i \in [n]}$ must be identical. 
This implies that if the input bid vectors are permutations of each other, then
the vector of expected utilities are permutations of each other too.

\subsubsection{Approximate Incentive Compatibility}
Because of the limitation shown in \Cref{thm:zero-user-welfare-informal}, we 
relax the notion to approximate  
incentive compatibility, and ask if we can achieve optimal miner
revenue in the finite block setting.
\ignore{
Indeed, we can design an MPC-assisted TFM that
satisfies Bayesian UIC, MIC, and SCP 
in $(h, \rho, c, d)$-environments for finite block size:

\begin{mdframed}
\begin{center}
{\bf MPC-assisted, diluted threshold-based mechanism}

{\it //~Let $m$ be the median of the distribution $\mcal{D}$ 
such that $\Pr_{x\sim\mcal{D}}[x \geq m] = 1/2$. }
\end{center}
\begin{itemize}[leftmargin=5mm]
\item 
Let all bids offering at least 
$m$ be the pool of candidates. If the pool size is less than $T = $\elaine{FILL}, 
we pad the pool with filler bids to a capacity of $T$.
\item 
Randomly select $\min(k, \frac{h}{4})$ out of the $T$ candidates to confirm. 
Every real (i.e., non-filler) bid selected  
pays $m$. 
\item 

\end{itemize}
\end{mdframed}
}
Consider the following TFM.
\begin{mdframed}
    \begin{center}
    {\bf MPC-assisted, diluted threshold-based Mechanism}
    \end{center}
    %\paragraph{Parameters:} the block size $k$, environment parameter $\environ$, the approximation parameter $\epsilon$, the distribution median $\m$, and the distribution maximum $T$.
    
    %\vspace{-1em}
    %\paragraph{Input:}a bid vector $\bfb = (b_1,\dots,b_n)$.

    %\vspace{-1em}
    %\paragraph{Mechanism:}

\noindent {\it /*~Let $k$ be the block size, let $m$ be the median of $\mcal{D}$ 
such that $\Pr_{x\sim\mcal{D}}[x \geq m] = 1/2$, let $T$ be the maximum value
of the distribution $\mcal{D}$.
*/
}
    \begin{itemize}[leftmargin=5mm,itemsep=1pt,topsep=1pt]
    \item 
    %{\it Confirmation rule.} 
    Let $R := \max \left(2c \sqrt{\frac{kT}{\eps}}, k\right)$. %\ke{check this parameter}
%            Given a bid vector $\bids$, let $\widetilde{\bids} = (\widetilde{b}_1,\dots, \widetilde{b}_{s})$ denote the bids that are at least $m$.
All bids offering at least $m$ are candidates. %, and let $s$ be the number of candidates. 
 If the number of candidates $s\leq R$, randomly select $\frac{k}{R}\cdot s$ candidates to confirm; 
else, randomly select $k$ candidates to confirm.
    %\item 
    %{\it Payment rule.} 
Every confirmed bid pays $\m$.
    \item 
%    {\it Miner revenue rule.} 
    % Let $\bico:=n-d$ and $\target = \frac{h}{4}\cdot \frac{k}{R}\cdot \m$.
    % Let $\minerrev=(\target, \target,\dots, \target)^T \in\R^{c+1}$.
    % Define $M\in\R^{(c+1)\times(n+1)}$ to be the following matrix: 
    % \begingroup
    %     \renewcommand*{\arraystretch}{1.5}
    %     \begin{equation*}
    %     M = \frac{1}{2^{\bico}}
    %     \begin{pmatrix}
    %     \binom{\bico}{0} & \binom{\bico}{1} & \binom{\bico}{2} & \dots &  \binom{\bico}{\bico} & 0 & 0 & \dots & 0\\
    %     0 & \binom{\bico}{0} & \binom{\bico}{1} & \dots & \binom{\bico}{\bico-1} & \binom{\bico}{\bico} & 0 & \dots & 0\\
    %     \vdots & \vdots & \vdots & \vdots & \vdots & \vdots & \vdots & \vdots & \vdots \\
    %     0 & 0 & 0 & \dots & \dots & \dots & \binom{\bico}{\bico-2}& \binom{\bico}{\bico-1}&\binom{\bico}{\bico} \\
    %     \end{pmatrix}
    %     \end{equation*}
    % \endgroup
    % That is, the $i$-th row of $M$ is $\left(\binom{\bico}{0}, \binom{\bico}{1},  \dots,\binom{\bico}{\bico}, 0 \dots,0\right)$ right-shifted by $i-1$ elements.
    % Let ${\bf y}=(y_0,\dots,y_n)$ be the solution to the following linear program $\mathcal{P}$:
    % \begin{align}
    % \label{eqn:lp-diluted}
    %     &\MatM{\bf y} = \minerrev,\\
    %     \text{s.t. }& 0\leq y_i\leq \m\cdot \min\left(i\cdot\frac{k}{R},\,k\right) \text{ for all } i.\nonumber
    % \end{align}
    % Recall that $s$ is the number of bids in $\bids$ larger than or equal to $\m$. Miner gets $y_s\cdot \m$.
    If $s\geq \frac{h}{4}$, then the total miner revenue is 
$\min(\frac{h}{4}\cdot \frac{k}{R}, k) \cdot \m$. Otherwise, the miners get nothing.
    \end{itemize}
\end{mdframed}

Intuitively, here are modifying the earlier threshold-based mechanism
to 1) make it compatible with finite block size, and 2) make sure
that up to $c$ users dropping out  can only minimally increase
their friend's probability of getting confirmed. 
In particular, resilience to drop-out 
is achieved by artificially diluting the probability that a user is confirmed
when the number of eligible bids (i.e., offering at least $m$) is small. 
With the dilution, we guarantee that a coalition of $c$ users   
cannot noticeably alter their own probability of getting confirmed, 
nor their friend's probability. 
This implies a strategic coalition has little influence
over the expected utility of all users in the coalition.
Moreover, we guarantee that a strategic coalition has very little influence
on the miner revenue as well: similar to the threshold-based mechanism,
except with $\exp(-\Omega(h))$ probability, 
the miner revenue 
is an a-priori fixed amount, that is, 
$\min(\frac{h}{4} \cdot \frac{k}{R}, k) \cdot m$.
Summarizing the above, we can show that the mechanism satisfies
ex post UIC, Bayesian $\epsilon$-MIC, and Bayesian $\epsilon$-SCP
in $(h, *, c, *)$-environments, as long as $\epsilon \geq \m\cdot \frac{h}{2}\cdot e^{-\frac{h}{16}}$.

Finally, for sufficiently large $h \geq \max(4k, 8c \sqrt{\frac{kT}{\epsilon}})$, the mechanism
achieves $k \cdot m$ total miner
revenue and $k \cdot C_{\mcal{D}}$ user social welfare where $C_{\mcal{D}}$
is defined in \Cref{thm:theta-h-minerrev-informal}.
For example, suppose we are willing to tolerate $\epsilon = 0.01 T$,  
then we just need $h \geq \max(4k, 80c \cdot \sqrt{k})$
to achieve asymptotic optimality in miner revenue and social welfare.
The full proof is deferred to \cref{sec:diluted-threshold-appendix}.
\elaine{TODO: double check}

%\elaine{TODO: discuss}

\subsection{Additional Results}

\elaine{change this to approx}
\paragraph{Limit on miner revenue.}
In \cref{sec:lb-strict}, we prove \Cref{thm:intro-limit-rev}.
Specifically, we prove 
that $\Theta(h)$ revenue is optimal in $(h, \rho, c, d)$-environments
for strict incentive compatibility; 
and further, we
generalize the bound to approximate incentive compatibility 
as well (see \cref{thm:revenue-approxIC}).
The proof is a generalization of the techniques
proposed by Shi, Chung, Wu~\cite{crypto-tfm}.
Specifically, they proved that 
any mechanism that satisfies Bayesian UIC, MIC, and SCP
in $(*, \rho, 1, *)$-environments
must suffer from 0 miner revenue. %\elaine{FILL}
In their proof, 
they argue that if we remove one bid, the miner  
revenue must be unaffected. 
In our case, because the mechanism is promised
a lower bound $h$ on the number of honest uesrs, we can repeat
this argument 
till there are $h$ honest bids left, and no more.
This gives rise to an $O(h)$ limit on  
miner revenue.
The proof for the approximate incentive compatibility
is also a generalization of 
\cite{crypto-tfm}'s techniques, but more technically involved since
even Myerson's lemma does not hold for approximate incentive compatibility.

\paragraph{Necessity of Bayesian equilibrium.}
As mentioned, our 
reasonable-world assumptions (formalized through the definition of an $(h, \rho, c, d)$-environment)
would not have helped had we insisted on ex post notions of equilibrium (for 
all of UIC, MIC, and SCP). 
In \cref{sec:expost}, we explain why 
for ex post notions of incentive compatibility, even 
mechanisms in 
the $(h, \rho, c, d)$-environment
are subject to the same miner-revenue limitations 
of universal mechanisms.

\section{Model and Definitions}
\label{sec:model}
\ignore{
Consider a {\it seller} who has some number of identical items to sell,
and a set of $n$ {\it bidders} each wanting to buy one item. 
Each bidder $i \in [n]$ has a 
true value denoted $v_i \in \R$ for the item.
The seller and the bidders 
carry out the auction with the help of an intermediary {\it platform}
(e.g., Google Ad Exchange or eBay),
who is rewarded with a percentage of the total sale revenue,
and the remaining revenue goes to the seller.
Throughout the paper, we consider a single-parameter environment
where each bidder's bid is expressed as a single non-negative real value.

\paragraph{The plain model.}
Today, the bidders send their bids 
in the clear to the third-party platform which implements
the rules of the auction, and informs the seller
and all bidders of the outcome. 
%In this model, it is impossible to achieve  

\paragraph{The MPC-assisted model.}
}
\elaine{TODO: rewrite the model}

Imagine that there are $n_0$ users, and each user has a transaction  
that wants to be confirmed.
For $i \in [n_0]$, let $v_i$ be user $i$'s true valuation 
of getting its transaction confirmed.
We want to design a transaction fee mechanism (TFM)
such that no individual user or miner or a coalition thereof
have any incentive to deviate from honest behavior.
Throughout the paper, we consider a single-parameter environment, i.e., 
each user's bid is represented by a single, non-negative real number.

Chung and Shi~\cite{foundation-tfm}'s results
ruled out the existence of interesting TFMs in the plain model 
without cryptography.
First, they show a {\it zero miner-revenue} bound:  
any TFM that guarantees incentive compatibility for 
each individual user as well as for a miner-user coalition 
must suffer from zero miner revenue. The zero miner-revenue
bound holds matter
whether the block size is infinite or finite, and even when
the miner is allowed to collude with only one user.
Second, they prove a {\it finite-block impossibility}:
assuming finite block size,
then no TFM can simultaneously guarantee incentive compatibility for  
each individual user as well as for a miner-user coalition (even when
the miner is allowed to collude with at most one user).

The subsequent work of Shi, Chung, and Wu~\cite{crypto-tfm}
considered how cryptography can help circumvent these 
strong impossibilities.
They proposed the {\it MPC-assisted model}, where the rules of the TFM 
are enforced through a multi-party computation (MPC) protocol jointly executed
among a set of miners. Unlike the plain model, 
the MPC-assisted model guarantees that a single miner 
cannot unilaterally decide which transactions to include in the block.
This ties the hands of the strategic player(s), 
and this model indeed results in interesting mechanisms 
that achieve properties that would otherwise be impossible in the plain model.
Below, we review the MPC-assisted model proposed by 
\cite{crypto-tfm}.
\elaine{note: may be able to shirnk after writing intro}

%\paragraph{MPC-assisted model.}
\subsection{MPC-Assisted Model}
\label{sec:MPC-model}

The definition of TFM has been given in \cref{sec:TFM}.
Here, we formalize the MPC-assisted model and the strategy spaces for users and miners.

\paragraph{Ideal-world game.}
Recently, blockchain projects such as Ethereum are developing 
``encrypted mempool'' techniques which can be viewed
as concrete protocols that realize 
an MPC-assisted model for TFM.
However,
for understanding the game theoretic landscape, it helps
to abstract out the cryptography and think of 
it as a trusted 
{\it ideal functionality} (henceforth denoted $\fmpc$) that  
always honestly implements the rules of the TFM.

%Just like \cite{crypto-tfm}, it suffices to consider
%an idealized model where we think of the MPC as a trusted 
%{\it ideal functionality} (henceforth denoted $\fmpc$) that  
%always honestly implements the rules of the TFM.
With the ideal functionality $\fmpc$, we can imagine the following game
that captures an instance of the TFM:
\begin{enumerate}[leftmargin=6mm]
\item  
Each user registers zero, one, or more identities with $\fmpc$,
and submits exactly one bid on behalf of each identity.
\item 
Using the vector of input bids 
as input, 
$\fmpc$ executes the rules of the TFM.  
$\fmpc$ now sends to all miners and users
the output of the mechanism, including 
the set of bids that are confirmed, how much each confirmed bid pays, and how
much revenue the miner gets.
\end{enumerate}

We make a couple of standard assumptions:
\begin{itemize}[leftmargin=6mm,itemsep=1pt]
\item {\it Individual rationality}:
each confirmed bid should pay no more than the bid itself;
\item {\it Budget feasibility}:
the miner revenue should not exceed the total payment from all confirmed bids. 
\end{itemize}

Using standard techniques in cryptography, we can instantiate
the ideal functionality $\fmpc$ using an actual cryptographic
protocol among the miners and users (see Appendix D of \cite{crypto-tfm}). 
Further, in the actual instantiation, the users only need to be involved
in the input phase: they only need to verifiably
secret-share their input bids among all miners, and they need not be involved
in the remainder of the protocol. 
The miners then jointly run some MPC protocols to compute securely 
the outcome of the auction.
We can use an MPC protocol
that retains security even when all but one miner are corrupt~\cite{gmw87}. 
Such protocols
achieve a security notion called ``security with abort'', i.e., 
an adversary controlling a majority coalition can cause the protocol to abort
without producing any outcome. 
Conceptually, one can imagine that in the ideal-world protocol
where parties interact with $\fmpc$, the adversary
is allowed to send $\bot$ to $\fmpc$, 
in which case $\fmpc$ will abort and output $\bot$ to everyone.
However, no strategic coalition should have an incentive 
to cause the protocol to abort --- in this case, no block
will be mined and the coalition 
has a utility of $0$.
Thus, without loss of generality, we need not explicitly capture aborting as a possible strategy in our ideal-world game mentioned above.

\paragraph{Strategy space and utility.}
An honest user will always register a single identity and submit only one bid
reflecting its true value. A strategic user or miner
(possibly colluding with others)
can adopt the following strategies or a combination thereof:
\begin{itemize}[leftmargin=5mm,itemsep=1pt]
\item 
{\it Bid untruthfully}: a strategic user can misreport its value;
\item   
{\it Inject fake bids}:  
a strategic user or miner can {\it inject
fake bids} by registering fake identities; 
\item 
{\it Drop out}:  
a strategic user can also {\it drop out} by not registering its real identity.
\end{itemize}

In the real-world cryptographic instantiation, strategic miners 
can also deviate from the honest MPC protocol. 
However, as mentioned, the MPC protocol retains security 
(i.e., can be simulated by the ideal-world game)
as long as at least one miner is honest. Therefore, we need not
explicitly capture
this deviation in the ideal-world game.
Finally, strategic miners can cause the MPC protocol
to abort without producing output, and 
as mentioned, this deviation never makes sense since it results in a utility of $0$;
thus we also need not explicitly 
capture it in the ideal-world game.

Let $v_i$ denote user $i$'s true value.
If user $i$'s transaction is confirmed and its payment is $p_i$, then its utility is defined as $v_i- p_i$.
The miner's utility is simply its revenue.
The joint utility of a coalition $\mcal{C}$ is defined as the sum of the utilities of all members in $\mcal{C}$.

\subsection{Defining Incentive Compatibility}
\label{sec:ic}

In the plain model without cryptography, users submit their bids 
in the clear over a broadcast channel, and a strategic coalition
can decide its strategy after observing the remaining honest users' bids.
By contrast, in the MPC-assisted model, bids are submitted 
to the ideal functionality $\fmpc$ (in the actual cryptographic instantiation,
the users verifiably secret-share their bids among the miners).
This means that the strategic coalition 
must now submit its bids without having observed other users' bids. 
Therefore, in the MPC model, it makes sense to consider a {\it Bayesian} notion of equilibrium rather than an {\it ex post} notion. 
In an {\it ex post} setting, we require that  
a strategic individual or coalition's best response is to act honestly
even after having observed others' actions.
In a {\it Bayesian} setting, 
we assume that every honest user's bid
is sampled independently from some distribution $\mcal{D}$, and 
we require that acting honestly maximizes
the strategic individual or coalition's expected utility where
the expectation is taken over not just the random coins 
of the TFM itself, but also over the randomness in sampling
the honest users' bids.

\paragraph{Notations.} Henceforth, we use the notation ${\bf b}$ 
to denote a bid vector.  
Since we allow strategic players to inject fake bids or drop out,
the length of ${\bf b}$ need not be the same as the number of users $n_0$. 
We use the notation $\mcal{C}$ to denote a coalition,
and we use ${\bf b}_{-\mcal{C}}$ to denote the bid vector
belonging to honest users outside $\mcal{C}$.
We use the notation $\mcal{D}_{-\mcal{C}}$
to denote the joint distribution 
of ${\bf b}_{-\mcal{C}}$, that is, 
$\mcal{D}_{-\mcal{C}} = \mcal{D}^{h_0}$
where $h_0$ denotes the number of honest users outside $\mcal{C}$.
Similarly, if $i$ is an individual strategic user, then the notation ${\bf b}_{-i}$
denotes the bid vector belonging the remaining honest users in $[n_0]\backslash \{i\}$.
We use the notation $\mcal{D}_{-i}$ to denote the joint distribution
of the honest bid vector 
${\bf b}_{-i}$.

%Jumping ahead, although our final results  
%can achieve {\it strict} incentive compatibility, \elaine{double check later}
%to get these results, 
For generality, we define 
%we use the notion of {\it approximate} incentive compatibility
%as a stepping stone.
{\it approximate} incentive compatibility parameterized by an additive slack $\epsilon$.
The case of {\it strict} incentive compatibility can be viewed
as the special case when $\epsilon = 0$. 

\ignore{
Therefore, in our formal definitions below, 
we define the more general version of $\epsilon$-incentive compatibiliy.
When $\epsilon = 0$, the notion reduces
to strict incentive compatibility, and for $\epsilon > 0$,
we get approximate incentive compatibility.
}

\iffullversion
\subsubsection{Bayesian Incentive Compatibility}
\else
\paragraph{Bayesian incentive compatibility.}
Here we formally define Bayesian incentive compatibilities.
\fi

\begin{definition}[Bayesian incentive compatibility]
We say that an MPC-assisted TFM satisfies Bayesian 
$\epsilon$-incentive compatibility 
for a coalition or individual $\mcal{C}$, 
iff for any ${\bf v}_{\mcal{C}}$
denoting the true values of users in $\mcal{C}$, 
sample ${\bf b}_{-\mcal{C}} \sim \mcal{D}_{-\mcal{C}}$, 
then, no strategy  
can increase 
%honest behavior maximizes the 
$\mcal{C}$'s expected
utility by more than $\epsilon$ 
in comparison with honest behavior, 
where the expectation is taken over randomness
of the 
honest users' bids 
${\bf b}_{-\mcal{C}}$, as well as random coins consumed by the TFM.
Specifically, 
we define the following notions depending on who is the strategic
individual or coalition:
\begin{itemize}[leftmargin=5mm]
\item 
{\it User incentive compatibility (UIC).}
We say that an MPC-assisted TFM satisfies
Bayesian $\epsilon$-UIC in some environment $\mcal{E}$, iff 
%under the conditions required by the environment $\mcal{E}$,  
%the following holds:
for any $n$, for any user $i \in [n]$, 
for any true value $v_i \in \R^{\geq 0}$
of user $i$, 
for any strategic bid vector ${\bf b}_i$
from user $i$ which could be empty or consist of multiple bids, 
the following holds as long as the conditions required by the environment $\mcal{E}$ are respected:
\[
%\E_{{\bf b}_{-i} \sim \mcal{D}^{n-1}}
\underset{{\bf b}_{-i}\sim \mcal{D}_{-i}}{\E}
\left[{\sf util}^i({\bf b}_{-i}, v_i)\right] \geq
\underset{{\bf b}_{-i}\sim \mcal{D}_{-i}}{\E} 
%\E_{{\bf b}_{-i} \sim \mcal{D}^{n-1}}
\left[{\sf util}^i({\bf b}_{-i}, {\bf b}_i)\right]
- \epsilon
\]
where ${\sf util}^i({\bf b})$ denotes the expected utility (taken over
the random coins of the TFM)
of user $i$ when the bid vector is ${\bf b}$.
\item 
{\it Miner incentive compatibility (MIC).}
We say that an MPC-assisted TFM satisfies
Bayesian $\epsilon$-MIC in some environment $\mcal{E}$, iff 
%under the conditions required by the environment $E$, 
%the following holds: 
for any miner coalition $\mcal{C}$, 
%controlling at most $\rho$  fraction of the miners, 
for any strategic bid vector ${\bf b}'$ injected
by the miner, 
the following holds as long as the conditions required by the environment $\mcal{E}$ are respected:
\[
\underset{{\bf b_{-\mcal{C}}\sim \mcal{D}_{-\mcal{C}}}}{\E}
\left[{\sf util}^{\mcal{C}}({\bf b}_{-\mcal{C}})\right]\geq
%\E_{{\bf b}_{-\mcal{C}} \sim \mcal{D}^{|-\mcal{C}|}}
\underset{{\bf b_{-\mcal{C}}\sim \mcal{D}_{-\mcal{C}}}}{\E}
\left[{\sf util}^{\mcal{C}}({\bf b}_{-\mcal{C}}, {\bf b}') \right]
- \epsilon
\]
where ${\sf util}^{\mcal{C}}({\bf b})$
denotes the expected utility (taken over the random coins
of the TFM) of the coalition $\mcal{C}$ when the input bid
vector is ${\bf b}$.
\item 
{\it Side-contract-proofness (SCP).}
We say that an MPC-assisted TFM satisfies
Bayesian $\epsilon$-SCP in some environment $\mcal{E}$, iff 
for any miner-user coalition, %consisting of at most $\rho$
%fraction of the miners and at most $c$ users, 
for any true value vector ${\bf v}_{\mcal{C}}$ 
of users in $\mcal{C}$, 
for any strategic bid vector ${\bf b}_{\mcal{C}}$ 
of the coalition 
(whose length may not be equal to the number of users in $\mcal{C}$),
the following holds as long as the requirements of the environment $\mcal{E}$ 
are respected:
\[
\underset{{\bf b_{-\mcal{C}}\sim \mcal{D}_{-\mcal{C}}}}{\E}
\left[{\sf util}^{\mcal{C}}({\bf b}_{-\mcal{C}}, {\bf v}_{\mcal{C}})\right]\geq
%\E_{{\bf b}_{-\mcal{C}} \sim \mcal{D}^{|-\mcal{C}|}}
\underset{{\bf b_{-\mcal{C}}\sim \mcal{D}^{-\mcal{C}}}}{\E}
\left[{\sf util}^{\mcal{C}}({\bf b}_{-\mcal{C}}, {\bf b}_{\mcal{C}}) \right]
- \epsilon
\]
\end{itemize}
\end{definition}
Henceforth, if a mechanism satisfies Bayesian $\epsilon$-UIC
%(or $\epsilon$-MIC, $\epsilon$-SCP) respectively 
for $\epsilon = 0$ (i.e., the {\it strict} incentive compatibility case), 
we often omit writing the $\epsilon$,
and simply say that the mechanism satisfies 
Bayesian UIC. 
The terms ``Bayesian MIC'', and ``Bayesian SCP'' are similarly defined.

Notice that we only require honest users' true values are i.i.d.~sampled, while the strategic players' true values can be arbitrary.

\iffullversion
\subsubsection{Ex Post Incentive Compatibility}
\begin{definition}[Ex post incentive compatibility]
We say that a TFM satisfies ex post 
$\epsilon$-UIC, 
$\epsilon$-MIC, 
and $\epsilon$-SCP 
respectively, for a coalition or individual $\mcal{C}$, 
iff the following conditions hold, respectively:
\begin{itemize}[leftmargin=5mm]
\item 
{\it User incentive compatibility (UIC).}
We say that an MPC-assisted TFM satisfies
ex post $\epsilon$-UIC in some environment $\mcal{E}$, iff 
for any $n$, for any user $i \in [n]$, 
for any bid vector ${\bf b}_{-i}$ 
denoting the bids of everyone else besides $i$, 
for any true value $v_i \in \R^{\geq 0}$
of user $i$, 
for any strategic bid vector ${\bf b}_i$
from user $i$ which could be empty or consist of multiple bids, 
the following holds as long as the conditions required by the environment $\mcal{E}$ are respected:
\[
%\underset{{\bf b}_{-i}\sim \mcal{D}_{-i}}{\E}
{\sf util}^i({\bf b}_{-i}, v_i) \geq
%\underset{{\bf b}_{-i}\sim \mcal{D}_{-i}}{\E} 
%\E_{{\bf b}_{-i} \sim \mcal{D}^{n-1}}
{\sf util}^i({\bf b}_{-i}, {\bf b}_i)
- \epsilon
\]
where ${\sf util}^i({\bf b})$ denotes the expected utility (taken over
the random coins of the TFM)
of user $i$ when the bid vector is ${\bf b}$.
\item 
{\it Miner incentive compatibility (MIC).}
We say that an MPC-assisted TFM satisfies
ex post $\epsilon$-MIC in some environment $\mcal{E}$, iff 
for any miner coalition $\mcal{C}$, 
for any bid vector ${\bf b}_{-\mcal{C}}$, 
for any strategic bid vector ${\bf b}'$ injected
by the miner, 
the following holds as long as the conditions required by the environment $\mcal{E}$ are respected:
\[
%\underset{{\bf b_{-\mcal{C}}\sim \mcal{D}_{-\mcal{C}}}}{\E}
{\sf util}^{\mcal{C}}({\bf b}_{-\mcal{C}}) \geq
%\E_{{\bf b}_{-\mcal{C}} \sim \mcal{D}^{|-\mcal{C}|}}
%\underset{{\bf b_{-\mcal{C}}\sim \mcal{D}_{-\mcal{C}}}}{\E}
{\sf util}^{\mcal{C}}({\bf b}_{-\mcal{C}}, {\bf b}') 
- \epsilon
\]
where ${\sf util}^{\mcal{C}}({\bf b})$
denotes the expected utility (taken over the random coins
of the TFM) of the coalition $\mcal{C}$ when the input bid
vector is ${\bf b}$.
\item 
{\it Side-contract-proofness (SCP).}
We say that an MPC-assisted TFM satisfies
ex post $\epsilon$-SCP in some environment $\mcal{E}$, iff 
for any miner-user coalition, %consisting of at most $\rho$
%fraction of the miners and at most $c$ users, 
for any bid vector $b_{-\mcal{C}}$ submitted by non-coalition-members,
for any true value vector ${\bf v}_{\mcal{C}}$ 
of users in $\mcal{C}$, 
for any strategic bid vector ${\bf b}_{\mcal{C}}$ 
of the coalition 
(whose length may not be equal to the number of users in $\mcal{C}$),
the following holds as long as the requirements of the environment $\mcal{E}$ 
are respected:
\[
%\underset{{\bf b_{-\mcal{C}}\sim \mcal{D}_{-\mcal{C}}}}{\E}
{\sf util}^{\mcal{C}}({\bf b}_{-\mcal{C}}, {\bf v}_{\mcal{C}})\geq
%\E_{{\bf b}_{-\mcal{C}} \sim \mcal{D}^{|-\mcal{C}|}}
%\underset{{\bf b_{-\mcal{C}}\sim \mcal{D}^{-\mcal{C}}}}{\E}
{\sf util}^{\mcal{C}}({\bf b}_{-\mcal{C}}, {\bf b}_{\mcal{C}}) 
- \epsilon
\]
\end{itemize}
\end{definition}
Henceforth, if a mechanism satisfies ex post $\epsilon$-UIC
%(or $\epsilon$-MIC, $\epsilon$-SCP) respectively 
for $\epsilon = 0$ (i.e., the {\it strict} incentive compatibility case), 
we often omit writing the $\epsilon$,
and simply say that the mechanism satisfies 
ex post UIC. 
The terms ``ex post MIC'', ``ex post SCP'' are similarly defined.

Recall that in the game representing the plain model, strategic players can choose their actions 
{\it after} having observed the bids submitted by honest users.  
This gives rise to the following fact which essentially says it does not make sense
to consider Bayesian notions of equilibrium in the plain model.

\begin{fact}
Any plain-model TFM that satisfies 
\emph{Bayesian} $\epsilon$-UIC (or \emph{Bayesian} $\epsilon$-MIC, \emph{Bayesian} $\epsilon$-SCP resp.)
in some environment $\mcal{E}$
must also satisfy 
\emph{ex post} $\epsilon$-UIC (or \emph{ex post} $\epsilon$-MIC, \emph{ex post} $\epsilon$-SCP resp.).
\label{fct:plainexpost}
\end{fact}
\else
\paragraph{Ex post incentive compatibility.}
In the interest of space, we define 
ex post incentive compatibility in \cref{sec:expost-ic-defn}.
\fi

%As mentioned, for the $\epsilon = 0$ special case, we also say
%that the mechanism satisfies {\it strict} incentive compatibility.

\ignore{
The Bayesian notions of incentive compatibility
do not make sense in the plain model, since in the plain model,
the strategic individual or coalition can decide
its move {\it after} having observed the remaining honest users' bids.
This is why we adopt only the ex post notion 
in the plain model.
Formally, it is easy to show that any mechanism that satisfies
Bayesian incentive compatibility in the plain model
also satisfies ex post incentive compatibility.

In the MPC-assisted model, both notions make sense and the ex post
notions are strictly stronger than the Bayesian counterparts. 
Jumping ahead, all of our impossibility results for the MPC-assisted
model work even for the Bayesian notions, 
and all of our mechanism designs  
in the MPC-assisted model 
work even for the ex post notions. This makes both our lower-
and upper-bounds
stronger.
}

\elaine{TODO: remark the relation to truthfulness?}

\section{Feasibility for Infinite Block Size}

\subsection{MPC-Assisted, Threshold-Based Mechanism}
\label{sec:thresholdPosted-appendix}
We assume that honest users' bids are drawn i.i.d.~from some distribution $\mathcal{D}$ 
with the median $\m$ such that $\Pr_{x\sim\mathcal{D}}[x\geq \m] = 1/2$ (see \cref{rmk:nomedian}).
For convenience, we repeat the MPC-assisted, threshold-based mechanism, which has been introduced in \cref{sec:roadmap-inf}.

\begin{mdframed}
    \begin{center}
    {\bf MPC-assisted, threshold-based mechanism}
    \end{center}
    \noindent\textbf{Parameters:} lower bound $h$ on the number of honest users, the distribution median $\m$.
    
    \noindent\textbf{Mechanism:}
    \begin{itemize}[leftmargin=5mm,itemsep=1pt]
    \item 
    {\it Confirmation rule.}
    Given a bid vector $\bfb = (b_1, \ldots, b_{\ell})$, 
    for each bid $b_i$, confirm $b_i$ if $b_i \geq \m$.
    \item 
    {\it Payment rule.}
    Each confirmed bid pays $\m$.
    \item 
    {\it Miner revenue rule.}
    Let $s$ be the number of confirmed bids. If $s\geq \frac{h}{4}$, miner gets $\frac{h}{4}\cdot \m$. Otherwise, the miner gets nothing.
    \end{itemize}
\end{mdframed}

\begin{theorem}
\label{thm:threshold-appendix}
Fix any $h\geq 1$.
The MPC-assisted, threshold-based mechanism satisfies ex post UIC, Bayesian $\epsilon$-MIC and Bayesian $\epsilon$-SCP in an $(h,*,*,*)$-environment, where $\eps = \frac{h}{4}\cdot \m \cdot e^{-\frac{h}{16}}$.
\end{theorem}

\begin{proof}
First, UIC follows from the same reasoning as in \cref{lem:parity}. 
We will focus on MIC and SCP in the rest of the proof.

\paragraph{$\eps$-MIC.} 
Recall that the only strategy that a strategic miner can apply is injecting some fake bids.
Because injecting fake bids smaller than $\m$ does not influence the colluding miners' utility, 
we only consider injecting bids at least $\m$.
Let $X$ denote the random variable representing the number of honest bids at least $\m$. 
The only situation where the colluding miners can increase their expected gain by injecting fake bids is when $X<\frac{h}{4}$.
By the following Chernoff Bound, 
\begin{lemma}[Chernoff bound, Corollary A.1.14~\cite{alon2016probabilistic}]
\label{lem:chernoff}
Let $X_1$,$\dots$, $X_n$ be independent Bernoulli random variables. Let $\mu = \E\left[\sum_{i=1}^n X_i\right]$. Then, for any $\epsilon\in(0,1)$, it holds that
\[\Pr\left[\sum_{i=1}^n X_i \leq (1-\epsilon)\mu\right]\leq e^{-{\epsilon^2}\mu/2}.\]
\end{lemma}
We have
\[\Pr\left[X < \frac{h}{4}\right]\leq e^{-\frac{h}{16}}.\]
Therefore, the colluding miners can gain at most $\frac{h}{4}\cdot \m\cdot e^{-\frac{h}{16}}$ more expected revenue by injecting fake bids.

\paragraph{$\epsilon$-SCP.} 
Since the confirmation and the payment of each bid are independent of other bids, and the mechanism is strict UIC, the coalition cannot increase colluding users' utilities.
Therefore, by deviating from the mechanism, the coalition can only try to increase the expected total miner revenue. 
By a similar argument as MIC, the coalition can only increase the expected total miner revenue when $X < \frac{h}{4}$, which happens with a probability no more than $e^{-\frac{h}{16}}$.
It follows that no matter how the coalition deviates, the expected miner's revenue can increase by at most $\rho\cdot \frac{h}{4}\cdot \m \cdot e^{-\frac{h}{16}}$.
\end{proof}

\subsection{Analysis of the LP-Based Mechanism}
\label{sec:infinite}

\subsubsection{Preliminaries: Linear Algebra Tools}
We first introduce some linear algebra tools needed
for analyzing the LP-based mechanism.

Throughout this section, all our indexing for vectors and matrices starts from $0$.
Given a vector $\bfb = (b_0, b_1,\dots,b_n)$ and two integers $i,j$ such that $i \leq j$, we define $\bfb[i:j]$ to be the subvector $(b_i,\dots,b_j)$.
We use $A =(a_{ij})\in\R^{n,m}$ to denote a matrix in which the entry of the $i$-th row and $j$-th column is $a_{ij}$.
Let $A^T$ denote the transpose of $A$, and $A^{-1}$ denote the inverse of $A$ if $A$ is non-singular.

\paragraph{Norm.} Define the \emph{infinity-norm} $\|\bfb\|_\infty$ of a vector $\bfb$ to be 
	$\|\bfb\|_\infty = \max \{|b_i|: 0 \leq i \leq n\}$.
For a square $n\times n$ matrix $A=(a_{ij})$, define the following matrix norms:
\begin{itemize}
    \item \emph{Infinity norm}: $\|A\|_{\infty} = \underset{\|x\|_{\infty} = 1}{\sup}\|Ax\|_{\infty} = \max_i\sum_{j=1}^n|a_{ij}|$.
    \item \emph{$\ell_2$-norm}: $\|A\|_{2} = \underset{\|x\|_{2} = 1}{\sup}\|Ax\|_{2}$.
    \item \emph{Frobenius norm}: $\|A\|_F = \left(\sum_{i,j=0}^{n-1}a_{ij}^2\right)^{1/2}$.
\end{itemize}
It is easy to check that $\|A\|_{\infty}\leq \|A\|_2$, and that $\|Ax\|_{\infty}\leq \|A\|_{\infty}\|x\|_{\infty}$.

\paragraph{Singular value.} For a square $n\times n$ matrix $A$, the singular values are the square roots of the eigenvalues of $A^TA$.
\begin{fact}
    \label{lem:singular-norm}
    Let $A\in \R^{n\times n}$ be non-singular.
    Let $\lambda_1\geq \dots\geq \lambda_n$ be the singular values of $A$.
    Then $\|A^{-1}\|_2 = \frac{1}{\lambda_n}$.
\end{fact}

\begin{lemma}[Yu and Gu \cite{yi1997note}]
\label{lem:singular}
Let $A\in \R^{n\times n}$ be non-singular and $\lambda$ be the smallest singular value of $A$. Then 
\[\lambda\geq |\det(A)|\cdot \left(\frac{n-1}{\|A\|_F^2}\right)^{(n-1)/2}>0.\]
\end{lemma}

\paragraph{Determinant.} The determinant of a matrix $A=(a_{ij})\in\R^{n\times n}$ is 
$\det(A) = \sum_{\sigma\in S_{n}} {\sf sgn}(\sigma) \prod_{i=1}^{n} a_{i,\sigma_i}$,
where $S_n$ is the set of all permutations $\sigma$ over the set $\{0,\dots,n-1\}$.
For each permutation $\sigma \in S_n$, let $\sigma_i$ denote the value of the $i$-th position after reordering by $\sigma$.
The signature ${\sf sgn}(\sigma)$ of a permutation $\sigma$ is $+1$ if the permutation can be obtained by an even number of swaps between two entries and $-1$ otherwise.
%\ke{Need to add prelim: norm, singular value, determinant, vector notations (indexing, slicing)}

\subsubsection{Proofs for the LP-Based Mechanism}
\label{sec:lp-proof}
We now prove that the MPC-assisted LP-based mechanism satisfies strict incentive compatibility in an $(h,*,c,d)$-environment.
Suppose that honest users’ values
are sampled i.i.d. from some distribution $\mcal{D}$.
Recall that $\m$
denotes the median of the bid distribution $\mcal{D}$,
and $C_{\mcal{D}} = \E_{x \sim \mcal{D}}[x-m| x\geq m]$
is another constant related to the distribution $\mcal{D}$.
Without loss of generality, we assume $\Pr[x \geq m] = \frac{1}{2}$
(see \Cref{rmk:nomedian}).

\begin{theorem}[\cref{thm:lp-based-informal} restated]
Suppose that the block size is infinite.
%and that $c \leq d \leq \frac18\sqrt{\frac{h}{2\log h}}$.
Fix any\footnote{For the special case $h =1$, we can just
use the parity-based mechanism of \Cref{sec:roadmap-inf}.}
 $h \geq 2$, %$d\geq c \geq 1$, 
and any $d \leq \frac18\sqrt{\frac{h}{2\log h}}$, 
the MPC-assisted, LP-based mechanism guarantees 
ex post UIC, Bayesian MIC, and Bayesian SCP 
in an $(h, *, *, d)$-environment, 
and meanwhile, the mechanism 
achieves $\Theta(h \cdot \m)$
expected miner revenue, and  
at least $\Theta(\widetilde{h} \cdot C_{\mcal{D}})$
expected social welfare for the users
where $\widetilde{h} \geq h$ is the 
the actual number of honest users
that show up.
\label{thm:lpmech}
\end{theorem}

We prove \Cref{thm:lpmech}
in two steps. 
First, we show that 
if the linear program defined in \cref{eqn:budget,eqn:expmu} has a feasible solution,
then the resulting 
mechanism satisfies incentive compatibility, as formally stated below:
%in $(h, *, c, d)$-environments.

\begin{lemma}
\label{thm:lp}
    When the linear program 
defined in \cref{eqn:budget,eqn:expmu}
has a feasible solution, the LP-based mechanism satisfies ex post UIC, Bayesian MIC, and Bayesian SCP in an $(h,*,*,d)$-environment.
    Moreover, the expected miner revenue is $\frac{h\cdot m}{4}$, and the user social welfare is $\Theta(\widetilde{h}\cdot C_{\mathcal{D}})$.
\end{lemma}
\iffullversion
\begin{proof}
    First, it is easy to see that the expected total miner revenue is $\frac{h\cdot \m}{4}$, as guaranteed by the linear program \cref{eqn:budget,eqn:expmu}.
    Moreover, since the expected utility of a user with true value $v$ is $v-m$ if $v\geq m$,    
    the expected user social welfare is at least
    \[\sum_{i\in H} \underset{v_i\sim\mathcal{D}}{\E} [v_i - m \mid v_i > m] = \widetilde{h}\cdot \underset{x\sim\mathcal{D}}{\E}[x-m\mid x \geq m],\]
    where $H$ is the set of all honest users.
    
    Next, we prove that the mechanism is strict incentive compatible if the linear program has a solution.
    UIC is easy to see.
    Next, we only prove SCP since MIC follows from the same reasoning. 

    \paragraph{SCP.} Since the confirmation and the payment of each bid are independent of other bids, and the mechanism is strict UIC, the coalition cannot increase colluding users' expected utilities.
    Therefore, we only need to show that the coalition cannot increase the expected total miner revenue by deviating from the mechanism. 
    Intuitively, the linear program \cref{eqn:budget,eqn:expmu} ensures that for arbitrary $d$ bids, the total miner revenue taking an expectation over the remaining $n-d$ bids always remains $\frac{h\cdot m}{4}$.
    
    Formally, let $\widetilde{h}$ denote the number of \emph{real honest bids} and $\bids_{-\mathcal{C}}$ denote the random variable of honest users' bids. 
    Then $\widetilde{h}\geq n-d = \bico$.
    For any bid $\bids_{\mathcal{C}}$ controlled by the coalition, the expected total miner revenue is
    \begin{equation}
    \label{eqn:expected-miner-rev-lp}
        \underset{\bids_{-\mathcal{C}}\sim \mathcal{D}^{\widetilde{h}}}{\E} \left[\mu(\bids_{-\mathcal{C}}, \bids_{\mathcal{C}})\right]
        = \underset{{\bf t}\sim\mathcal{D}^{\widetilde{h}-\gamma}}{\int} \underset{\bids\sim \mathcal{D}^{\bico}}{\E} \left[\mu(\bids, {\bf t}, \bids_{\mathcal{C}})\right] f({\bf t}) d{\bf t},
    \end{equation}
    where $f(\cdot)$ is the p.d.f. for $\mathcal{D}^{\widetilde{h}-\bico}$.
    For any fixed $({\bf t}, \bids_{\mathcal{C}})$, let $I$ denote the number of bids that are larger than or equal to $\m$ in $({\bf t}, \bids_{\mathcal{C}})$.
    Since the probability of an honest bid being at least $\m$ is exactly $\frac{1}{2}$,
    \begin{equation*}
        \underset{\bids\sim \mathcal{D}^{\bico}}{\E} \left[\mu(\bids, {\bf t}, \bids_{\mathcal{C}})\right]
        =\sum_{i=0}^{\bico} \frac{1}{2^{\bico}} \binom{\bico}{i} y_{i+I},
    \end{equation*}
    which is exactly $\frac{h\cdot m}{4}$ as guaranteed by \cref{eqn:expmu}.
    Substituting back into~\eqref{eqn:expected-miner-rev-lp}, for any bid $\bids_{\mathcal{C}}$, we have that
    \begin{align*}
        \underset{\bids_{-\mathcal{C}}\sim \mathcal{D}^{\widetilde{h}}}{\E} \left[\mu(\bids_{-\mathcal{C}}, \bids_{\mathcal{C}})\right]= \underset{{\bf t}\sim\mathcal{D}^{\widetilde{h}-\gamma}}{\int} \frac{h\cdot m}{4} \cdot f({\bf t}) d{\bf t} = \frac{h\cdot m}{4}.
    \end{align*}
    Therefore, for any $d$ bids controlled by the coalition, the expected miner revenue remains $\frac{h\cdot m}{4}$. 
\end{proof}
\else
The proof of \Cref{thm:lp}
requires an exhaustive case-by-case analysis, and a key observation 
is that \cref{eqn:expmu} ensures that the expected miner revenue is always $\frac{h\cdot m}{4}$ as long as no more than $d$ bids are contributed by the strategic coalition.
\ignore{
We separate the proof of this Theorem into two parts. 
First, in \cref{thm:lp}, we show that the mechanism satisfies Bayesian UIC, MIC, and SCP, when the linear program defined in \cref{eqn:budget,eqn:expmu} has a feasible solution. 
This is because \cref{eqn:expmu} ensures that the expected miner revenue is always $\frac{h\cdot m}{4}$ when there are at least $n-d$ number of honest users.
}
We defer the formal proof of \Cref{thm:lp} to \cref{sec:lp-appendix}.
\fi
\ignore{
\begin{lemma}
\label{thm:lp}
    When the linear program $\mathcal{P}$ has a feasible solution, the linear programming mechanism satisfies ex post UIC, Bayesian MIC, and Bayesian SCP in an $(h,*,*,d)$-environment.
    Moreover, the expected miner revenue is $\frac{h\cdot m}{4}$, and the user social welfare is $\Theta(\widetilde{h}\cdot C_{\mathcal{D}})$.
\end{lemma}
}

In the main body, we focus on proving the more challenging step,
that is, 
as long as $d \leq \frac18\sqrt{\frac{h}{2\log h}}$, 
the linear program indeed has a feasible solution, formally stated below.
%the linear program \cref{eqn:budget,eqn:expmu} is guaranteed to have a feasible solution.
\begin{lemma}
\label{lem:delta-sol}
    For $h\geq 2$ and $d\leq \frac{1}{8}\sqrt{\frac{h}{2\log h}}$, the linear program 
specified by \cref{eqn:budget,eqn:expmu} is guaranteed to have a feasible solution.
    % If $\bico\geq 2$, \ke{fill}, and moreover, $\target < \frac{\bico}{4}$, then the linear program $\mathcal{P}$ has a feasible solution $(y_0,\dots,y_{n})$ that satisfies the following property:
    % \[
    % y_i =
    % \begin{cases}
    % 0, & \text{ if } 0\leq i\leq \lfloor\frac{\bico}{4}\rfloor,\\
    % \target + \delta_i \text{ for some } |\delta_i|\leq \target, & \text{ if } \lfloor\correct\rfloor \leq i \leq \lfloor\correct\rfloor + c,\\
    % \target, & \text{ otherwise}.
    % \end{cases}
    % \]
    % \begin{itemize}
    %     \item For any $0\leq i\leq \lfloor\frac{\bico}{4}\rfloor$, $y_i = 0$;
    %     \item For any $0 \leq i\leq c$, $y_{\lfloor\correct\rfloor+i} = \target + \delta_i$, where $|\delta_i|\leq \target$;
    %     \item For all other indices $i$, $y_i = \target$.
    % \end{itemize}
\end{lemma}
\begin{proof}
    We will give a constructive solution to the linear program \cref{eqn:budget,eqn:expmu}. 
    Let $\bico:= n-d$ denote the number of bids that are sampled randomly from $\mathcal{D}$. 
    Let $\thresh = \lfloor\frac{\bico}{4}\rfloor$, and $\target$ be our target expected miner revenue $\frac{m\cdot h}{4}$.
    We start from an ``approximate'' solution $\threshold=(\widehat{y}_0,\dots,\widehat{y}_n)\in\R^{n+1}$ such that $\widehat{y}_i = 0$ for any $i \leq \thresh$, and $\widehat{y}_i = \target$ for any $i > \thresh$.
    Our goal is to find a correction  $\err = (e_0,\dots,e_n)\in\R^{n+1}$ that is zero everywhere except for the indices $i\in[\correct+d, \correct+2d]$ for some $z\geq \frac{\bico}{2}$ such that $\threshold + \err$ is a feasible solution to the linear program \cref{eqn:budget,eqn:expmu}.
    Henceforth, let $\correction:= \err[z+d, z+2d]/\target$ be the non-zero coordinates of the correction, scaled by $\target$.
    Then $\correction$ must satisfy the linear system $\MatA(z)\correction = {\bf \Delta}$, where $A(z)$ and ${\bf \Delta}$ are defined as follows:
    \begingroup
        \renewcommand*{\arraystretch}{1.5}
        \begin{equation*}
        \MatA(z) = 
        \begin{pmatrix}
        \binom{\bico}{\correct+d} & \binom{\bico}{\correct+d+1} &  \dots &  \binom{\bico}{\correct+2d}\\
        \binom{\bico}{\correct+d-1} & \binom{\bico}{\correct+d}  & \dots &  \binom{\bico}{\correct+2d-1}\\
        \vdots & \vdots & \ddots & \vdots \\
         \binom{\bico}{\correct} & \binom{\bico}{\correct+1} & \dots &  \binom{\bico}{\correct+d}\\
        \end{pmatrix}, \qquad
        {\bf \Delta} = 
        \begin{pmatrix}
            \sum_{i=0}^{\thresh} \binom{\bico}{i}\\
            \sum_{i=0}^{\thresh-1} \binom{\bico}{i}\\
            \vdots\\
            \sum_{i=0}^{\thresh-d} \binom{\bico}{i}
        \end{pmatrix}.
        \end{equation*}
    \endgroup
    If there exists a $z^*\in[\lceil\frac{\bico}{2}\rceil, \lceil\frac{\bico}{2}\rceil+2d^2]$ such that this linear system $\MatA(z^*)\correction = {\bf \Delta}$ has a solution $\correction$, then choosing $\err$ such that $\err[z^*+d: z^*+2d] = \target\cdot \correction$ gives a solution $\threshold + \err$ that satisfies \cref{eqn:expmu}. 
    % We only need to show that the solution $\correction$ to $\MatA\correction = {\bf \Delta}$ indeed exists and it satisfies that $0\leq \widehat{y}_i+e_i\leq i$.
    \begin{claim}
    \label{clm:infty-norm}
        There exists a $z^*\in[\lceil\frac{\bico}{2}\rceil, \lceil\frac{\bico}{2}\rceil+2d^2]$ such that the matrix $\MatA(z^*)$ is non-singular, and 
        \begin{equation}
        \label{eqn:inf_norm}
            \|\MatA(z^*)^{-1}\|_{\infty} \leq \frac{(z^*+2d)^{2d(d+1)}}{\binom{\bico}{z^*}}\cdot\left(\frac{d+1}{\sqrt{d}}\right)^d.
        \end{equation}
    \end{claim}
    \noindent When choosing this $z^*$, we have a unique solution $\correction = \MatA(z^*)^{-1}{\bf \Delta}$.
    Moreover, under the given parameter range, the solution $\correction$ has bounded infinity norm:
    \begin{claim}
    \label{clm:param}
    For $h\geq 2$ and $d\leq \frac{1}{8}\sqrt{\frac{h}{2\log h}}$, we have $\|\correction\|_{\infty}\leq 1$.
    \end{claim}
    \noindent For now, we assume that \Cref{clm:infty-norm}
and \Cref{clm:param} are true, and we show how they lead to \Cref{lem:delta-sol}. 
    The proofs of the two claims appear right afterward.
To prove \Cref{lem:delta-sol}, 
    it suffices to show that $\widehat{\bf y} + \err$ indeed satisfies the budget feasibility specified by \cref{eqn:budget}.
    Since for all $i\notin [z^*+d,z^*+2d]$, we have $\widehat{y}_i+e_i = \widehat{y}_i\leq i\cdot \m$, so we only need to show that for the correction position $z^*+d,\dots,z^*+2d$, the budget feasibility is satisfied. 
    Substituting $\|\correction\|_{\infty}\leq 1$, for each $i\in [z^*+d,z^*+2d]$, we have $|e_i|\leq \target$.
    This implies that $\widehat{y}_i+e_i \geq \target - \target = 0$.
    Moreover,
    \[\widehat{y}_i+e_i\leq 2\target \leq \frac{\bico}{2}\cdot m\leq i\cdot m.\]
    \Cref{lem:delta-sol} thus follows.
    \end{proof}

\begin{proof}[Proof of \cref{clm:infty-norm}]
    We separate the proof in two parts: we first show that there exists a 
   $z^*\in[\lceil\frac{\bico}{2}\rceil, \lceil\frac{\bico}{2}\rceil+2d^2]$ such that $\MatA(z^*)$ is non-singular;
    then we show that the infinity norm of the inverse of $A(z^*)$ satisfies Equation \eqref{eqn:inf_norm}.
    
    \paragraph{Non-singularity.} We show that there exists $z^*$ in the given range such that $\det(\MatA(z^*))\neq 0$. 
    Define \[B(z) = \frac{A(z)}{\binom{\bico}{z}}\cdot \prod_{i=1}^{2d}(z+i).\]
    Since
    \begin{align*}
        \frac{\binom{\bico}{z+j}}{\binom{\bico}{z}}\cdot \prod_{i=1}^{2d}(z+i) &= \frac{(\bico-z-j+1)\dots (\bico-z)}{(z+1)\dots (z+j)}\cdot \prod_{i=1}^{2d}(z+i)\\
        &=\prod_{i=1}^{j}(\bico-z-j+i)\cdot \prod_{i=j+1}^{2d}(z+i),
    \end{align*}
    % \begin{align*}
    %     &\frac{\binom{\bico}{z+j}}{\binom{\bico}{z}}\cdot \prod_{i=1}^{2d}(z+i) = \frac{(\bico-z-j+1)\dots (\bico-z)}{(z+1)\dots (z+j)}\cdot \prod_{i=1}^{2d}(z+i)\\
    %     =&\prod_{i=1}^{j}(\bico-z-j+i)\cdot \prod_{i=j+1}^{2d}(z+i),
    % \end{align*}

    \iffullversion
    $B(z)$ is equal to the following matrix:
    
    \begingroup
    \renewcommand*{\arraystretch}{1.5}
        \begin{align*}
        \resizebox{\textwidth}{!}{
        $\begin{pmatrix}
            \prod_{i=1}^{d}(\bico-z-d+i)\prod_{i=d+1}^{2d}(z+i) & \prod_{i=1}^{d+1}(\bico-z-d-1+i)\prod_{i=d+2}^{2d}(z+i)& \dots & \prod_{i=1}^{2d}(\bico-z-2d+i)\\
            \prod_{i=1}^{d-1}(\bico-z-d+1+i)\prod_{i=d}^{2d}(z+i) & \prod_{i=1}^{d}(\bico-z-d+i)\prod_{i=d+1}^{2d}(z+i) & \dots & \prod_{i=1}^{2d-1}(\bico-z-d+1+i)(z+2d)\\
            \vdots & \vdots & \ddots & \vdots \\
            \prod_{i=1}^{2d}(z+i) & (\bico-z)\prod_{i=2}^{2d}(z+i) &  \dots &  \prod_{i=1}^{d}(\bico-z-d+i)\prod_{i=d+1}^{2d}(z+i)
        \end{pmatrix}$
        }
    \end{align*}
    \endgroup
    
    \else
    $B(z) = (b(z)_{\sf row, col})$ is a $d+1$ by $d+1$ matrix, where the entry of row ${\sf row}=0,\dots,d$ and column ${\sf col} = 0,\dots,d$ equals
    \[b(z)_{\sf row, col} = \prod_{i=1}^{d+{\sf col} - {\sf row}}(\bico - z - d + i +{\sf row} - {\sf col})\prod_{d+{\sf col} - {\sf row}+1}^{2d}(z+i).\]
    For better understanding, we write out $B(z)$ in the matrix form in \cref{fig:matrix-B} in \cref{sec:matrix-fig} at the end of the appendix.
    \fi

    It is sufficient to show that there exists a $z^*\in[\lceil\frac{\bico}{2}\rceil, \lceil\frac{\bico}{2}\rceil+2d^2]$ such that $\det(B(z^*))\neq 0$.
    To show this, note that the determinant of $B(z)$ is a polynomial $q(z)$ of $z$ with a degree at most $2d^2$. 
    As long as $q(z)$ is not a zero polynomial, $q(z)$ has at most $2d^2$ roots. 
    That means, there must exist a $z^*\in[\lceil\frac{\bico}{2}\rceil, \lceil\frac{\bico}{2}\rceil+2d^2]$ such that $q(z^*)\neq 0$.
    The non-singularity of $A(z^*)$ thus follows.
    
    Hence, it suffices to show that $q(z)$ is not a zero polynomial.
    \iffullversion
    Indeed, when $z=\bico-d$, the matrix $B(z)$ becomes the following lower triangle matrix, which has a positive determinant.
    \begingroup  
    \renewcommand*{\arraystretch}{1.5}
        \begin{align*}
        \resizebox{0.8\textwidth}{!}{
        $\begin{pmatrix}
            \prod_{i=1}^{c}i\cdot \prod_{i=c+1}^{2c}(z+i) & 0& \dots & 0\\
            \prod_{i=1}^{c-1}(i+1)\cdot \prod_{i=c}^{2c}(z+i) & \prod_{i=1}^{c}i\cdot \prod_{i=c+1}^{2c}(z+i) & \dots & 0\\
            \vdots & \vdots & \ddots & \vdots \\
            \prod_{i=1}^{2c}(z+i) & c\cdot \prod_{i=2}^{2c}(z+i) &  \dots &  \prod_{i=1}^{c}i\cdot \prod_{i=c+1}^{2c}(z+i)
        \end{pmatrix}$
        }
        \end{align*}
    \endgroup 
    \else
    Indeed, when $z=\bico-d$, the matrix $B(z)$ becomes a lower triangle matrix where the diagonal elements equals 
    $\prod_{i=1}^{d} i \prod_{d+1}^{2d}(\bico - d +i)$.
    For better understanding, we write out $B(\bico-d)$ in the matrix form in \cref{fig:triangle-matrix} in \cref{sec:matrix-fig} at the end of the appendix.
    Thus, when $z = \bico - d$, the matrix $B(z)$ has a non-zero determinant.
    \fi
    This implies that $q(z)$ is not a zero polynomial.
    
    \paragraph{Infinity norm.} For simplicity, we use $A := A(z^*)$ in this part. 
    By \cref{lem:singular-norm}, $\|\MatA^{-1}\|_2 = \frac{1}{\lambda}$, where $\lambda$ is the smallest singular value of $\MatA$. 
    By \cref{lem:singular}, the smallest singular value $\lambda$ satisfies 
    \[\lambda \geq |\det(\MatA)|\cdot\left(\frac{d}{\|\MatA\|_F^2}\right)^{\frac{d}{2}}.\]
    By the definition of Frobenius norm and the fact that the largest term in $\MatA$ is 
    $\binom{\bico}{z^*}$,
    \[\|\MatA\|_F^2 = \sum_{i=0}^{c}\sum_{j=0}^{d} a^2_{ij}\leq (d+1)^2\cdot\binom{\bico}{z^*}^2.\]
    We only need to bound the determinant of $\MatA$.  
    % For any $0\leq i\leq d$, we have 
    % \[\frac{\binom{\bico}{z^*+i}}{\binom{\bico}{z^*}} = \frac{(\bico-z^*-i+1)\dots (\bico-z^*)}{(z^*+1)\dots (z^*+i)}.\]
    Let $\MatA' = (a'_{i,j})_{(d+1)\times(d+1)}$ where $a'_{i,j} = \frac{a_{i,j}}{\binom{\bico}{z^*}}$.
    Then we have that $|\det(\MatA)| = \binom{\bico}{z^*}^{(d+1)} \cdot |\det(\MatA')|$, 
    \iffullversion
    where
    \begingroup  
    \renewcommand*{\arraystretch}{1.8}
        \begin{align*}
        \MatA'= %\left\vert\det
        \begin{pmatrix}
            \frac{(\bico-z^*-d+1)\dots(\bico-z^*)}{(z^*+1)\dots(z^*+d)} & \frac{(\bico-z^*-d)\dots(\bico-z^*)}{(z^*+1)\dots(z^*+d+1)}& \dots & \frac{(\bico-z^*-2d+1)\dots(\bico-z^*)}{(z^*+1)\dots(z^*+2d)}\\
            \frac{(\bico-z^*-d+2)\dots(\bico-z^*)}{(z^*+1)\dots(z^*+d-1)} & \frac{(\bico-z^*-d+1)\dots(\bico-z^*)}{(z^*+1)\dots(z^*+d)} & \dots & \frac{(\bico-z^*-2d+2)\dots(\bico-z^*)}{(z^*+1)\dots(z^*+2d-1)}\\
            \vdots & \vdots & \ddots & \vdots \\
            1 & \frac{\bico-z^*}{z^*+1} &  \dots & \frac{(\bico-z^*-d+1)\dots(\bico-z^*)}{(z^*+1)\dots(z^*+d)}
        \end{pmatrix}
%        \right\vert
        \end{align*}
    \endgroup     
    \else
    For better understanding, we write out $A'$ in the matrix form in \cref{fig:matrix-B} in \cref{sec:matrix-fig} at the end of the appendix.
    \fi
    % \begingroup
    % \renewcommand*{\arraystretch}{1.5}
    %     \begin{align*}
    %     |\det(\MatA)| = \binom{\bico}{z^*}^{(c+1)} \cdot \left\lvert\det
    %     \begin{pmatrix}
    %         1 & \frac{z}{z+1} & \frac{(z-1)z}{(z+1)(z+2)} & \dots & \frac{(z-c+1)\dots z}{(z+1)\dots (z+c)}\\
    %         \frac{z}{z+1} & 1 & \frac{z}{z+1} & \dots & \frac{(z-c)\dots z}{(z+1)\dots (z+c-1)}\\
    %         \vdots & \vdots & \vdots & \ddots & \vdots \\
    %         \frac{(z-c+1)\dots z}{(z+1)\dots (z+c)} & \frac{(z-c)\dots z}{(z+1)\dots (z+c-1)} & \frac{(z-c-1)\dots z}{(z+1)\dots (z+c-2)} & \dots & 1
    %     \end{pmatrix}
    %     \right\rvert.
    %     \end{align*}
    % \endgroup
    
    By the definition of determinant, we have
    \[\det(\MatA') = \sum_{\sigma\in S_{d+1}} {\sf sgn}(\sigma) \prod_{i=1}^{d+1} a'_{i,\sigma_i}.\]
    For each $\sigma\in S_{d+1}$, let $q_\sigma$ denote the product $\prod_{i=1}^{d+1} a'_{i,\sigma_i}$.
    Since all the entries in $\MatA'$ are rational numbers, for each $\sigma$, the product $q_\sigma$ is also a rational number.
    Note that the denominator of each entry is a factor of $\prod_{i=1}^{2d} (z^*+i)$, we can write each $q_\sigma = \frac{p_{\sigma}}{\prod_{i=1}^{2d} (z^*+i)^{(d+1)}}$ for an integer $p_{\sigma}$.
\elaine{it's not so easy to see?}
    Thus, 
    \begin{align*}
        |\det (\MatA')| &= \left\vert\sum_{\sigma\in S_{d+1}} {\sf sgn}(\sigma) \prod_{i=1}^{d+1} a'_{i,\sigma_i}\right\vert\\
        &= \left\vert\sum_{\sigma\in S_{d+1}} {\sf sgn}(\sigma) \frac{p_{\sigma}}{\prod_{i=1}^{2d} (z^*+i)^{(d+1)}}\right\vert\\
        & = \frac{|\sum_{\sigma\in S_{d+1}}{\sf sgn}(\sigma) p_{\sigma}|}{\prod_{i=1}^{2d} (z^*+i)^{(d+1)}}\geq \frac{1}{\prod_{i=1}^d (z^*+i)^{(d+1)}},
    \end{align*}
    where the last step follows from the fact that $\MatA'$ is non-singular; thus the absolute value of the nominator is at least $1$.
    Therefore,
    \begin{align*}
        \lambda &\geq |\det(\MatA)|\cdot\left(\frac{d}{\|\MatA\|_F^2}\right)^{\frac{d}{2}}\\
        &\geq \binom{\bico}{z^*}^{(d+1)}\cdot \frac{1}{\prod_{i=1}^{2d} (z^*+i)^{(d+1)}} \cdot \left(\frac{d}{(d+1)^2\binom{\bico}{z^*}^2}\right)^{\frac{d}{2}}\\
        & \geq \binom{\bico}{z^*} \cdot \frac{1}{(z^*+2d)^{2d(d+1)}}\cdot\left(\frac{\sqrt{d}}{d+1}\right)^d.
    \end{align*}
    The claim thus follows from the fact that $\|\MatA^{-1}\|_{\infty}\leq \|\MatA^{-1}\|_2= \frac{1}{\lambda}$.
\end{proof}

\begin{proof}[Proof of \cref{clm:param}]
    Since $\MatA(z^*)$ is non-singular, we have $\correction = \MatA(z^*)^{-1}{\bf \Delta}$.
By properties of matrix norms, we have that 
\begin{equation}
\label{eqn:norm}
    \|\correction\|_{\infty}\leq \|\MatA(z^*)^{-1}\|_{\infty}\cdot\|{\bf \Delta}\|_{\infty}.
\end{equation}
By \cref{clm:infty-norm} and note that $\|{\bf \Delta}\|_{\infty}\leq  t \cdot \binom{\bico}{t}$, we have 
\[
    \|\correction\|_{\infty}\leq \|\MatA(z^*)^{-1}\|_{\infty}\cdot\|{\bf \Delta}\|_{\infty}
    \leq \frac{(z^*+2d)^{2d(d+1)}}{\binom{\bico}{z^*}}\cdot\left(\frac{d+1}{\sqrt{d}}\right)^d \cdot t \cdot \binom{\bico}{t}.
\]
Because $z^*+2d \leq \bico$, $\frac{d+1}{\sqrt{d}} \leq \bico$ and $t \leq \bico$, we have 
\begin{equation}\label{eq:delta-1}
    \|\correction\|_{\infty}\leq \frac{(z^*+2d)^{2d(d+1)}}{\binom{\bico}{z^*}}\cdot\left(\frac{d+1}{\sqrt{d}}\right)^d \cdot t \cdot \binom{\bico}{t}
    \leq  \bico^{2d^2+3d+1}\cdot \frac{\binom{\bico}{t}}{\binom{\bico}{z^*}}.
\end{equation}
% Equation (\ref{eq:delta-1})

By the assumption that $d\leq \frac{1}{8}\sqrt{\frac{h}{2\log h}}$, we have
\[
(2d^2+3d+1)\cdot\log(h-d)+\frac{\log\frac{6}{5}}{4}\cdot d \leq 8d^2\cdot \log h\leq 
\frac{h}{4}\cdot \log\frac{6}{5}.
\]
Re-arrange the inequality and notice that $h-d \leq \bico$, we have
\begin{equation}\label{eq:delta-2}
2d^2+3d+1 \leq \frac{(h-d)\log\frac{6}{5}}{4\log (h-d)} \leq \frac{\bico\log\frac{6}{5}}{4\log \bico},
\end{equation}
therefore,
\begin{equation}\label{eq:delta-3}
    \bico^{2d^2+3d+1} \leq \left(\frac{6}{5}\right)^{\frac{\bico}{4}}.
\end{equation}
% Equation (\ref{eq:delta-3})

Next, note that for any integers $a,b$ such that $a < b$, we have 
$\frac{\binom{\bico}{a}}{\binom{\bico}{b}} = \frac{(a+1)(a+2)\cdots b}{(\bico - b + 1)(\bico - b + 2)\cdots (\bico - a)} \leq (\frac{b}{\bico - a})^{b-a}$.
Because $z^* \in [\lceil\frac{\bico}{2}\rceil, \lceil\frac{\bico}{2}\rceil+2d^2]$, we have $\binom{\bico}{z^*} \geq \binom{\bico}{\lceil\frac{\bico}{2}\rceil + 2d^2}$.
% Recall that $t = \lfloor \frac{\bico}{4}\rfloor$ and $z^* \leq \lceil \lceil\frac{\bico}{2}\rceil\rceil + 2d^2$, so we have 
Thus,
\begin{align}
    \frac{\binom{\bico}{t}}{\binom{\bico}{z^*}} &\leq  \frac{\binom{\bico}{t}}{\binom{\bico}{\lceil\frac{\bico}{2}\rceil + 2d^2}}
  \leq \left(\frac{\lceil\frac{\bico}{2}\rceil + 2d^2}{\bico - t}\right)^{\lceil\frac{\bico}{2}\rceil + 2d^2 - t}\nonumber\\
  &\leq \left(\frac{\frac{\bico}{2}+2d^2+1}{\frac{3}{4}\bico}\right)^{\lceil\frac{\bico}{2}\rceil + 2d^2 - t}.
  \label{eq:delta-4}
\end{align}

Because $\bico \geq h - d$, for any $h\geq 2$ and $d\leq \frac{1}{8}\sqrt{\frac{h}{2\log h}}$, it must be $\frac{\log\frac{6}{5}}{\log \bico} < \frac{1}{2}$.
By Equation (\ref{eq:delta-2}), we have \[
    2d^2+1 \leq 2d^2+3d+1 \leq\frac{\bico\log\frac{6}{5}}{4\log \bico} \leq \frac{\bico}{8}.
\]
By $2d^2+1 \leq \frac{\bico}{8}$ and Equation (\ref{eq:delta-4}), we have
\begin{equation}\label{eq:delta-5}
  \frac{\binom{\bico}{t}}{\binom{\bico}{z^*}}
  \leq \left(\frac{\frac{\bico}{2}+2d^2+1}{\frac{3}{4}\bico}\right)^{\lceil\frac{\bico}{2}\rceil + 2d^2 - t}
  \leq \left(\frac{5}{6}\right)^{\lceil\frac{\bico}{2}\rceil + 2d^2 - t}
  \leq \left(\frac{5}{6}\right)^{\frac{\bico}{4}}.
\end{equation}
Combining Equations (\ref{eq:delta-1}), (\ref{eq:delta-3}), and (\ref{eq:delta-5}),
we have that $\|\correction\|_{\infty} \leq 1.$
\end{proof}

\section{Characterization for Finite Block Size}
\label{sec:finite}
\subsection{Characterization for Strict IC}
In this section, we give a characterization for strict incentive compatibility for finite block size.
In an $\environ$-environment, we can indeed circumvent the $0$-miner revenue impossibility result in ~\cite{crypto-tfm}.
However, it turns out that for $c = 1$ and $c \geq 2$, the mechanisms are different.
Specifically, for $c\geq 2$, each user's utility has to be $0$.
Therefore, we separately give the mechanisms for $c = 1$ and $c \geq 2$.

\subsubsection{Feasibility for $c=1$}
\label{sec:finc1}
For $c = 1$, the mechanism is simply the LP-based mechanism in \cref{sec:infinite} with a random selection process.
Still, we assume that honest users' values are sampled i.i.d. from some distribution $\mathcal{D}$, and the median $m$ of the distribution satisfies that $\Pr[x\geq m] = \frac{1}{2}$ (see \cref{rmk:nomedian}).
For convenience, we repeat the MPC-assisted, LP-based mechanism with random selection, which has been introduced in \cref{sec:roadmap-finite}.
\begin{mdframed}
    \begin{center}
    {\bf MPC-assisted, LP-based mechanism with random selection}
    \end{center}
    \paragraph{Parameters:} the block size $k$, the environment parameter $(h,*,1,d)$, the distribution median $\m$. 
    
    \paragraph{Input:} a bid vector $\bfb = (b_1,\dots,b_n)$.
    
    \paragraph{Mechanism:}
    \begin{itemize}[leftmargin=5mm,itemsep=1pt]
    \item 
    {\it Confirmation Rule.}
    Let $\widetilde{\bids} = (\widetilde{b}_1,\dots,{\widetilde{b}_{s}})$ denote the bids that are at least $\m$.
    If $s \leq k$, confirm all bids in $\widetilde{\bids}$.
    Otherwise, randomly select $k$ bids from $\widetilde{\bids}$ to confirm.
    \item 
    {\it Payment rule.}
    Each confirmed bid pays $\m$.
    \item
    {\it Miner revenue rule.}
    % Let $\bico:=n-d$ and $\target = \frac{\min(h,k)}{4}\cdot \m$.
    % Let $\minerrev=(\target, \target,\dots, \target)^T \in\R^{c+1}$.
    % Define $M\in\R^{(c+1)\times(n+1)}$ to be the following matrix: 
    % \begingroup
    %     \renewcommand*{\arraystretch}{1.5}
    %     \begin{equation*}
    %     M = \frac{1}{2^{\bico}}
    %     \begin{pmatrix}
    %     \binom{\bico}{0} & \binom{\bico}{1} & \binom{\bico}{2} & \dots &  \binom{\bico}{\bico} & 0 & 0 & \dots & 0\\
    %     0 & \binom{\bico}{0} & \binom{\bico}{1} & \dots & \binom{\bico}{\bico-1} & \binom{\bico}{\bico} & 0 & \dots & 0\\
    %     \vdots & \vdots & \vdots & \vdots & \vdots & \vdots & \vdots & \vdots & \vdots \\
    %     0 & 0 & 0 & \dots & \dots & \dots & \binom{\bico}{\bico-2}& \binom{\bico}{\bico-1}&\binom{\bico}{\bico} \\
    %     \end{pmatrix}
    %     \end{equation*}
    % \endgroup
    % That is, the $i$-th row of $M$ is $\left(\binom{\bico}{0}, \binom{\bico}{1},  \dots,\binom{\bico}{\bico}, 0 \dots,0\right)$ right-shifted by $i-1$ elements.
    % Let ${\bf y}=(y_0,\dots,y_n)$ be the solution to the following linear program $\mathcal{P}$:
    % \begin{align}
    % \label{eqn:lp-finite}
    %     &\MatM{\bf y} = \minerrev,\\
    %     \text{s.t. }& 0\leq y_i\leq \min(i,k) \text{ for all } i.\nonumber
    % \end{align}
    % Recall that $s$ is the number of bids in $\bids$ larger than or equal to $\m$. Miner gets $y_s\cdot \m$.
    Let 
${\bf y} := (y_0, y_1, \ldots, y_n)$ 
be any feasible solution to the following linear program:
\begin{alignat}{2}
\forall i \in [n]: & \ \   0 \leq y_i \leq \min(i, k)\cdot m
\label{eqn:lp-finite-budget-appendix}
\\
\forall 0 \leq j \leq d:  
& \ \ \sum_{i = 0}^{n-d} q_i \cdot y_{i + j} 
= \frac{m \cdot \min(h, k)}{4} 
\label{eqn:lp-finite-appendix}
\end{alignat}
where $q_i = \frac{1}{2^{n-d}}\binom{n-d}{i}$ is the probability 
of observing $i$ heads if we flip $n-d$ independent fair coins.
The total miner revenue is $y_t$ where $t$ is the number of confirmed bids in the block.
\end{itemize}
\end{mdframed}
\begin{theorem}
\label{thm:lp-finite}
    Suppose the block size is $k$.
    Fix any\footnote{For the special case $h =1$, we can just
    use the parity-based mechanism of \Cref{sec:roadmap-inf} with the random selection.}
    $h \geq 2$,
    \hao{Double check whether parity-based mech with the random selection is fine.}
    and any $d \leq \frac18\sqrt{\frac{h}{2\log h}}$.
    The MPC-assisted, LP-based mechanism with random selection is ex post UIC, Bayesian MIC, and Bayesian SCP in an $(h,*,1,d)$-environment.
    Moreover, the expected miner revenue is $\Theta(\min\{h,k\})$.
\end{theorem}
\begin{proof}
First, \cref{eqn:lp-finite-budget-appendix} guarantees that total miner revenue is at most the total payment of the confirmed users,
so the mechanism satisfies budget feasibility.
% The mechanism confirms $\min(s,k)$ number of bids, where $s$ is the number of bids at least $\m$. x

Next, we show that when the linear program \cref{eqn:lp-finite-budget-appendix,eqn:lp-finite-appendix} has a solution, the mechanism satisfies all three incentive-compatible properties.

\begin{itemize}
\item 
\textbf{UIC:}
By the same reasoning as in \cref{lem:parity}, overbidding or underbidding does not increase the user's utility.
Injecting bids cannot increase the user's utility either: it may only decrease the probability that the user gets confirmed.
Moreover, dropping out can only give the user zero utility.
Therefore, a user cannot increase its utility by deviating.
\item \textbf{SCP:}
By the same reasoning as in the proof of \cref{thm:lp}, the linear program \cref{eqn:lp-finite-appendix} guarantees that no matter how the coalition chooses the $d$ bids it controls, the expected total miner revenue remains unchanged.
    Meanwhile, the coalition cannot increase the colluding user's utility by UIC. 
    Therefore, this mechanism is SCP.
\item \textbf{MIC:}
Follows by the same reasoning as SCP.
\end{itemize}
It remains to show that the linear program indeed has a feasible solution. 
We will give a constructive solution.
Let $\widetilde{\bf y} = (\widetilde{y}_1,\dots, \widetilde{y}_n)$ denote the constructive solution given in the proof of \cref{lem:delta-sol} that satisfies 
    \[\forall 0 \leq j \leq d:  
 \ \ \sum_{i = 0}^{n-d} q_i \cdot \widetilde{y}_{i + j} 
= \frac{m \cdot h}{4}.\]
In the proof of \cref{lem:delta-sol}, $\widetilde{\bf y}$ satisfies that $0\leq \widetilde{y}_i\leq \min(i,h) \cdot m$ for any $0\leq i\leq n$.
There are two possible cases.
\begin{itemize}
\item $h\leq k$.
We have $0\leq \widetilde{y}_i\leq \min(i,h) \cdot m \leq \min(i,k) \cdot m$.
Thus, $\widetilde{\bf y}$ is a feasible solution to the linear program in this case.
\item $h > k$.
Let ${\bf y} = (y_0,\dots,y_n) = \frac{k}{h}\cdot \widetilde{\bf y}$. 
    Then $y_i$ satisfies that $0\leq y_i \leq \frac{k}{h}\min(i,h)\cdot m\leq \min(i,k)\cdot m$.
    Moreover, for any $0\leq j\leq d$, 
    \[\sum_{i = 0}^{n-d} q_i \cdot y_{i + j} 
= \frac{k}{h}\cdot \sum_{i = 0}^{n-d} q_i \cdot \widetilde{y}_{i + j} = \frac{m \cdot k}{4}.\]
Thus, ${\bf y} = \frac{k}{h}\cdot\widetilde{\bf y}$ is a feasible solution to the linear program if $h > k$.

\end{itemize}
    
\end{proof}

\subsubsection{Zero Social Welfare for Users When $c\geq 2$} 
\label{sec:zero-usw}
Unfortunately, the above MPC-assisted, LP-based mechanism with random selection only works for $c = 1$. 
When $c\geq2$, although deviating cannot increase the expected total miner revenue, the coalition 
can increase a colluding user's utility.
Imagine that the coalition consists of some colluding miners and two users $i$ and $j$, where user $i$ has true value $\m$ and user $j$ has a large true value.
Then user $i$ may choose to drop out to increase the probability of user $j$ getting confirmed.
This strictly increases the expected joint utility of the coalition.

Therefore, to construct a Bayesian SCP mechanism in an $\environ$-environment for $d\geq c\geq 2$, we need to make sure that deviating cannot increase a colluding user's utility.
Indeed, for some (contrived) distributions, 
we can construct a mechanism that generates optimal miner revenue
and achieves UIC, MIC, and SCP in an $\environ$-environment for $d\geq c\geq 2$.
However, the total social welfare for all users is $0$. 
For example, imagine that honest users' true values are drawn i.i.d. from ${\rm Bernoulli}(\frac{1}{2})$.
Now, if we run the 
MPC-assisted, LP-based mechanism with random selection
(see \Cref{sec:finc1}) and set $m = 1$, 
the resulting mechanism achieves ex post UIC, Bayesian MIC, and Bayesian SCP
in $(h, *, c, d)$-environments, even when $c \geq 2$ (as long 
as the condition $d \leq \frac18\sqrt{\frac{h}{2\log h}}$ is satisfied).
This is because setting $m = 1$ makes sure that every user's utility is always $0$.
Thus, no matter how the coalition deviates, it cannot increase the strategic users' joint utility.
Moreover, as long as the linear program \cref{eqn:lp-finite-budget-appendix,eqn:lp-finite-appendix} has a feasible solution, the coalition cannot increase the expected total miner revenue either.
The mechanism achieves $\Theta(m)$ expected miner revenue
but unfortunately, the total user social welfare
is always $0$.
%Unfortunately, the above mechanism degenerates in some sense: the users' social welfare is always $0$.
It turns out that this zero user social welfare limitation is intrinsic,
as stated below.
\begin{theorem}[Restatement of \cref{thm:zero-user-welfare-informal}]
\label{thm:zero-usw}
Suppose that the block size is finite, and fix any $h \geq 1$, any $d \geq c \geq 2$,
and any $\rho \in (0, 1)$.
Then, 
any MPC-assisted TFM 
that simultaneously satisfies 
Bayesian UIC, MIC and SCP 
in an $(h, \rho, c, d)$-model
must suffer from $0$ social welfare for the users 
when there actually are more than $h$ honest bids. 
Equivalently, for any $\ell > h$, 
\begin{equation}
\label{eqn:finite-LB}
    %\underset{\bids_{-i}\sim\mathcal{D}^{\ell}}{\E}[{\sf util}^i(v,\bids_{-i})] = 0.
    \underset{\bids\sim\mathcal{D}^{\ell}}{\E}[{\sf USW}(\bids)] = 0.
\end{equation}
In the above, ${\sf USW}(\bids)$
denotes the expected total user social welfare under the bid vector $\bids$
where the expectation is taken over the randomness of the mechanism.
\end{theorem}

\noindent The proof is similar to the proof of Theorem 5.2 of \cite{crypto-tfm}. 
% Intuitively speaking, we want to guarantee that injecting fake bids and dropping out cannot increase the colluding users' joint utility. 
% This implies that each user's utility should remain the same when the number of input bids to the mechanism changes.
% That is to say, for any input length, each user's utility should be equivalent to its utility when there are infinitely many bids, under which each user's utility is just $0$.
% % We defer the full proof to \ke{fill}
We will use the following lemma of~\cite{crypto-tfm} to prove this theorem.
Although the original lemma considers a universal MPC-assisted mechanism, the proof also holds for MPC-assisted TFM in an $\environ$-environment for $d\geq c\geq 2$.
Henceforth, we use ${\sf util}^i(\bids)$ to denote the utility of identity $i$ when the input bid vector is $\bids$.
In the proof, we use $v_{\textsf{id}}$ ($b_{\textsf{id}}$) to denote a bid $v$ ($b$) coming from identity $\textsf{id}$.
\begin{lemma}
\label{lem:hartline}
Fix any $h\geq 1$, any $d\geq c\geq 2$, any $\rho\in(0,1)$.
%Suppose each user's true value is drawn i.i.d. from a distribution $\mathcal{D}$.
Given any (possibly random) MPC-assisted mechanism that is Bayesian UIC, MIC and SCP in an $\environ$-environment,
for any identity $i$ and identity $j$, for any bid $b_j$ and $b'_j$, it must be that for any $\ell\geq h$,
\begin{equation}
\label{eqn:hartline}
    \underset{(v_i,\bids_{-i,j})\sim\mathcal{D}^{\ell+1}}{\E}[{\sf util}^i(v_i,b_j,\bids_{-i,j})] = \underset{(v_i,\bids_{-i,j})\sim\mathcal{D}^{\ell+1}}{\E}[{\sf util}^i(v_i,b'_j,\bids_{-i,j})],
\end{equation}
where $\bids_{-i,j}$ represents all except identity $i$ and $j$'s bids. Moreover, it must be that 
\begin{align}
\label{eqn:util-unchange}
    \underset{(v_i,\bids_{-i,j})\sim\mathcal{D}^{\ell+1}}{\E}[{\sf util}^i(v_i,b_j,\bids_{-i,j})] = \underset{(v_i,\bids_{-i})\sim\mathcal{D}^{\ell+1}}{\E}[{\sf util}^i(v_i,\bids_{-i})].
\end{align}
\end{lemma}
\begin{proof}
The proof to this lemma is the same as in Lemma 5.2 and 5.3 of \cite{crypto-tfm}, 
except that now we need to guarantee that at least $h$ bids 
are sampled randomly from $\mathcal{D}$.
\end{proof}

\begin{corollary}
\label{cor:uihv}
Fix any $h\geq 1$, any $d\geq c\geq 2$, any $\rho\in(0,1)$.
%Suppose each user's true value is drawn i.i.d. from a distribution $\mathcal{D}$.
Given any (possibly random) MPC-assisted mechanism  that is Bayesian UIC, MIC and SCP in an $\environ$-environment,
for any two sets $H$ and $H'$ consisting of at least $h$ identities, let $\bids_H$ ($\bids_{H'}$) denote the bids from identities in $H$ ($H'$). 
For any $i\notin H\cup H'$, 
it must be that 
\[\underset{(v_i,\bids_{H})\sim\mathcal{D}^{|H|+1}}{\E}[{\sf util}^i(v_i,\bids_{H})] = \underset{(v_i,\bids_{H'})\sim\mathcal{D}^{|H'|+1}}{\E}[{\sf util}^i(v_i,\bids_{H'})],\]
where $v_i$ denotes that identity $i$ bids $v$.
% define \[U_i^{H,v}:= \underset{\bids_{H}\sim\mathcal{D}^{|H|}}{\E}[{\sf util}^i(v_i,\bids_{H})],\]
% where $v_i$ means that user $i$ bids $v$.
% For any set $S$ consisting of users not in $H\cup\{i\}$,  it must be that
% \begin{align}
% \label{eqn:util-U-i-H-v}
% \underset{\bids_S,\bids_{H}\sim\mathcal{D}^{|S|+|H|}}{\E}[{\sf util}^i(v_i,{\bf b}_{S},\bids_{H})]=
%     U_i^{H,v}.
% \end{align}
\end{corollary}
\begin{proof}
    % First, we show that \[\underset{\bids_{H}\sim\mathcal{D}^{|H|}}{\E}[{\sf util}^i(v_i,{\bf v}_S,\bids_{H})] = \underset{(\bids_S,\bids_H)\sim\mathcal{D}^{K+|H|}}{\E}[{\sf util}^i(v_i,\bids_S,\bids_{H})].\]
    Let $S = H'\setminus H$.
    Without loss of generality, we assume that $S$ consists of identities $1,\dots, |S|$.
    By the definition of the total expectation, we have
    \iffullversion
    \begin{align*}
    \label{eqn:cor-uihv}
    &\underset{(v_i,\bids_S,\bids_H)\sim\mathcal{D}^{|S|+|H|+1}}{\E}[{\sf util}^i(v_i,\bids_S,\bids_{H})]\nonumber\\
        = &\int_{0}^{\infty} \underset{(v_i,\bids_{S\setminus\{1\}},\bids_H)\sim\mathcal{D}^{|S|+|H|}}{\E}[{\sf util}^i(v_i,z_1,\bids_{S\setminus\{1\}},\bids_{H})]f(z_1) dz_1\nonumber\\
        = & \underset{(v_i,\bids_{S\setminus\{1\}},\bids_H)\sim\mathcal{D}^{|S|+|H|}}{\E}[{\sf util}^i(v_i,b_1,\bids_{S\setminus\{1\}},\bids_{H})]\int_{0}^{\infty} f(z_1) dz_1 &\text{By \cref{eqn:hartline}}\nonumber\\
        = & \underset{(v_i,\bids_{S\setminus\{1\}},\bids_H)\sim\mathcal{D}^{|S|+|H|}}{\E}[{\sf util}^i(v_i,b_1,\bids_{S\setminus\{1\}},\bids_{H})]\\
        =& \underset{(v_i,\bids_{S\setminus\{1\}},\bids_H)\sim\mathcal{D}^{|S|+|H|}}{\E}[{\sf util}^i(v_i,\bids_{S\setminus\{1\}},\bids_{H})] &\text{By \cref{eqn:util-unchange}}\nonumber\\
        =& \dots = \underset{(v_i,\bids_H)\sim\mathcal{D}^{|H|+1}}{\E}[{\sf util}^i(v_i,\bids_H)].
    \end{align*}
    \else
    \begin{align*}
    \label{eqn:cor-uihv}
    &\underset{(v_i,\bids_S,\bids_H)\sim\mathcal{D}^{|S|+|H|+1}}{\E}[{\sf util}^i(v_i,\bids_S,\bids_{H})]\nonumber\\
        = &\int_{0}^{\infty} \underset{(v_i,\bids_{S\setminus\{1\}},\bids_H)\sim\mathcal{D}^{|S|+|H|}}{\E}[{\sf util}^i(v_i,z_1,\bids_{S\setminus\{1\}},\bids_{H})]f(z_1) dz_1\nonumber\\
        = & \underset{(v_i,\bids_{S\setminus\{1\}},\bids_H)\sim\mathcal{D}^{|S|+|H|}}{\E}[{\sf util}^i(v_i,b_1,\bids_{S\setminus\{1\}},\bids_{H})]\int_{0}^{\infty} f(z_1) dz_1 \\
        = & \underset{(v_i,\bids_{S\setminus\{1\}},\bids_H)\sim\mathcal{D}^{|S|+|H|}}{\E}[{\sf util}^i(v_i,b_1,\bids_{S\setminus\{1\}},\bids_{H})]\\
        =& \underset{(v_i,\bids_{S\setminus\{1\}},\bids_H)\sim\mathcal{D}^{|S|+|H|}}{\E}[{\sf util}^i(v_i,\bids_{S\setminus\{1\}},\bids_{H})]\\
        =& \dots = \underset{(v_i,\bids_H)\sim\mathcal{D}^{|H|+1}}{\E}[{\sf util}^i(v_i,\bids_H)],
    \end{align*}
    where the second equality comes from \cref{eqn:hartline} and the fourth equality follows from \cref{eqn:util-unchange}.
    \fi
    By the same reasoning, consider $S' = H\setminus H'$.
    Then it must be that 
    \iffullversion
    \[\underset{(v_i,\bids_{S'},\bids_{H'})\sim\mathcal{D}^{|S'|+|H'|+1}}{\E}[{\sf util}^i(v_i,\bids_{S'},\bids_{H'})] = \underset{(v_i,\bids_{H'})\sim\mathcal{D}^{|H'|+1}}{\E}[{\sf util}^i(v_i,\bids_{H'})].\]
    \else
    \begin{align*}     &\underset{(v_i,\bids_{S'},\bids_{H'})\sim\mathcal{D}^{|S'|+|H'|+1}}{\E}[{\sf util}^i(v_i,\bids_{S'},\bids_{H'})] \\
    = &\underset{(v_i,\bids_{H'})\sim\mathcal{D}^{|H'|+1}}{\E}[{\sf util}^i(v_i,\bids_{H'})].
    \end{align*}
    \fi
    Note that $S' \cup H' = S\cup H = H'\cup H$.
    Hence, 
    \iffullversion
    \[\underset{(v_i,\bids_{S'},\bids_{H'})\sim\mathcal{D}^{|S'|+|H'|+1}}{\E}[{\sf util}^i(v_i,\bids_{S'},\bids_{H'})] = \underset{(v_i,\bids_S,\bids_H)\sim\mathcal{D}^{|S|+|H|+1}}{\E}[{\sf util}^i(v_i,\bids_S,\bids_{H})].\]
    \else
    \begin{align*}
&\underset{(v_i,\bids_{S'},\bids_{H'})\sim\mathcal{D}^{|S'|+|H'|+1}}{\E}[{\sf util}^i(v_i,\bids_{S'},\bids_{H'})] \\
= &\underset{(v_i,\bids_S,\bids_H)\sim\mathcal{D}^{|S|+|H|+1}}{\E}[{\sf util}^i(v_i,\bids_S,\bids_{H})].
    \end{align*}
    \fi
    Combining the equalities, we have
    \iffullversion
    \begin{align*}
\underset{(v_i,\bids_{H})\sim\mathcal{D}^{|H|+1}}{\E}[{\sf util}^i(v_i,\bids_{H})]\\
& = \underset{(v_i,\bids_S,\bids_H)\sim\mathcal{D}^{|S|+|H|+1}}{\E}[{\sf util}^i(v_i,\bids_S,\bids_{H})]\\
        & = \underset{(v_i,\bids_{S'},\bids_{H'})\sim\mathcal{D}^{|S'|+|H'|+1}}{\E}[{\sf util}^i(v_i,\bids_{S'},\bids_{H'})]\\
        & = \underset{(v_i,\bids_{H'})\sim\mathcal{D}^{|H'|+1}}{\E}[{\sf util}^i(v_i,\bids_{H'})]
    \end{align*}
    \else
        \begin{align*}
        &\underset{(v_i,\bids_{H})\sim\mathcal{D}^{|H|+1}}{\E}[{\sf util}^i(v_i,\bids_{H})]
        = \underset{(v_i,\bids_S,\bids_H)\sim\mathcal{D}^{|S|+|H|+1}}{\E}[{\sf util}^i(v_i,\bids_S,\bids_{H})]\\
        =& \underset{(v_i,\bids_{S'},\bids_{H'})\sim\mathcal{D}^{|S'|+|H'|+1}}{\E}[{\sf util}^i(v_i,\bids_{S'},\bids_{H'})]\\
        =& \underset{(v_i,\bids_{H'})\sim\mathcal{D}^{|H'|+1}}{\E}[{\sf util}^i(v_i,\bids_{H'})]
    \end{align*}
    \fi
    % Meanwhile, by \cref{eqn:util-unchange}, we have
    % \begin{align*}
    % \underset{(\bids_S,\bids_H)\sim\mathcal{D}^{|S|+|H|}}{\E}[{\sf util}^i(v_i,\bids_S,\bids_{H})] &= \underset{(\bids_{S\setminus\{1\}},\bids_H)\sim\mathcal{D}^{K+|H|-1}}{\E}[{\sf util}^i(v_i,b_1,\bids_{S\setminus\{1\}},\bids_{H})]\\
    % &= \underset{(\bids_{S\setminus\{1\}},\bids_H)\sim\mathcal{D}^{K+|H|-1}}{\E}[{\sf util}^i(v_i,\bids_{S\setminus\{1\}},\bids_{H})].
    % \end{align*}
    % Applying this argument for $|S|$ times, we get 
    % \[\underset{(\bids_S,\bids_H)\sim\mathcal{D}^{|S|+|H|}}{\E}[{\sf util}^i(v_i,\bids_S,\bids_{H})] = \underset{\bids_H\sim\mathcal{D}^{|H|}}{\E}[{\sf util}^i(v_i,\bids_{H})].\]
    % The corollary thus follows by combining the above equality with \cref{eqn:cor-uihv}.
\end{proof}

This corollary implies that when identity $i$'s bid is sampled from $\mcal{D}$
in a world with $h$ or more random bids,
its expected utility only depends on its identity $i$. 
%Thus, for any identity $i$, any value $v$, for any set $H$ of at least $h$ number of identities other than $i$, we define
Henceforth we will use the following notation to denote this utility (where the notation $v_i$ means identity $i$ is bidding
the value $v$):
\begin{equation}
    \label{eqn:defn-uiv}
    U_i:= \underset{(v_i,\bids_H)\sim\mathcal{D}^{|H|+1}}{\E}[{\sf util}^i(v_i,\bids_H)].
\end{equation}

\begin{lemma}
\label{lem:equal-util}
Fix any $h\geq 1$, any $d\geq c\geq 2$, any $\rho\in(0,1)$.
%Suppose each user's true value is drawn i.i.d. from a distribution $\mathcal{D}$.
Given any (possibly random) MPC-assisted mechanism that is Bayesian UIC, MIC and SCP in an $\environ$-environment,
for any user $i,j$, 
it must be that 
\begin{equation}
\label{eqn:equal-util}
    U_i = U_j.
\end{equation}
\end{lemma}
\begin{proof}
    % Let $l$ be any identity other than $i, j$ that is not in $H$, and $H' = H\cup \{l\}$.  
    % Combining the inequality above and \cref{cor:uihv}, we have 
    % for any $l\notin H$, let $H' = H\cup \{l\}$, then
    % \begin{align*}
    %     \underset{\bids_{H'}\sim\mathcal{D}^{|H|+1}}{\E}[{\sf util}^i(v_i,\bids_{H'})] = \int_{0}^{\infty}\underset{\bids_{H}\sim\mathcal{D}^{|H|}}{\E}[{\sf util}^i(v_i,b_{\ell},\bids_{H})]f(b_{l})db_{l} = \underset{\bids_{H}\sim\mathcal{D}^{|H|+1}}{\E}[{\sf util}^i(v_i,\bids_{H})].
    % \end{align*}
    % The same argument holds for $\underset{\bids_{H}\sim\mathcal{D}^{|H|}}{\E}[{\sf util}^j(v_j,\bids_{H})]$.
    % Therefore, this implies that
    % \[\underset{\bids_{H'}\sim\mathcal{D}^{|H|+1}}{\E}[{\sf util}^i(v_i,\bids_{H'})] > \underset{\bids_{H'}\sim\mathcal{D}^{|H|+1}}{\E}[{\sf util}^j(v_j,\bids_{H'})].\]
    Fix any set $H$ of at least $h+1$ number of users.
    By our symmetric assumption, it must be that 
    \begin{equation}
    \label{eqn:equal-usw}
        \underset{\bids_{H}\sim\mathcal{D}^{|H|}}{\E}\left[{\sf USW}(v_{i},\bids_{H})\right] =
    \underset{\bids_{H}\sim\mathcal{D}^{|H|}}{\E}\left[{\sf USW}(v_{j},\bids_{H})\right],
    \end{equation}
    where $v_i$ ($v_j$) denotes that identity $i$ ($j$) bids $v$, and ${\sf USW}(\bids)$ denotes the expected social welfare for all users when the input bid vector is $\bids$.
    For any identity $l\in H$, for any $v_i$ from identity $i$, let $H' = H\setminus\{l\}$. 
    It must be
    \iffullversion
    \begin{align*}
        \underset{(b_l,\bids_{H'})\sim\mathcal{D}^{|H|}}{\E}[{\sf util}^{l}(b_l, v_{i},\bids_{H'})] &= \underset{(b_l,\bids_{H'})\sim\mathcal{D}^{|H|}}{\E}[{\sf util}^{l}(b_l, \bids_{H'})] &\text{By \cref{eqn:util-unchange}}\\
        &= U_l &\text{ By \cref{eqn:defn-uiv}}
    \end{align*}
    \else
    \begin{align*}
        &\underset{(b_l,\bids_{H'})\sim\mathcal{D}^{|H|}}{\E}[{\sf util}^{l}(b_l, v_{i},\bids_{H'})]\\
        =& \underset{(b_l,\bids_{H'})\sim\mathcal{D}^{|H|}}{\E}[{\sf util}^{l}(b_l, \bids_{H'})] &\text{By \cref{eqn:util-unchange}}\\
        =& U_l &\text{ By \cref{eqn:defn-uiv}}
    \end{align*}
    \fi
    By the same reasoning, $\underset{(b_l,\bids_{H'})\sim\mathcal{D}^{|H|}}{\E}[{\sf util}^{l}(b_l,v_j,\bids_{H'})] = U_l$.
    Thus, for any value $v$, the sum of the expected utility of every user in $H$ is
    \begin{align*}
        \sum_{l\in H} \underset{\bids_H\sim \mathcal{D}^{|H|}}{\E}[{\sf util}^l(v_i,\bids_H)] = \sum_{l\in H} U_l = \sum_{l\in H} \underset{\bids_H\sim \mathcal{D}^{|H|}}{\E}[{\sf util}^l(v_j,\bids_H)].
    \end{align*}
    Combining this with \cref{eqn:equal-usw}, it must be that for any $v_i$ and $v_j$ (which denote that identity $i$ and $j$ bid value $v$, respectively),
    \[\underset{\bids_H\sim \mathcal{D}^{|H|}}{\E}[{\sf util}^i(v_i,\bids_H)] = \underset{\bids_H\sim \mathcal{D}^{|H|}}{\E}[{\sf util}^j(v_j,\bids_H)].\]
    The lemma follows by taking expectations over $v$ on both sides.
    % To see this,
    % \begin{align*}
    % \underset{\bids_{H'}\sim\mathcal{D}^{|H|-1}}{\E}[{\sf util}^{i^*}(v^*, v_{i},\bids_{H'})] &=  \underset{\bids_{H'}\sim\mathcal{D}^{|H|-1}}{\E}[{\sf util}^{i^*}(v^*, \bids_{H'})] &\text{By \cref{eqn:util-unchange}}\\
    % &=U_{i^*}^{v^*} &\text{By \cref{eqn:defn-uiv}}.
    % \end{align*}
    % Similarly, 
    % \[\underset{\bids_{H'}\sim\mathcal{D}^{|H|-1}}{\E}[{\sf util}^{i^*}(v^*,v_j,\bids_{H'})] = U_{i^*}^{v^*}.\]
    % Thus, we reach a contradiction.
\end{proof}

\begin{lemma}
    \label{lem:zero-util}
    Fix any $h\geq 1$, any $d\geq c\geq 2$, any $\rho\in(0,1)$,
and suppose that the distribution $\mcal{D}$ has bounded support.
%    Suppose each user's true value is drawn i.i.d.~from a distribution $\mathcal{D}$.
    Given any (possibly random) MPC-assisted mechanism that is Bayesian UIC, MIC and SCP in an $\environ$-environment,
    for any identity $i$, %for any value $v$, 
it must be that 
%    \[U_i^v = 0.\]
\[U_i = 0.\]
\end{lemma}
\begin{proof}
Consider a crowded world with many users 
and all of their bids are sampled independently at 
random from $\mcal{D}$. 
Let $K$ be the total number of users.
By \Cref{cor:uihv,lem:equal-util}, every user's expected
utility is the same where the expectation is taken over the random
coins for sampling all bids as well as random coins of the mechanism.
On the other hand, since there are  
$K$ total bids, there must exist a user whose confirmation
probability is at most $k/K$, and thus its expected utility 
is at most $\max(\mcal{D}) \cdot k/K$ where $k$ is the block size.
The lemma follows by taking $K$ to be arbitrarily large.
\end{proof}

    \paragraph{Proof of \cref{thm:zero-usw}} 
    Fix any set $H$ of size at least $h+1$.
    Then \[\underset{\bids\in \mathcal{D}^{|H|}}{\E}[{\sf USW}(\bids)] = \sum_{i\in H} \underset{\bids\in \mathcal{D}^{|H|}}{\E}{\sf util}^i(\bids).\]
    By \cref{lem:zero-util}, for each identity $i$ in $H$, 
    \begin{align*}
        \underset{\bids\in \mathcal{D}^{|H|}}{\E}{\sf util}^i(\bids) = U_i = 0.
    \end{align*}
    Therefore, the user social welfare is $0$.
    % Without loss of generality, assume $1\in S$ and let $S' = S\setminus\{1\}$.
    % Then by \cref{lem:hartline}, we have
    
    % Thus, there must exist a user $j$, whose utility is no more than $v\cdot \frac{k}{K} < u^*$ when there are $K+h$ number of users not including $j$.
    % By \cref{eqn:util-unchange}, we have that
    % \begin{align*}
    %     \underset{\bids_{-j}\sim\mathcal{D}^{K+h}}{\E}[{\sf util}^{j}(v_{j},\bids_{-j})] = \underset{\bids_{-j,1}\sim\mathcal{D}^{K+h-1}}{\E}[{\sf util}^{j}(v,v,\bids_{-j,1})] = \underset{\bids_{-j}\sim\mathcal{D}^{K+h-1}}{\E}[{\sf util}^{j}(v_{j},\bids_{-j})].
    % \end{align*}
    % Apply this argument for $K+h-L$ times,
    % \[\underset{\bids_{-j^*}\sim\mathcal{D}^{L}}{\E}[{\sf util}^{j^*}(v_{j^*},\bids_{-j^*})] = u^* > \underset{\bids_{-j}\sim\mathcal{D}^{K+h}}{\E}[{\sf util}^j(v_j,\bids_{-j})] = \underset{\bids_{-j}\sim\mathcal{D}^{L}}{\E}[{\sf util}^j(v_j,\bids_{-j})].\]
    % However, this contradicts \cref{lem:equal-util}.

\subsection{Feasibility for Approximate IC: Diluted Threshold-Based Mechanism}
\label{sec:diluted-threshold-appendix}
Although there is no interesting mechanism for strict incentive compatibility when $c\geq 2$, there are meaningful mechanisms if we allow approximate incentive compatibility.
Still, we assume that honest users’ values
are sampled i.i.d. from some bounded distribution  $\mathcal{D}$, and $m$ is the median of $\mathcal{D}$ such that $\Pr[x\geq m] = \frac{1}{2}$ (see \cref{rmk:nomedian}).
In addition, we assume that there is an upper bound $T$ on users' true values: $\Pr[x \leq T] = 1$.
Without loss of generality, we assume $T\geq \eps$.

\begin{mdframed}
    \begin{center}
    {\bf MPC-assisted, diluted threshold-based Mechanism}
    \end{center}
    \paragraph{Parameters:} the block size $k$, the environment parameter $(h,*,c,*)$, the approximation parameter $\epsilon$, the distribution median $\m$, 
    and the upper bound $T$ of the distribution.

    \paragraph{Input:} a bid vector $\bfb = (b_1,\dots,b_n)$.
    
    \paragraph{Mechanism:}
    \begin{itemize}[leftmargin=5mm,itemsep=1pt]
    \item 
    {\it Confirmation rule.} 
    Let $R := \max \left(2c \sqrt{\frac{kT}{\eps}}, k\right)$.
            Given a bid vector $\bids$, let $\widetilde{\bids} = (\widetilde{b}_1,\dots, \widetilde{b}_{s})$ denote the bids that are at least $m$.
            If $s\leq R$, randomly select $\frac{k}{R}\cdot s$ bids from $\widetilde{\bids}$ to confirm; otherwise, randomly select $k$ bids from $\widetilde{\bids}$ to confirm.
    \item 
    {\it Payment rule.} Every confirmed bid pays $\m$.
    \item 
    {\it Miner revenue rule.} 
    % Let $\bico:=n-d$ and $\target = \frac{h}{4}\cdot \frac{k}{R}\cdot \m$.
    % Let $\minerrev=(\target, \target,\dots, \target)^T \in\R^{c+1}$.
    % Define $M\in\R^{(c+1)\times(n+1)}$ to be the following matrix: 
    % \begingroup
    %     \renewcommand*{\arraystretch}{1.5}
    %     \begin{equation*}
    %     M = \frac{1}{2^{\bico}}
    %     \begin{pmatrix}
    %     \binom{\bico}{0} & \binom{\bico}{1} & \binom{\bico}{2} & \dots &  \binom{\bico}{\bico} & 0 & 0 & \dots & 0\\
    %     0 & \binom{\bico}{0} & \binom{\bico}{1} & \dots & \binom{\bico}{\bico-1} & \binom{\bico}{\bico} & 0 & \dots & 0\\
    %     \vdots & \vdots & \vdots & \vdots & \vdots & \vdots & \vdots & \vdots & \vdots \\
    %     0 & 0 & 0 & \dots & \dots & \dots & \binom{\bico}{\bico-2}& \binom{\bico}{\bico-1}&\binom{\bico}{\bico} \\
    %     \end{pmatrix}
    %     \end{equation*}
    % \endgroup
    % That is, the $i$-th row of $M$ is $\left(\binom{\bico}{0}, \binom{\bico}{1},  \dots,\binom{\bico}{\bico}, 0 \dots,0\right)$ right-shifted by $i-1$ elements.
    % Let ${\bf y}=(y_0,\dots,y_n)$ be the solution to the following linear program $\mathcal{P}$:
    % \begin{align}
    % \label{eqn:lp-diluted}
    %     &\MatM{\bf y} = \minerrev,\\
    %     \text{s.t. }& 0\leq y_i\leq \m\cdot \min\left(i\cdot\frac{k}{R},\,k\right) \text{ for all } i.\nonumber
    % \end{align}
    % Recall that $s$ is the number of bids in $\bids$ larger than or equal to $\m$. Miner gets $y_s\cdot \m$.
    If $s\geq \frac{h}{4}$, the total miner revenue is $\target := \m\cdot \min\left(\frac{h}{4}\cdot \frac{k}{R}, k\right)$. Otherwise, the total miner revenue is $0$.
    \end{itemize}
\end{mdframed}

\begin{theorem}
\label{thm:approx-posted-price-finite}
Suppose the block size is $k$.
For any $h\geq 1$, $c\geq 1$, and $\eps \geq \m\cdot \frac{h}{2}\cdot e^{-\frac{h}{16}}$,
the diluted threshold posted price auction satisfies strict ex post UIC, Bayesian $\eps$-MIC, and Bayesian $\eps$-SCP in an $(h,*,c,*)$-environment.
Moreover, the expected total miner revenue is $\m\cdot \min\left(\frac{h\sqrt{k\epsilon}}{8c\sqrt{T}}, \frac{h}{4}, k\right)$, where $T$ is the upper bound of users' true values. 
\end{theorem}

\begin{proof}
    We first show that the budget feasibility is satisfied. 
    Since the mechanism confirms $\min\left(s\cdot \frac{k}{R}, k\right)$ number of bids that are at least $\m$, the total payment is $\m\cdot \min\left(s\cdot \frac{k}{R}, k\right)$.
    When $s\geq \frac{h}{4}$, the total miner revenue is at most $\m\cdot \frac{h}{4}\cdot \frac{k}{R} \leq \m\cdot s\cdot \frac{k}{R}$, which is no more than the total payment of the users.
    Next, we prove UIC, MIC, and SCP separately.
    
    \paragraph{UIC.} Since the mechanism is posted price auction from each user's perspective, each user's best response is to follow the protocol honestly, as in the proof of \cref{thm:lp-finite}.
    
    \paragraph{MIC.} By the same reasoning as in \cref{thm:threshold-appendix}, by injecting fake bids, the miner can only increase its expected miner revenue if the number of bids that are at least $m$ from honest users is less than $\frac{h}{4}$.
    This happens with a probability at most $e^{-\frac{h}{16}}$.
    Thus, the expected total miner revenue increases by no more than 
    \[\target\cdot e^{-\frac{h}{16}}\leq \m\cdot e^{-\frac{h}{16}} \cdot \frac{h}{4}\leq \frac{\eps}{2}.\]
    
    \paragraph{SCP.} 
    By the same reasoning as in MIC, the expected increase of the miner revenue is at most $\eps/2$ by any deviation.
    Thus, to show that the mechanism is Bayesian $\eps$-SCP, it suffices to show that the coalition cannot increase the joint utility of the ``users'' in the coalition by more than $\frac{\eps}{2}$.
    
    % no matter how the coalition chooses their $d$ bids, they can only increase the total expected miner revenue  with a probability 
    % no more than $e^{-\frac{h}{16}}$.
    % Thus, the coalition can increase the utility of the colluding miners by at most $\eps/2$.
    % Henceforth, in the rest of the proof, we focus on showing that by deviating, the coalition cannot increase the joint utility of the users by more than $\frac{\eps}{2}$.

    % First, injecting bids does not increase the joint users' utility. 
    Because injecting bids smaller than $\m$ does not change the confirmation probability and the payment of each confirmed bid is fixed, injecting bids smaller than $\m$ does not  increase the users' utilities.
    On the other hand, injecting bids at least $\m$ will only decrease the probability of each colluding user getting confirmed, which does not increase the users' utilities.
    % Therefore, injecting fake bids does not increase the joint users' utility.

    Now, it suffices to show that overbidding and underbidding do not increase the coalition's joint utility since dropping out is equivalent to underbidding to some value less than $\m$.
    Let $s$ be the number of bids $\geq m$ when every user bids truthfully.
    % Suppose that when bidding honestly, the number of bids larger than or equal to $\m$ is $s$. 
    Each bid is confirmed with probability $\frac{k}{R}$ if $s\leq R$, and $\frac{k}{s}$ if $s > R$.
    Let $s'$ be the number of bids $\geq m$ when the colluding users bid strategically.
    % Assume that by bidding untruthfully, the coalition changes the number of bids larger than or equal to $\m$ to $s'$. \
    The colluding users can be partitioned into four groups:
    \begin{itemize}
        \item $S_1$: Those whose true values are less than $\m$ but overbid to values larger than or equal to $\m$;
        \item $S_2$: Those whose true values are less than $\m$ and bid values less than $\m$;
        \item $S_3$: Those whose true values are at least $\m$ but underbids to values less than $\m$;
        \item $S_4$: Those whose true values are at least $\m$ and still bid values at least $\m$.
    \end{itemize}

    % For each colluding user $i$, there are four possibilities:
    % \begin{enumerate}
    %     \item $i$'s true value is $< \m$ but $i$ overbids $b_i \geq \m$.
    %     \item $i$'s true value is $< \m$ but $i$ bids $b_i < \m$.
    %     \item $i$'s true value is $\geq \m$ but $i$ underbids $b_i < \m$.
    %     \item $i$'s true value is $\geq \m$ and $i$ bids $b_i \geq \m$.
    % \end{enumerate}
    % There are four groups of colluding users:
    When the coalition bids strategically, only the utilities of the users in $S_4$ increase compared to the honest case.
    Consider a colluding user in $S_4$ with the true value $v\geq m$.
    Its utility increases by at most
    \begin{align}
    \label{eqn:util-inc-dilut}
        (v - \m) \cdot \frac{k}{\max\{s',R\}} - (v - \m) \cdot \frac{k}{\max\{s,R\}}.
    \end{align}
    Note that \cref{eqn:util-inc-dilut} is positive only when $s' < s$ and $s > R$.
    In this case, \cref{eqn:util-inc-dilut} can be upper bounded by
    \begin{align*}
        &(v - \m) \left[\frac{k}{\max\{s',R\}} - \frac{k}{s}\right]
        \leq (T - \m) \left[\frac{k}{s'} - \frac{k}{s}\right]\\
        \leq & (T-\m) \cdot \frac{ck}{s(s-c)} & \text{By $s' \geq s-c$.}\\
        \leq & T \cdot \frac{ck}{R(R-c)}.
    \end{align*}
    Since by the choice of $R$, $R(R-c) \geq \frac{1}{2}R^2$, we have
    \[\text{Equation } \eqref{eqn:util-inc-dilut}\leq T\cdot \frac{2ck}{R^2}\leq \frac{\eps}{2c}.\]
    Therefore, by bidding untruthfully, each user's utility can increase by at most $\frac{\eps}{2c}$.
    Therefore, the joint utility of the users in the coalition by more than $\frac{\eps}{2}$.
    % In total, joint utility of the coalition can only increase by no more than $\eps$.
\end{proof}

% \section{Characterization for Approximate Incentive Compatibility --- Infinite Block Size}
% \label{sec:infinite}

\section{Bounds on Miner Revenue}
\label{sec:LB}

In this section, we prove bounds on the miner revenue under different settings.
Henceforth, let $\mu(\bids)$ denote the 
expected total miner revenue when the input bid vector is $\bids$, where the expectation
is taken over the mechanism's randomness.

%\subsection{Optimality of $\Theta(h)$-Miner Revenue in the Known-$h$ Model}
\subsection{Known-$h$ Model}
\label{sec:lb-strict}

In this section, we prove 
limits on miner revenue in the known-$h$ model (i.e., \Cref{thm:intro-limit-rev}).

\label{section:strict-IC-known-h}

% \begin{lemma}
% 	\label{lem:miner-eps-UIC}
% 	Fix any $\rho\in(0,1]$ and any natural number $h$.
% 	For any (possibly randomized) MPC-assisted TFM in an $(h,\rho,1,*)$-environment that is 
% 	Bayesian $\eps_u$-UIC and 
% 	Bayesian $\eps_s$-SCP,
% 	for any user $i$, for any value $r$ and for any natural number $n > h$, it must be that
% 	\begin{equation}
%         \underset{\bids_{-i}\sim\mathcal{D}^{n-1}}{\E}[\mu(\bids, r)]
%         -
%         \underset{\bids_{-i}\sim\mathcal{D}^{n-1}}{\E}[\mu(\bids, 0)] \leq
% 		% \mu(\bids,r) - \mu(\bids,0) \leq
% 		\begin{cases}
% 		\frac{2}{\rho}(\epsilon_s+\eps_u), &\text{if } r\leq \epsilon_s+\eps_u;\\
% 		\frac{2}{\rho}(\sqrt{r(\eps_s+\eps_u)}), &\text{if } r>\epsilon_s+\eps_u.
% 		\end{cases}
% 	\end{equation}
% \end{lemma}
% \begin{proof}
% When $n > h$, at least one bid is possible to be submitted by a colluding user.
% The proof of Lemma 3.3 in \cite{crypto-tfm} only relies on the fact 
% The proof is the same as the proof of Lemma 3.3 in \cite{crypto-tfm}.
% \end{proof}

\begin{theorem}[Limit on miner revenue for approximate incentive compatibility, 
\cref{thm:theta-h-minerrev-informal} restated]
\label{thm:revenue-approxIC}
Fix any $h \geq 1$, $d\geq c \geq 1$, and $\rho \in (0, 1)$.
Given any MPC-assisted mechanism 
that is Bayesian $\eps_u$-UIC, Bayesian $\eps_m$-MIC and Bayesian $\eps_s$-SCP 
in an $(h,\rho,c,d)$-environment, 
for all $n \geq h$,
it must be that
\begin{equation}\underset{\bids\sim\mathcal{D}^{n}}{\E}[\mu(\bids)]\leq  h \cdot \E(\mcal{D})
+  \frac{2(n-h)}{\rho}\left(\epsilon+C_{\mcal{D}}\sqrt{\epsilon}\right),
\label{eqn:minerrevlimit}
\end{equation}
where $\eps = \eps_u + \eps_m+ \eps_s$, 
$\E(\mcal{D}) = \E_{X\sim\mcal{D}}[X]$ and $C_{\mcal{D}} = \E_{X\sim\mcal{D}}[\sqrt{X}]$ are the terms
	that depend on the ``scale'' of the distribution $\mcal{D}$.

As a special case, for strict incentive compatibility where $\epsilon = 0$, 
we have that 
\begin{equation}\underset{\bids\sim\mathcal{D}^{n}}{\E}[\mu(\bids)]
\leq  h \cdot \E(\mcal{D}) %C_{\mcal{D}}  
%+  \frac{2(n-h)}{\rho}\left(\epsilon+C'_{\mcal{D}}\sqrt{\epsilon}\right),
\end{equation}

\end{theorem}

To prove 
\Cref{thm:revenue-approxIC}, we need 
the following lemma.

\begin{lemma}[Lemma 3.3 of \cite{crypto-tfm}]
    \label{lem:miner-lb-strict}
    Fix any $h\geq 1$, $d\geq c\geq 1$ and $\rho\in(0,1)$. 
    Given any (possibly randomized) MPC-assisted TFM 
that is Bayesian $\eps_u$-UIC and Bayesian $\eps_s$-SCP
in an $(h,\rho,c,d)$-environment, for any user $i$ and any value $v$, for any $\ell \geq h$ it must be that
    \begin{equation}
    \label{eq:miner-lb-strict}
    \underset{\bids\sim \mathcal{D}^{\ell}}{\E}\left[\mu(\bids, v)\right] - \underset{\bids\sim \mathcal{D}^{\ell}}{\E}\left[\mu(\bids, 0)\right]\leq
        \begin{cases}
        \frac{2}{\rho}(\epsilon_s+\eps_u), &\text{if } v \leq \epsilon_s+\eps_u;\\
        \frac{2}{\rho}(\sqrt{v(\eps_s+\eps_u)}), &\text{if } v > \epsilon_s+\eps_u.
        \end{cases}
    \end{equation}
\end{lemma}
\begin{proof}
\cite{crypto-tfm} stated for any $\rho \in (0,1)$ and any $c \geq 1$,
given any mechanism 
that is Bayesian $\eps_u$-UIC and Bayesian $\eps_s$-SCP in an $(*,\rho,c,*)$-environment, 
for any user $i$ and any value $v$, for any $\ell \geq 0$,
\cref{eq:miner-lb-strict} must hold.
\elaine{FILL in their lemma using our language}

In their proof, they showed that if for some $\ell$ and some $v$, \cref{eq:miner-lb-strict} is violated, and moreover assuming that the mechanism satisfies Bayesian $\eps_s$-SCP, then there exists some value $v'$ such that if a user of value $v'$ colludes with a subset of the miners, the coalition can play strategically and jointly benefit when the rest of the world contains $\ell$ honest users. 
We can apply the exactly same argument, but because it is promised that there are at least $h$ honest users, the argument only holds for $\ell \geq h$. 
Observe also the coalition's strategy requires only $d=1$.

%stated the same lemma for 
%MPC-assisted mechanisms that work in $(*, \rho, c, *)$-environments
%for any $\rho \in (0, 1)$ and any $c \geq 1$. 
%One can verify that their 
%exact proof holds for MPC-assisted mechanisms in $(h, \rho, c, d)$-environment.
\end{proof}

Now, we are ready to prove \Cref{thm:revenue-approxIC}.
\begin{proof}[Proof of \Cref{thm:revenue-approxIC}]
The proof mainly follows the proof of Theorem 3.4 in \cite{crypto-tfm}.
% We will prove the case for approximate incentive compatibility.
% Then, by taking $\eps = 0$, we obtain the bound for the strict incentive compatibility.
Since the TFM is Bayesian $\eps_m$-MIC in an $(h,\rho,c,d)$-environment, it must be that
for any $\ell \geq h$, 
\begin{equation}
    \underset{\bids\sim\mathcal{D}^{\ell}}{\E}[\rho\mu(\bids, 0 )]\leq \underset{\bids\sim\mathcal{D}^{\ell}}{\E}[\rho\mu(\bids)]+\eps_m.
    \label{eqn:eps-MIC}
\end{equation}
Otherwise, %a $\rho$-sized 
% the strategic miner can collude with any user, and ask the user to inject a bid $0$ and increase its miner revenue by 
% strictly more than $\eps_s$, while it does not need pay anything for injecting this $0$-bid. This violates Bayesian $\eps_s$-SCP. 
when there are $\ell$ honest bids,  
a strategic $\rho$-sized miner coalition 
can inject a bid $0$ and increase its miner revenue by strictly more than $\eps_m$, while it does not need to pay anything for injecting this $0$-bid. 
This violates Bayesian $\eps_m$-MIC.

Let $f(\cdot)$ be the p.d.f.~of distribution $\mcal{D}$. 
By the law of total expectation, for all $\ell \geq 1$, we have
\[\quad\underset{\bids\sim\mcal{D}^{\ell}}{\E} [\mu(\bids)]= \int_{0}^{\infty} \underset{\bids'\sim\mcal{D}^{\ell-1}}{\E} [\mu(\bids', r)] f(r) dr.\]
Let $\eps' = \eps_s+\eps_u$. 
Since the mechanism is Bayesian $\eps_u$-UIC and Bayesian $\eps_s$-SCP in an $(h,\rho,c,d)$-environment, by~\cref{lem:miner-lb-strict}, for all $\ell \geq h + 1$, it must be that 
\iffullversion
\begin{align*}
    \int_{0}^{\eps'} \underset{\bids'\sim\mcal{D}^{\ell-1}}{\E} [\mu(\bids', r)] f(r) dr &\leq \int_{0}^{\eps'} \left[\underset{\bids'\sim\mcal{D}^{\ell-1}}{\E} [\mu(\bids', 0)]+\frac{2\eps'}{\rho}\right]f(r)dr;\\
    \int_{\eps'}^{\infty} \underset{\bids'\sim\mcal{D}^{\ell-1}}{\E} [\mu(\bids', r)] f(r) dr &\leq \int_{\eps'}^{\infty} \left[\underset{\bids'\sim\mcal{D}^{\ell-1}}{\E} [\mu(\bids', 0)]+\frac{2\sqrt{r\eps'}}{\rho}\right]f(r)dr.
\end{align*}
\else
\begin{align*}
    &\int_{0}^{\eps'} \underset{\bids'\sim\mcal{D}^{\ell-1}}{\E} [\mu(\bids', r)] f(r) dr\\
    \leq &\int_{0}^{\eps'} \left[\underset{\bids'\sim\mcal{D}^{\ell-1}}{\E} [\mu(\bids', 0)]+\frac{2\eps'}{\rho}\right]f(r)dr;\\
    \text{and }&\int_{\eps'}^{\infty} \underset{\bids'\sim\mcal{D}^{\ell-1}}{\E} [\mu(\bids', r)] f(r) dr\\
    \leq& \int_{\eps'}^{\infty} \left[\underset{\bids'\sim\mcal{D}^{\ell-1}}{\E} [\mu(\bids', 0)]+\frac{2\sqrt{r\eps'}}{\rho}\right]f(r)dr.
\end{align*}
\fi
Summing up the two inequalities above, we can bound the expected miner revenue with
\begin{align*}
    &\quad\underset{\bids\sim\mcal{D}^{\ell}}{\E} [\mu(\bids)]\\ 
    &= \int_{0}^{\eps'} \underset{\bids'\sim\mcal{D}^{\ell-1}}{\E} [\mu(\bids', r)] f(r) dr +  \int_{\eps'}^{\infty} \underset{\bids'\sim\mcal{D}^{\ell-1}}{\E} [\mu(\bids', r)] f(r) dr \\
    &\leq \int_{0}^{\eps'} \left[\underset{\bids'\sim\mcal{D}^{\ell-1}}{\E} [\mu(\bids', 0)]+\frac{2\eps'}{\rho}\right]f(r)dr\\
    &\qquad+\int_{\eps'}^{\infty} \left[\underset{\bids'\sim\mcal{D}^{\ell-1}}{\E} [\mu(\bids', 0)]+\frac{2\sqrt{r\eps'}}{\rho}\right]f(r)dr\\
    &\leq \underset{\bids'\sim\mcal{D}^{\ell-1}}{\E} [\mu(\bids', 0)]+\frac{2\eps'}{\rho}\int_{0}^{\eps'} f(r) dr + \frac{2\sqrt{\eps'}}{\rho} \int_{\eps'}^{\infty}\sqrt{r}f(r)dr
\end{align*}
By~\eqref{eqn:eps-MIC}, we have that $\underset{\bids'\sim\mcal{D}^{\ell-1}}{\E} [\mu(\bids',0)]\leq \underset{\bids'\sim\mcal{D}^{\ell-1}}{\E} [\mu(\bids')] + \frac{\eps_s}{\rho}$.
Therefore, for all $\ell \geq h + 1$,
\begin{align}
     &\quad\underset{\bids\sim\mcal{D}^{\ell}}{\E} [\mu(\bids)]\nonumber\\ 
     & \leq \underset{\bids'\sim\mcal{D}^{\ell-1}}{\E} [\mu(\bids', 0)]+\frac{2\eps'}{\rho}\int_{0}^{\eps'} f(r) dr + \frac{2\sqrt{\eps'}}{\rho} \int_{\eps'}^{\infty}\sqrt{r}f(r)dr\nonumber\\
     & \leq \underset{\bids'\sim\mcal{D}^{\ell-1}}{\E} [\mu(\bids')]+\frac{\eps_s}{\rho}+\frac{2\eps'}{\rho}+\frac{2\sqrt{\eps'}}{\rho}\E_{X\sim\mcal{D}}[\sqrt{X}]\nonumber\\
     &\leq \underset{\bids'\sim\mcal{D}^{\ell-1}}{\E} [\mu(\bids')]+\frac{2\epsilon}{\rho}+\frac{2C_{\mcal{D}}\sqrt{\epsilon}}{\rho},
	 \label{eq:induction}
\end{align}
where the last step comes from the fact that $\eps = \eps_u + \eps_m + \eps_s$.

Finally, for any bid vector $\bids$ such that at least $h$ bids submitted by honest users, we can represent $\bids$ as $\bids = \bids_H + \bids_{-H}$, where $\bids$ are submitted all from honest users and $|\bids_H| = h$.
Notice that $\bids_{-H}$ might or might not contain the bids submitted by honest users.
With this notation, for all $\ell \geq h$, we can write \[
	\underset{\bids\sim\mcal{D}^{\ell}}{\E} [\mu(\bids)] = \underset{\bids_H\sim\mcal{D}^{h}, \bids_{-H}\sim\mcal{D}^{\ell-h}}{\E} [\mu(\bids_H + \bids_{-H})].
\]
By applying \cref{eq:induction} $(\ell - h)$ times, we have \[
	\underset{\bids\sim\mcal{D}^{\ell}}{\E} [\mu(\bids)] \leq \underset{\bids_H\sim\mcal{D}^{h}}{\E} [\mu(\bids_H)] + \frac{2(\ell-h)}{\rho} \left(\epsilon + \sqrt{\epsilon} \cdot \E_{X\sim\mcal{D}}[\sqrt{X}] \right).
\]
Because honest users always submit their true values and the miner revenue is bounded by the sum of the payments, if the bids are submitted by $h$ honest users, it must be \[
	\underset{\bids\sim\mathcal{D}^{h}}{\E}[\mu(\bids)]\leq  h \cdot \E_{X\sim\mcal{D}}[X].
\]
Therefore, we conclude that \[
	\underset{\bids\sim\mcal{D}^{\ell}}{\E} [\mu(\bids)] \leq h \cdot \E_{X\sim\mcal{D}}[X] + \frac{2(\ell-h)}{\rho} \left(\epsilon + \sqrt{\epsilon} \cdot \E_{X\sim\mcal{D}}[\sqrt{X}] \right).
\]
% The theorem follows by induction on $n$, where in each induction step we repeat the argument above.
\end{proof}

% \begin{corollary}
%     Fix any 
%     $h \geq 1$, $d\geq c \geq 1$, and $\rho \in (0, 1)$.
%     No MPC-assisted TFM 
%     that simultaneously satisfies 
%     Bayesian UIC, MIC, and SCP 
%     in an $(h, \rho, c, d)$-environment
%     can achieve more than $h \cdot \E(\mcal{D})$
%     expected miner revenue where $\E(\mcal{D})$
%     denotes the expectation of the bid distribution $\mcal{D}$.
%     % \hao{Do we need MIC here?}
% \end{corollary}

\subsection{Necessity of Bayesian Incentive Compatibility}
\label{sec:expost}
If we insist on ex post incentive compatibility, the known-$h$ model will not help us 
overcome 
the limit on miner revenue 
for mechanisms that are universal in $h$ in \cite{crypto-tfm}.

\begin{theorem}
\label{cor:lb-expost}
Fix any $h\geq 1$, $d\geq c\geq 1$ and $\rho\in(0,1)$.
Given any (possibly randomized) MPC-assisted TFM that is ex post $\eps_u$-UIC and ex post $\eps_s$-SCP in an $(h,\rho,c,d)$-environment,
it must be that for any $\bids=(b_1,\dots,b_n)$ of length $n>h$,
\begin{equation}
    \mu(\bids)\leq\frac{2n\eps}{\rho}+\frac{2\sqrt{\eps}}{\rho}\sum_{i=1}^n \sqrt{b_i},
\label{eqn:minerrevlimit-expost}
\end{equation}
where $\eps = \eps_s + \eps_u$.

As a special case, for strict incentive compatibility where $\epsilon = 0$, 
we have that for any bid $\bids$,
\[\mu(\bids)=0.\]
\end{theorem}
\begin{proof}
In the proof we make use of the following lemma.
\begin{lemma}
\label{lem:expost}
Fix any $h\geq 1$, $d\geq c\geq 1$ and $\rho\in(0,1)$.
Given any (possibly randomized) MPC-assisted TFM that is ex post $\eps_u$-UIC and ex post $\eps_s$-SCP in an $(h,\rho,c,d)$-environment,   
for any value $v$ and any $\bids$ of length at least $h$, it must be that
    \begin{equation}
    \mu(\bids, v) - \mu(\bids, 0) \leq
    \begin{cases}
    \frac{2}{\rho}(\epsilon_s+\eps_u), &\text{if } v\leq \epsilon_s+\eps_u\\
    \frac{2}{\rho}(\sqrt{v(\eps_s+\eps_u)}), &\text{if } v>\epsilon_s+\eps_u.
    \end{cases}
    \label{eqn:miner-rev-expost}
\end{equation}
\end{lemma}
For now, we assume that the lemma holds, and we show how the theorem follows.
The proof to \cref{lem:expost} appears afterward.
For any $\bids = (b_1,\dots,b_n)$ of length $n> h$, it must be that
    \iffullversion
    \begin{align*}
        &\mu(\bids) =\mu(b_1,b_2,\dots,b_n)\\
        \leq& \mu(b_1,\dots,b_{n-1},0) + \frac{2}{\rho}\eps+\frac{2}{\rho}\sqrt{b_n\epsilon} &\text{By \cref{lem:expost}}\\
        \leq& \mu(b_1,\dots,b_{n-2},0,0) + \frac{4}{\rho}\eps+\frac{2}{\rho}\sqrt{b_n \eps}+\frac{2}{\rho}\sqrt{b_{n-1} \eps}\\
        \leq& \dots \leq \mu(0\dots,0) + \frac{2n\eps}{\rho}+\frac{2\sqrt{\eps}}{\rho}\sum_{i=1}^n \sqrt{b_i}\\
        \leq& \frac{2n\eps}{\rho}+\frac{2\sqrt{\eps}}{\rho}\sum_{i=1}^n \sqrt{b_i}.
    \end{align*}
    \else
    \begin{align*}
        &\mu(\bids)
        = \mu(b_1,b_2,\dots,b_n)\\
        \leq& \mu(b_1,\dots,b_{n-1},0) + \frac{2}{\rho}\eps+\frac{2}{\rho}\sqrt{b_n\epsilon} &\text{By \cref{lem:expost}}\\
        \leq& \mu(b_1,\dots,b_{n-2},0,b_{n-1}) + \frac{2}{\rho}\eps+\frac{2}{\rho}\sqrt{b_n\epsilon} &\text{By symmetry}\\
        \leq& \mu(b_1,\dots,b_{n-2},0,0) + \frac{4}{\rho}\eps+\frac{2}{\rho}\sqrt{b_n \eps}+\frac{2}{\rho}\sqrt{b_{n-1} \eps}\\
        \leq& \dots \leq \mu(0\dots,0) + \frac{2n\eps}{\rho}+\frac{2\sqrt{\eps}}{\rho}\sum_{i=1}^n \sqrt{b_i}\\
        \leq& \frac{2n\eps}{\rho}+\frac{2\sqrt{\eps}}{\rho}\sum_{i=1}^n \sqrt{b_i}.
    \end{align*}
    \fi
\end{proof}

\paragraph{Proof of \cref{lem:expost}.}
The proof is similar to the proof of Lemma 3.3 in \cite{crypto-tfm}, while we consider the ex post setting here.
To prove this lemma, we introduce the following notations.
For the $i$-th user, we define
$x_i(\bids,v)$ to be the probability of bid $v$ being confirmed, and $p_i(\bids,v)$ to be the expected payment of bid $v$, when user $i$ bids $v$ and other users bid $\bids$.
\begin{lemma}
\label{lem:expost-sandwich}
Fix any $h\geq 1$, $d\geq c\geq 1$ and $\rho\in(0,1)$.
Given any (possibly randomized) MPC-assisted TFM that is ex post $\eps_u$-UIC and ex post $\eps_s$-SCP in an $(h,\rho,c,d)$-environment, 
it must be that 
for any bid vector $\bids$ of length at least $h$, 
for any user $i$,
for any $y \leq z$, 
\begin{align}
    &z\cdot[x_i(\bids,z)-x_i(\bids,y)]+\eps\nonumber\\
    \geq& p_i(\bids,z)- p_i(\bids,y)\nonumber\\
    \geq& y\cdot [x_i(\bids,z)-x_i(\bids,y)]-\epsilon.
    \label{eqn:payment-sandwich}
\end{align}
\end{lemma}
\begin{proof}
The proof is similar to the proof of Myerson's Lemma. 
For any bid vector $\bids$ submitted by all users other than user $i$, 
user $i$'s utility is $v\cdot x_i(\bids, r) - p_i(\bids, r)$ if its true value is $v$ and its bid is $r$.
Because the mechanism is $\eps_u$-UIC in an $(h,\rho,c,d)$-environment, for all $\bids$ such that $|\bids| \geq h$, it must be that
\[z\cdot x_i(\bids,z) - p_i(\bids,z)+\epsilon\geq z\cdot x_i(\bids,y) - p_i(\bids,y).\]
Otherwise, if user $i$'s true value is $z$, bidding $y$ can bring it strictly more than $\epsilon$ utility compared to bidding truthfully, which contradicts $\eps$-UIC of the mechanism in $\environ$-environment. 
By the same reasoning, for all $\bids$ such that $|\bids| \geq h$, we have
\[y\cdot x_i(\bids,y) - p_i(\bids,y)+\epsilon \geq y\cdot x_i(\bids,z) - p_i(\bids,z).\]
The lemma thus follows by combining these two inequalities.
\end{proof}
\begin{lemma}
\label{lem:miner-step-eps-UIC}
Fix any $h\geq 1$, $d\geq c\geq 1$ and $\rho\in(0,1)$.
Given any (possibly randomized) MPC-assisted TFM that is ex post $\eps_u$-UIC and ex post $\eps_s$-SCP in an $(h,\rho,c,d)$-environment,    
it must be that 
for any user $i$,
for any bid vector $\bids$ of length at least $h$, for any $y \leq z$,
\begin{equation}
    \mu(\bids, z) - \mu(\bids, y)\leq \frac{1}{\rho}(\eps_u + \eps_s + S_{\bids}(y,z)),
    \label{eqn:eps-UIC-miner-revenue}
\end{equation}
where $S_{\bids}(y,z) = (z-y)[x_i(\bids,z)-x_i(\bids,y)]$.
\end{lemma}
\begin{proof}
The utility of user $i$ is $v\cdot x_i(\bids,r)-p_i(\bids,r)$ if its true value is $v$ and it bids $r$, while the rest of the world bids $\bids$ of length at least $h$. 
Henceforth, we fix an arbitrary bid vector $\bids$ such that $|\bids| \geq h$.
Imagine that the user $i$'s true value is $y$. 
If user $i$ overbids $z > y$ instead of its true value $y$, then its expected utility decreases by
\begin{align*}
    \Delta &= y\cdot x_i(\bids,y)-p_i(\bids,y) - [y\cdot x_i(\bids,z)-p_i(\bids,z)]\\
    &= -y\cdot[x_i(\bids,z) - x_i(\bids,y)] + (p_i(\bids,z)-p_i(\bids,y))\\
    &\leq -y\cdot[x_i(\bids,z) - x_i(\bids,y)] + z\cdot[x_i(\bids,z) - x_i(\bids,y)] +\eps_u &\text{By
\cref{lem:expost-sandwich}}\\
    &= (z-y)\cdot[x_i(\bids,z)-x_i(\bids,y)] + \eps_u = S_{\bids}(y,z) + \eps_u.
\end{align*}
By $\epsilon_s$-SCP, 
it must be that $\rho\mu(\bids,z) - \rho\mu(\bids,y)\leq \Delta+\eps_s$; otherwise, a user $i$ with true value $y$ can collude with a set of miners and overbid $z$ instead of its true value $y$, while the rest honest users bid $\bids$.
This strategy involves only one colluding user and one bid controlled by the coalition, and it increases the coalition's utility by 
strictly more than $\eps_s$ compared to the honest strategy.
This contradicts $\eps_s$-SCP of the mechanism in the $\environ$-environment.
\end{proof}
Now we proceed to prove \cref{lem:expost}. 
Let $\eps' = \eps_s+\eps_u$. 
For any user $i$,
fix an arbitrary $\bids$ of length at least $h$ from other users.
Consider the following two cases.
\begin{itemize}
\item 
\textbf{Case 1: If $v\leq \epsilon'$.} In this case, by Lemma~\ref{lem:miner-step-eps-UIC}, we have that
\[\mu(\bids,v)-\mu(\bids,0)\leq \frac{1}{\rho}\left(\eps_u+\eps_s+S_{\bids}(0,v)\right)\leq\frac{1}{\rho}\left( \epsilon_u+\eps_s+v\right)\leq \frac{2\eps'}{\rho}.\]

\item 
\textbf{Case 2: If $v > \epsilon'$.} We choose a sequence of points that partitions the interval $[0,v]$ as follows.
Let $L = \lfloor\sqrt{\frac{v}{\epsilon'}}\rfloor$.
Set $v_0 = 0$ and $v_{L+1} = v$.
For $l = 1,\dots,L$, we set $v_l = l\cdot\sqrt{v\epsilon'}$.
Each segment except the last one is of length $\sqrt{v\epsilon'}$, while the last one has a length no more than $\sqrt{v\epsilon'}$.

Now we proceed to bound $\mu(\bids,v) - \mu(\bids,0)$. 
Note that 
\begin{align*}
    &\mu(\bids,v) - \mu(\bids,0)\\ 
    =& \sum_{l=0}^{L} [\mu(\bids,v_{l+1}) - \mu(\bids,v_{l})]\\
    \leq& \sum_{l=0}^L\frac{1}{\rho}[\eps'+S_{\bids}(v_l,v_{l+1})] &\text{By~\cref{lem:miner-step-eps-UIC}}\\
    =& \frac{L\eps'}{\rho} + \frac{1}{\rho}\sum_{l=0}^L(v_{l+1}-v_{l})\cdot[x_i(\bids,v_{l+1})-x_i(\bids,v_l)]\\
    \leq&  \frac{L\eps'}{\rho} + \frac{1}{\rho}\sqrt{r\eps'}\sum_{l=0}^L[x_i(\bids,v_{l+1})-x_i(\bids,v_l)] &\text{By choice of }v_l\\
    \leq& \frac{L\eps'}{\rho} + \frac{1}{\rho}\sqrt{v\epsilon'} &\text{By }x_i(\bids,v)\leq 1
\end{align*}
Since $L = \lfloor\sqrt{\frac{v}{\epsilon'}}\rfloor\leq \sqrt{\frac{v}{\epsilon'}}$, we have that
\[\mu(\bids,v) - \mu(\bids,0)\leq \frac{2\sqrt{v\epsilon'}}{\rho}.\]
\Cref{lem:expost} thus follows.
\end{itemize}

\subsection{Honest Majority of Bids}
\label{section:revenue-bound-honest-majority}

\ignore{
Conceptually, the proof in \cite{crypto-tfm} only focuses on a coalition between miners and a single user, and the miner should not extract any revenue from the colluding user's bid.
In other words, the miner revenue can only come from the bids which is guaranteed to be submitted by honest users.
However, given a bid vector of length $n$, the mechanism cannot differentiate the following two cases: 1) it comes from $n$ honest users; 2) it comes from $n-1$ honest users and a colluding user.
Thus, the miner revenue of the first case should not exceed the second case.
Applying the argument above repeatedly, we conclude that the miner revenue is always upper bounded by the case where $n = 2$, and it comes from one honest user and one colluding user.
The concept is formally stated as the following theorem.
}

As mentioned in 
\Cref{sec:intro}, 
we consider a ``sufficient honesty'' assumption but the precise
statement of the assumption matters. 
In particular, had we assumed that the majority of bids are submitted by honest
users (referred to as the ``honest majority bids'' assumption), then 
we would not be able to overcome the severe limitation on miner revenue.
The theorem below states that under the honest majority bids
assumption, we should still suffer from a constant miner revenue limitation.

\begin{theorem}
% Let $\Pi$ be any MPC-assisted TFM which takes an arbitrary number of bids as input.
Assuming that a majority number of bids are submitted by honest users.
If a mechanism is Bayesian UIC, Bayesian MIC, and Bayesian SCP 
(even for $c = 1$)
under the ``honest majority bids'' assumption, it must be that $\underset{\bids\sim\mcal{D}^{n}}{\E} [\mu(\bids)] \leq 2\E[\mcal{D}]$.
\hao{Strict IC case.}
\end{theorem}
\begin{proof}
The proof is based on the following lemma.
\begin{lemma}
\label{lem:inject-induction}
Assuming that a majority of bids are submitted by honest users.
Suppose that a TFM is Bayesian UIC, Bayesian MIC, and Bayesian SCP (even for $c = 1$) 
under the ``honest majority bids'' assumption. Then, as long
as the number of bids $n\geq 3$, it must be that
	$\underset{\bids\sim\mcal{D}^{n}}{\E} [\mu(\bids)] \leq \underset{\bids'\sim\mcal{D}^{n-1}}{\E} [\mu(\bids')]$.
 %even if the miner colludes with only one user.
\end{lemma}
For now assume that \cref{lem:inject-induction} holds, and we will show how the theorem follows.
The proof of \cref{lem:inject-induction} appears afterwards.
By induction on $n$, for any $n\geq 3$, 
\[
\underset{\bids\sim\mcal{D}^{n}}{\E} [\mu(\bids)] \leq \underset{\bids'\sim\mcal{D}^{2}}{\E} [\mu(\bids')]\leq 2\E[\mcal{D}],
\]
where the last inequality follows from the fact that the miner 
revenue must be upper bounded by the bids.
%In the above, essentially we are removing the bids one by one until $n = 2$. 
%At this moment, under the ``honest majority bids'' assumption, it must be that
%both bids are submitted by honest users. 

% Notice that \cref{lem:inject-induction} holds whenever the miner is able to collude with one user.
% Since it is only guaranteed that at least half of the input bids are submitted by honest users, when $n \geq 2$, at least one bid might be submitted by a colluding user.
% Thus, we can apply \cref{lem:inject-induction} repeatedly, and we obtain that for all $n \geq 2$, \[
% 	\underset{\bids\sim\mcal{D}^{n}}{\E} [\mu(\bids)] \leq \underset{\bids'\sim\mcal{D}^{(1)}}{\E} [\mu(\bids')].
% \]
% Because honest users always submit their true values and the miner revenue is bounded by the sum of the payments, we have \[
% 	\underset{\bids\sim\mcal{D}^{n}}{\E} [\mu(\bids)] \leq \underset{\bids'\sim\mcal{D}^{(1)}}{\E} [\mu(\bids')] \leq \underset{\bids'\sim\mcal{D}}{\E} [\bids'].
% \]
\end{proof}

\begin{proof}[Proof of \cref{lem:inject-induction}]
    Fix a mechanism that is Bayesian UIC, MIC and SCP under the ``honest majority bids'' assumption.
    We first prove that for any $n\geq 3$, for any value $v$, it must be that
    \begin{equation}
        \underset{\bids\sim\mcal{D}^{n-1}}{\E} [\mu(\bids, v)] = \underset{\bids\sim\mcal{D}^{n-1}}{\E} [\mu(\bids, 0)].
        \label{eqn:miner-rev-honest-majority}
    \end{equation}
%    In other words,  when one user changes its bid to $0$, the miner revenue should not change, assuming all other bids are sampled from the distribution $\mcal{D}$.
The proof of this is similar to that of 
Lemma C.4 of \cite{crypto-tfm}.
Suppose that the number of bids is some arbitrary $n$.
\cite{crypto-tfm}
show that if \Cref{eqn:miner-rev-honest-majority} does not hold, 
that is, if the expected miner revenue changes when some specific user 
lowers its bid from $v$ to $0$, 
and moreover, assuming that the mechanism satisfies Bayesian UIC, 
then, it must be that there exists some $v'$ such that 
a coalition involving 
 a single user whose true value is $v'$  
and a subset of the miners can
play strategically to benefit themselves (when there are $n-1$ other honest bids). 
Their proof still holds here under the ``honest majority bids''
assumption as long as $n\geq 3$, since when $n\geq 3$, there 
can be at least one  
strategic user 
colluding with the miners.

\elaine{TODO: add utility sharing among miners in roadmap}

\ignore{
    Note that when $n\geq 3$, the number of honest bids $n-1$ is strictly larger than half.
    Therefore, if the mechanism is Bayesian UIC and SCP in the ``honest majority bids'' assumption,
    \cref{eqn:miner-rev-honest-majority} must hold by the same proof as in Lemma C.4 of \cite{crypto-tfm}.
}
    Now, for any $n\geq 3$, we have
    \begin{align*}
    \underset{\bids\sim\mcal{D}^{n}}{\E} [\mu(\bids)] &= \int_{0}^{+\infty} \underset{\bids'\sim\mcal{D}^{n-1}}{\E} [\mu(\bids', r)] f(r) dr & \\
    &=\int_{0}^{+\infty} \underset{\bids'\sim\mcal{D}^{n-1}}{\E} [\mu(\bids', 0)] f(r) dr &\text{By \cref{eqn:miner-rev-honest-majority}}\\
    &=\underset{\bids'\sim\mcal{D}^{n-1}}{\E} [\mu(\bids', 0)].
\end{align*}
By Bayesian MIC, it must be that $\underset{\bids'\sim\mcal{D}^{n-1}}{\E} [\mu(\bids', 0)]\leq \underset{\bids'\sim\mcal{D}^{n-1}}{\E} [\mu(\bids')]$.
Otherwise, when there are $n-1$ honest bids, 
the miners can inject a $0$ and increase the miner revenue while it does not need to pay anything for injecting the $0$-bid. 
This violates MIC.  
Notice that since $n-1 \geq 2$, 
the miners injecting one bid 
does not violate the ``honest majority bids'' assumption.
%under the ``honest majority bids'' assumption since $n-1$ bids are honest.
\end{proof}

\section*{Acknowledgments}
This work is in part supported by NSF awards 2212746, 2044679, 1704788, a Packard Fellowship, a generous gift from the late Nikolai Mushegian, a gift from Google, and an ACE center grant from Algorand Foundation. 
The authors would like to thank the anonymous reviewers for their helpful comments.
We also thank Matt Weinberg
for helpful technical discussions regarding how to 
efficiently instantiate our MPC-assisted mechanisms.

\newpage
\bibliographystyle{alpha}
\bibliography{refs,crypto,gametheory}
\clearpage

\appendix

\end{document}